\documentclass[12pt,a4paper]{article}

\usepackage{amsmath,amssymb,xcolor,pstool,tikz,tensor,amsthm,mathtools,soul,braket,amsbsy}
\usepackage{wasysym}

\usepackage{mathtools}
\usepackage{amsfonts}
\usepackage{graphicx}
\usepackage{color}
\usepackage{subcaption}
\usepackage{booktabs}
\usepackage{inputenc}
\usepackage[T1]{fontenc}
\usepackage{mathrsfs}
\usepackage{enumerate}
\usepackage{xcolor,url}
\usepackage{cancel,dsfont}
\usepackage{tikz-3dplot}
\usepackage[normalem]{ulem}
\usetikzlibrary{positioning, arrows.meta, decorations.markings, calc, intersections,patterns}
\tikzset{arrowmark/.style={postaction={decorate, decoration={markings, mark=at position #1 with {\arrow{>};}}}},
    arrowmark/.default={.5}
}

\usepackage{cite}
\usepackage{physics}

\usetikzlibrary{decorations.markings,positioning}
\usetikzlibrary{decorations.pathmorphing}
\usetikzlibrary{decorations.markings,positioning}

\usepackage[colorlinks = true,
            linkcolor = blue!70!black,
            urlcolor  = red!70!black,
            citecolor = green!55!black,
            anchorcolor = blue!70!black,bookmarks]{hyperref}

\usepackage[top=0.5in, bottom=0.6in, left=0.5in, right=0.5in]{geometry}

\usepackage{cleveref}

\numberwithin{equation}{section}

\graphicspath{{figures/}}

\usepackage{tikz}
\usepackage{esint}

\makeatletter

\let\DOTSI\relax
\newcommand*{\letteronint}[1]{%
  \DOTSI
  \mathop{%
    \mathpalette\@LetterOnInt{#1}%
  }%
  \mkern-\thinmuskip
  \int
}
\newcommand*{\@LetterOnInt}[2]{%
  \sbox0{$#1\int\m@th$}%
  \sbox2{$%
    \ifx#1\displaystyle
      \textstyle
    \else
      \scriptscriptstyle
    \fi
    #2%
  \m@th$}%
  \dimen@=.4\dimexpr\ht0+\dp0\relax
  \ifdim\dimexpr\ht2+\dp2\relax>\dimen@
    \sbox2{\resizebox*{!}{\dimen@}{\unhcopy2}}%
  \fi
  \dimen@=\wd0 %
  \ifdim\wd2>\dimen@
    \dimen@=\wd2 %
  \fi
  \rlap{\hbox to \dimen@{\hfil
    $#1\vcenter{\copy2}\m@th$%
  \hfil}}%
  \ifdim\dimen@>\wd0 %
    \kern.5\dimexpr\dimen@-\wd0\relax
  \fi
}

\newcommand*{\curvearrowbotright}{\mathpalette\rotmath@internal\curvearrowleft}
\newcommand*\rotmath@internal[2]{\rotatebox{180}{$\m@th#1#2$}}

\makeatother

\usepackage{caption}
\usepackage[export]{adjustbox}

\numberwithin{equation}{section}

\definecolor{ultramarine}{rgb}{0.07, 0.04, 0.56}
\definecolor{cadmiumgreen}{rgb}{0.0, 0.42, 0.24}
\definecolor{indigo(dye)}{rgb}{0.0, 0.25, 0.42}
\definecolor{dgren}{rgb}{0.2, 0.6, 0.2}

\newcommand{\p}{{\partial}}

\newcommand{\eq}[1]{\begin{equation}#1\end{equation}}
\newcommand{\eqa}[1]{\begin{align}#1\end{align}}
\newcommand{\spl}[1]{\begin{split} #1 \end{split}}
\newcommand{\fg}[1]{\begin{figure}[tbp]\centering #1 \end{figure}}

\newcommand{\vxi}{\vec{\xi}}
\newcommand{\vx}{\vec{x}}
\newcommand{\vk}{\vec{k}}

\newcommand{\I}{{\rm Im}\,}

\newcommand{\tG}{\hat{G}}

\newcommand{\tF}{\hat{F}}

\newcommand{\td}{\tilde{d}}

\newcommand{\z}{\zeta}

\newcommand{\oo}{\mathcal{O}}

\newcommand{\bk}[2]{\langle #1, #2\rangle}

\begin{document}

\begin{center}

{\LARGE \bf Analyticity and positivity of Green's functions\\[0.3cm] without Lorentz}  \\[0.7cm]

{Paolo Creminelli$^{\,\rm a, \rm b, \rm c}\footnote{creminel@ictp.it}$, Alessandro Longo$^{\,\rm c, \rm d, \rm e}\footnote{alongo@apc.in2p3.fr}$, Borna Salehian$^{\,\rm f}\footnote{bsalehian@phys.ethz.ch}$, and Ahmadullah Zahed$^{\,\rm a}\footnote{azahed@ictp.it}$}
\\[0.5cm]

\small{
\textit{$^{\rm a}$
ICTP, International Centre for Theoretical Physics\\ Strada Costiera 11, 34151, Trieste, Italy}}
\vspace{.2cm}

\small{
\textit{$^{\rm b}$
IFPU, Institute for Fundamental Physics of the Universe\\ Via Beirut 2, 34014, Trieste, Italy }}
\vspace{.2cm}

\small{
\textit{$^{\rm c}$
INFN, National Institute for Nuclear Physics, 34127 Trieste, Italy}}
\vspace{.2cm}

\small{
\textit{$^{\rm d}$
SISSA, International School for Advanced Studies, 34136 Trieste, Italy}}
\vspace{.2cm}

\small{
\textit{$^{\rm e}$
Université Paris Cité, CNRS, Astroparticule et Cosmologie\\10 Rue Alice Domon et Léonie  Duquet, F-75013 Paris, France}}
\vspace{.2cm}

\small{
\textit{$^{\rm f}$ Institut f\"ur Theoretische Physik, ETH Z\"urich, 8093 Z\"urich, Switzerland}}

\vspace{.6cm}
\end{center}

\hrule \vspace{0.3cm}
{\small  \noindent \textbf{Abstract} \noindent \vspace{0.1cm}

\noindent 

We study the properties imposed by microcausality and positivity on the retarded two-point Green's function in a theory with spontaneous breaking of Lorentz invariance. We assume invariance under time and spatial translations, so that the Green's function $G$ depends on $\omega$ and $\vec k$. We discuss that in Fourier space microcausality is equivalent to the analyticity of $G$ when $\Im (\omega,\vec k)$ lies in the forward light-cone, supplemented by bounds on the growth of $G$ as one approaches the boundaries of this domain. Microcausality also implies that the imaginary part of $G$ (its spectral density) cannot have compact support for real $(\omega,\vec k)$. Using analyticity, we write multi-variable dispersion relations and show that the spectral density must satisfy a family of integral constraints. Analogous constraints can be applied to the fluctuations of the system, via the fluctuation-dissipation theorem. A stable physical system, which can only absorb energy from external sources, satisfies $\omega \cdot \Im G(\omega,\vec k) \ge 0$ for real $(\omega,\vec k)$. We show that this positivity property can be extended to the complex domain: $\Im [\omega\, G(\omega,\vec k)] >0$ in the domain of analyticity guaranteed by microcausality. Functions with this property belong to the Herglotz-Nevanlinna class. This allows to prove the analyticity of the permittivities $\epsilon(\omega,k)$ and $\mu^{-1}(\omega,k)$ that appear in Maxwell equations in a medium. We verify the above properties in several examples where Lorentz invariance is broken by a background field, e.g. non-zero chemical potential, or non-zero temperature. We study subtracted dispersion relations when the assumption $G \to 0$ at infinity must be relaxed.
\vspace{0.3cm}}
\hrule

\vspace{0.3cm}
\newpage

\tableofcontents

\vspace{0.3cm}

\hrule

\section{Introduction}

It is of considerable importance to understand how the fundamental principles of the underlying theory manifest in the observables. What are these principles? One is Lorentz invariance: in relativity, events that are space-like separated cannot influence each other—more technically, operators must commute outside the lightcone. This property, known as microcausality, implies analyticity when passing from position to momentum space. The second key principle is unitarity, the conservation of probability in quantum mechanics, which implies the positivity of certain quantities.  The combination of analyticity and unitarity is remarkably powerful. It leads to theoretical constraints on Effective Field Theories (EFTs), restricting the coefficients in the Lagrangian to lie within certain regions of parameter space. These so-called positivity bounds (see, for instance, \cite{Adams:2006sv,Bellazzini:2020cot,Caron-Huot:2020cmc,Tolley:2020gtv,Sinha:2020win,Arkani-Hamed:2020blm,deRham:2022hpx}) and the broader $S$-matrix bootstrap program (see e.g.~\cite{Paulos:2016fap,EliasMiro:2022xaa,Kruczenski:2022lot}) are currently the focus of intense research activity.

Lorentz invariance plays a prominent role in particle physics, where the main focus is on scattering particles in empty Minkowski space. This is, however, a quite restricted setup. In most areas of physics one is interested in systems where ``something'' is present and Lorentz invariance is {\em spontaneously} broken. In cosmology, there is always a preferred frame where ``matter'' is at rest and, moreover, on larger scales one is sensitive to the curvature of spacetime. In condensed matter, one is interested in the excitations of a system whose presence spontaneously breaks Lorentz invariance. The EFTs that are needed in this context do not enjoy (linearly realised) Lorentz symmetry.
Our understanding of QFT in this context is rather primitive, yet advancing it is crucial, since many fundamental phenomena at the frontiers of current research involve spontaneous Lorentz breaking. Let us emphasize that we are concerned with the common case where the state of the system breaks Lorentz invariance, as happens constantly in real-world situations. We are not referring to any ``fundamental'' breaking of Lorentz invariance. 

In this paper we focus on the simplest setup of Lorentz breaking. We assume that time and spatial translations are unbroken as well as rotational symmetry (the breaking of rotational symmetry is deferred to \cref{norot}). Lorentz boosts are (spontaneously) broken. This setup is typical of condensed matter, and when we consider theories at finite temperature or chemical potential. It is also relevant in cosmology when we focus on scales much shorter than the Hubble scale. In this setup the $S$-matrix is in general an ill-defined object, without the analytic properties of the Lorentz-invariant case \cite{Creminelli:2023kze,Hui:2023pxc}. We here study a simpler object, the retarded Green's function of a local operator, which in most of the discussion we will assume to be a scalar. 

Since we are assuming translational invariance, it is convenient to study the retarded Green's function in Fourier space: $G(\omega,k)$. The paper focuses on understanding the analyticity and positivity properties of this function of two complex variables, in analogy with the properties of the $S$-matrix in a Lorentz invariant setup, which also depends on two (complexified) variables. 

Our paper builds on previous works about microcausality and positivity when Lorentz invariance is broken. A first relevant result \cite{Leon} is the generalisation of the Kramers-Kronig relations to include microcausality (and not only retardation). Ref.~\cite{Dubovsky:2007ac,y5cg-1c7g} verified the persistence of microcausality beyond Lorentz invariance. Ref. \cite{Heller:2022ejw,Heller:2023jtd} studied the implications of microcausality on the dispersion relations of fluids, while \cite{Hui:2025aja} focused on the mixed representation $t,\vec k$, which is useful in the absence of time-translation invariance, as in cosmology. Ref.~\cite{Creminelli:2022onn} studied the EFT constraints one can derive using microcausality and positivity of retarded Green's functions. Recently, the relation between the retarded Green's function and the S-matrix on black hole backgrounds has been investigated in \cite{Correia:2025enx}.

After introducing the physical meaning and some basic properties of the retarded Green's function $G$ in \cref{sec: Gintro}, in \cref{Bogo} we explore the consequences of microcausality in Fourier space. We show, under the standard assumption that $G$ is a tempered distribution in real space, that microcausality is equivalent to the analyticity of $G(\omega,k)$ when $\Im (\omega,k)$ lies in the forward light-cone, supplemented by polynominal bounds on the growth of $G$ as $\Im (\omega, k)$ approaches the light-cone and when $\Im (\omega,k)$ goes to infinity. See the theorem that includes \cref{estimateddims}. We think it is worth emphasizing this important result, even though the statement can be straightforwardly extracted from classic results of the Lorentz-invariant literature, see \cite{bogolubov2012general}. (We review some necessary mathematical background and the proof of the theorems in \cref{AppA}.)

Based on these properties of analyticity, in \cref{muldisp} we write 2-variable dispersion relations, which allow to write the whole function in terms of its imaginary part (spectral density). Contrary to the case of a single variable, analyticity imposes stringent constraints on $\Im G(\omega,k)$ for real $\omega$ and $k$: this is the focus of \cref{sec:imconst}. We will show that $\Im G(\omega,k)$ cannot have compact support for real $\omega$ and $k$. Assuming that $G \to 0$ at infinity, we will also prove that the imaginary part must satisfy a family of integral constraints
 \eq{
\displaystyle\int_{-\infty}^{\infty}\dd{\z}{\rm Im}G(\z,k+\xi \z)=0\,,
}
for any $k$ and $|\xi|<1$.
We check these properties in several examples including the response mediated by a single mode and full-fledged models where the breaking of Lorentz is due a background field, e.g. non-zero chemical potential, or non-zero temperature. The constraints above also applies to the fluctuations of the system (Wightman functions), via the fluctuation-dissipation theorem.  We also briefly discuss how analyticity also constrains the {\em non-linear} response of a system.

In most cases $\Im G(\omega,k) \ge 0$ for $\omega>0$ (and $\le 0$ for $\omega<0$). This inequality guarantees that the system can only absorb energy from an external field, a property closely related to the stability of the system. A system of this kind is called passive and this property holds in the vacuum, in the presence of chemical potential and at non-zero temperature. This positivity property generalizes the one due to quantum mechanical unitarity when the system needs to be described by a density matrix. In \cref{pos} we show that in a passive system positivity is not only valid for real $\omega$ and $k$, but it extends to the whole region of analyticity guaranteed by microcausality. Using the dispersion relations of \cref{muldisp}, we show that the function $\omega G(\omega,k)$ has strictly positive imaginary part for $\Im (\omega, k)$ in the forward light-cone. Complex functions with this kind of properties are called Herglotz-Nevanlinna functions and are well studied in mathematics: this connection is explored in \cref{herg}. Besides their mathematical interest, these positivity properties have important physical consequences. For instance, they allow to prove the analyticity of the object that appears in the Maxwell equations in media, i.e.~the 1PI retarded Green's functions of currents. This object is usually parametrized in terms of the electric and magnetic responses $\epsilon(\omega,k)$ and $\mu^{-1}(\omega,k)$.

The rest of the paper is devoted to study how some of the assumptions made above can be relaxed. In \cref{subtract} we study subtracted dispersion relations, necessary when one cannot assume that $G \to 0$ at infinity. In \cref{norot} (supplemented by \cref{AppMapsNonRot}) we show that virtually all the properties discussed above hold also when rotational invariance is broken. Conclusions and future directions are discussed in \cref{sec:conc}.

\section{\label{sec: Gintro}Retarded Green's function}

The central object of interest in this work is the retarded two-point function, which encodes the linear response of a system to an external perturbation. Concretely, consider an observable $\oo_1$ whose expected value we wish to measure in the presence of an external source $J$. The source is coupled to the system through a deformation of the Hamiltonian, $H\to H-\int\dd[d-1]{\vx}\oo_2(\vx)J(t,\vx)$, where $H$ denotes the Hamiltonian of the unperturbed system, and $\oo_2$ is another operator determined by the coupling. Standard perturbation theory in the interaction picture gives the Kubo formula \cite{doi:10.1143/JPSJ.12.570}
\eq{
\ev{\oo_1(x)}_J=\ev{\oo_1(x)}_{J=0}+\int\dd[d]{x'}G(x,x')J(x')+\dots\,,
\label{kubo}
}
where $x\equiv(t,\vec{x})$ and the retarded Green's function, known also as the linear response, is defined as\footnote{A similar formula applies for classical statistical systems, described by a phase-space distribution function $f(q_a,p_b)$. In that case, the retarded Green's function takes the form $G=-\theta(t-t')\ev{\{\oo_1(t),\oo_2(t')\}}$, consistent with the usual replacement of commutators by Poisson brackets, and the expectation values given by $\ev{\cdots}=\int (\cdots)f$ with the integral taken over phase space. See \cite{doi:10.1143/JPSJ.12.570} for details.}
\eq{
G(x,x')\equiv i\theta(t-t')\ev{[\oo_1(x),\oo_2(x')]}\,.
\label{retG}
}
Expectation values are always computed with respect to the state of the system, which in general is described by a density matrix, i.e. $\ev{\cdots} = \Tr(\rho\cdots)$. Although $\oo_1$ and $\oo_2$ may in general be non-local operators, in this work we restrict to local operators, for which causality will be manifest. From now on we will focus on a single operator operator $\oo$ and thus suppress the indices $1,2$. Most of the results of the paper are valid also for different operators, except the positivity properties of \cref{pos}. 

Moreover, we restrict to systems that are time-translation invariant and spatially homogeneous. In this case, the Green's function depends only on the difference $x-x'$, and it is convenient to work in Fourier space
\eq{
G(k^\mu)=\int\dd[d]{x}\,e^{-ik\cdot x}G(x)\,,
\label{F1}
}
where $k^\mu \equiv (\omega,\vk)$ and $k\cdot x$ denoting the Minkowski inner product with mostly-plus metric signature $(-,+,\dots,+)$. We use the same notation for the Green's function in both position and momentum space; the intended meaning should always be evident from the argument and the context. For Hermitian operators, the Green's function is real in position space, which implies 
\eq{
G(k^\mu)^*=G(-k^\mu)\,,
\label{reality}
}
in Fourier space. We will refer to this as the reality condition. Throughout most of this work, we further assume rotational invariance, so that $G$ depends on the spatial momentum only through its magnitude $k \equiv |\vk|$. However, many of the results do not rely on this assumption, as will become explicit in \cref{Bogo} and \cref{norot}.

As we will see below, the imaginary part of the retarded Green's function in Fourier space, $\I G(k^\mu)$ plays an important role. In particular, the imaginary part controls the amount of source energy absorbed by the system (see for instance \cite{Creminelli:2024lhd} for a derivation)
\eq{
\Delta E=\int\frac{\dd[d]{k}}{(2\pi)^d}\,\omega\,\I G(k^\mu)\, |J(k^\mu)|^2\,,
\label{diss}
}
in which $J(k^\mu)$ is the Fourier transform of the source. We will also see that, knowing the imaginary part is (almost) enough to construct the full linear response. 

It should be mentioned that, despite their name, Green's functions are not in general ordinary functions. Rather, they belong to a sub-class of generalized functions, known as \textit{tempered distributions}. Intuitively, tempered distributions cannot be evaluated pointwise, but only through their action on smooth, rapidly decaying test functions (Schwartz functions; see \cref{AppA}). The set of Schwartz functions is usually indicated with $\cal{S}$ and tempered distributions with $\cal{S}'$. Tempered distributions can grow at most like a polynomial at infinity, a property referred as polynomial boundedness. Most importantly, they admit a well-defined Fourier transform, which maps a tempered distribution into a tempered distribution. Indeed, being able to take Fourier transform, is one of the main reasons to consider Green's functions (or, more generally, any $n-$point correlation function) to be a tempered distribution. This is the underlying assumption for a vast majority of systems, and we shall adopt it throughout this work (as noted in  \cref{FootTemp}, instabilities manifest as exponentially growing response functions. Such functions are not polynomially bounded and do not belong to the space of tempered distribution). For instance, in the axiomatic approach to quantum field theory, all Wightman functions are assumed to be tempered distributions \cite{bogolubov2012general}. The next section is devoted to explore the implications of micro-causality for the retarded Green's function.\footnote{Most operations on tempered distributions are simple extensions of the corresponding definition for ordinary functions. However, there are a few cases that require care, in particular the product and convolution of two tempered distributions. This is relevant for instance in the definition of the retarded Green's function in \cref{retG}, where both the step function and the expected value of the field commutator are distributions. Writing \cref{retG} as $\theta(t) G_c(t)$ (suppressing $\vec{x}$ here) one way to make sense of this product is to express $G_c(t)=f^{(n)}(t)$ as a finite-order derivative of a continuous function of polynomial growth (this is possible for any tempered distribution, as explained for instance in \cite{Nussenzveig:1972tcd}). Then the retarded product can be defined as 
\begin{displaymath}
    G(t)\equiv \frac{d^n}{dt^n}(\theta f)\,,
\end{displaymath}a well-behaved tempered distribution. This definition is unique up to contact terms, i.e.~$\delta(t)$ and its derivatives, see \cite{dütsch2019classical}. Following the steps in sec.~1.8 of \cite{Nussenzveig:1972tcd}, one can take the Fourier transform of the above equation and deduce that in Fourier space $G$ and $G_c$ are related by a dispersion relation with subtractions (see also \cref{subtract}).
}

\section{Causality and Analyticity}\label{Bogo}

The causal structure of Minkowski spacetime is encoded in the retarded Green’s function \cref{retG} through retardation and microcausality.
Retardation is the fact that the future cannot influence the past. In special relativity this is meaningful only if there is no effect from spacelike separated points---the notion of microcausality. As a quick reminder of why this holds (see \cite{Hartman:2015lfa,Dubovsky:2007ac}), consider the Kubo formula \cref{kubo} with a delta-function source, $J(x)=\lambda\,\delta^{4}(x)$:
\eq{
\ev{\oo_1(x)}_\lambda=\ev{\oo_1(x)}_{\lambda=0}+\lambda\theta(t)\ev{[\oo_1(x),\oo_2(0)]}+\dots
\label{kubodelta}
}
Consider a boosted observer, for whom the system is described by a new density matrix $\rho' = U^{-1} \rho U$, while operators transform as $\oo'_{1,2}(x') = U^{-1} \oo_{1,2}(x) U$, with $x$ and $x'$ related by a Lorentz boost and $U$ the corresponding unitary operator. Lorentz invariance requires that both observers must agree on their measurements (i.e. on the linear response term).  However, for the linear response term in \cref{kubodelta}, the step functions $\theta(t)$ and $\theta(t')$ associated with the two observers generally differ unless $x$ lies within or on the lightcone of the origin. Therefore, agreement is possible only if the linear response vanishes for spacelike separations. In summary, retardation is enforced by the step function in \cref{retG}, while microcausality follows from the vanishing of commutators outside the lightcone. We emphasize that microcausality is a statement about the operator algebra itself and must hold independently of the 
state. Hence, it remains true even in cases where Lorentz invariance is spontaneously broken.


In the remainder of this section, we discuss the implications of retardation and microcausality for the retarded Green's function in Fourier space.

\subsection*{Retardation}
As a warm-up, let us first consider cases in which the Green's function only depends on time. This is a quite common situation in which the response can be treated as local in space but non-local in time---a good approximation for many condensed-matter systems. It is well-known that retardation is related to analyticity in Fourier space. Indeed, it is easy to see from \cref{F1}, setting $d=1$, that the Fourier transform of the retarded Green's function
\begin{equation}
G(\omega)=\int_0^{+\infty} \dd{t}~ e^{i \omega t}G(t)\,,
\end{equation}
is an analytic function of complex $\omega=\omega_R+i\,\omega_I$ in the upper half plane: $\mathrm{UHP} \equiv \{\omega \in \mathbb{C}|\omega_I>0\}$. This can be argued by noticing that $G(t)$ has support only for $t>0$, and therefore the phase factor provides an exponential suppression when $\omega \in \mathrm{UHP}$, improving the convergence of the integral. One can then verify that also the complex derivative of $G(\omega)$ exists in the UHP, i.e.~it satisfies the Cauchy-Riemann conditions. Crucially, we have used the fact that $G(t)$ grows no faster than a polynomial at large $t$, a consequence of being a tempered distribution.\footnote{Physically, polynomial boundedness serves as the mathematical signature of stability, ensuring that a system's response to a localized perturbation does not experience runaway growth over time. This conditions is crucial for the Fourier transform $G(\omega)$ to be well-defined for $\omega\in \mathbb{R}$. By contrast, an unstable system violates this condition. As an example, the retarded Green's function of a simple harmonic oscillator with $m^2<0$
diverges exponentially with time. In this case, the system becomes extremely sensitive to external perturbations: it can exponentially amplify an initially small disturbance, leading to an instability.\label{FootTemp}}

Conversely, if one starts from a function $G(\omega)$ which is analytic in the UHP, it is possible to show that its inverse Fourier transform $G(t)$ will be retarded, provided that $G(\omega)$ is polynomially bounded as $|\omega|\to \infty$. To see this, one can compute
\begin{equation}
    G(t)=\int_{-\infty}^{\infty} \frac{\dd{\omega}}{2\pi} ~e^{-i \omega t}G(\omega)\,,
\end{equation}
and notice that, when $t<0$,  it is possible to deform the integration contour to the UHP, with no contribution from the semi-circle at infinity. It follows that $G(t)$ is supported only for $t\geq 0$, hence it is retarded.\footnote{If $|G(\omega)|\to 0$  when $|\omega|\to \infty$, Jordan's lemma guarantees that the integral over the infinite semicircular arc in the UHP vanishes for $t<0$, yielding $G(t)=0$. However, when $G(\omega)$ does not decay, $G(t)$ is not a function but only a tempered distribution, and pointwise evaluation at $t<0$ is ill-defined. In this case, the retarded nature of $G(t)$ can be verified by showing that the action of the distribution $G(t)$ on any test function supported at $t<0$ vanishes.}

The following theorem gives the necessary and sufficient conditions that characterize Fourier transforms of retarded tempered distributions:

\begin{center}
\rule{\textwidth}{0.4mm}
\end{center}
\paragraph{Retardation-Analyticity Theorem}\emph{
A tempered distribution $G(t)\in S'(\mathbb{R})$ is retarded if and only if its Fourier transform is the boundary value, $\omega_I\to0^+$, of an analytic function in the UHP which admits the bound 
\eq{
|G(\omega)|\leq A \left(1+|\omega|\right)^n\left(1+\frac{1}{|\omega_I|^m}\right)\>,\>\omega \in \mathrm{UHP}\,,
\label{Ttbound}
}
where the positive constant $A$ and the powers $n,m\in \mathbb{N}$ depend on $G$.}
\begin{center}
\rule{\textwidth}{0.4mm}
\end{center}

We refer the reader to \cref{AppA} for a proof, and to \cite{bogolubov2012general,Nussenzveig:1972tcd} for a more extensive discussion. The theorem states that if a tempered distribution is retarded, then its Fourier transform can be extended to an analytic \emph{function} in the UHP. Moreover, the bound ensures that $G(\omega)$ is polynomially bounded not only on the real line (a consequence of temperedness) but also on the whole UHP. Conversely, any analytic function in the UHP that satisfies \cref{Ttbound} can be identified as the analytic continuation of the Fourier transform of a retarded tempered distribution.

Let us discuss the physical meaning of the two bounds, parametrized by the two integers $n$ and $m$. The first one bounds the growth of $G(\omega)$ as $|\omega|\to \infty$ in the UHP. This corresponds to how singular is $G(t)$ at short-distances (not necessarily around $t=0$). For instance a singular behaviour of the kind $\delta^{(n)}(t-t_0)$, the $n$-th derivative of the $\delta$-function at $t_0>0$, gives a Fourier space power-law growth $|\omega|^{n-1}$. Loosely speaking, the smaller $n$ is the more regular $G(t)$ is. On the other hand the divergence for $\omega_I \to 0$, bounded by the integer $m$, has to do with the long-time behaviour of $G(t)$. An asymptotic behaviour $\sim t^{m-1}$ corresponds to $G(\omega)$ growing as $|\omega_I|^{-m}$. (Indeed $\int_0^{\infty}e^{i \omega t}t^{m-1} dt=\Gamma[m](-i\omega)^{-m}$, for $m>0$.) The smaller $m$ is, the milder is the growth of $G(t)$ at late times.  In short $n$ has to do with the UV behaviour of $G$, while $m$ with the IR one.


We should emphasize that retardation of $G(t)$ is not enough to ensure that ${G}(\omega)$ will satisfy \cref{Ttbound}. A \emph{super-polynomial} growth of $G(t)$ will violate the bounds we just discussed. An illustrative example is
\begin{equation}
G(t)=\theta(t)e^{2\sqrt{t}}\quad\to \quad {G}(\omega)=\frac{(i-1) \sqrt{\pi} \left(1+\text{erf}\left(\frac{1+i}{\sqrt{2\omega }}\right)\right)e^{\frac{i}{\omega }}}{\sqrt{2}\omega ^{3/2}}+\frac{i}{\omega }.
\end{equation}
In this case, the super-polynomial growth of $G(t)$ is responsible for the presence of an essential singularity at the origin of the complex $\omega$-plane. The leading singular behaviour at the origin is governed by the $e^{i/\omega}$ factor, and ${G}$ is not polynomially bounded as $\omega_I \to 0$ as required by the theorem. Indeed, the assumption that the Green's function is a tempered distribution excludes super-polynomial behaviour.\footnote{In this case one has to work with non-tempered distributions. They act on Schwarz functions with compact support. It is possible to prove a weaker version of the above theorem for non-tempered distributions \cite{bogolubov2012general}.}

Notice that the theorem above does not exclude the possibility of an essential singularity on the real axis. Consider 
\begin{equation}
    G(\omega)=e^{-i/\omega}\quad \to \quad
     G(t)=\delta(t)-\frac{J_1\left(2 \sqrt{t}\right)}{\sqrt{t}}\theta(t)\,,
\end{equation}
which has an essential singularity at $\omega=0$. In contrast to the case $e^{i/\omega}$ discussed above, it admits a limit consistent with \cref{Ttbound} for $\omega_I\to0^+$, and indeed it corresponds to a retarded tempered distribution $G(t)$. 

\subsection*{Microcausality}
Let us consider retarded Green’s function in general $d$ spacetime dimensions. As mentioned above, we assume spatial homogeneity (though not necessarily rotational invariance), while Lorentz boosts are spontaneously broken. By microcausality and retardation, the retarded Green's function is supported within the forward lightcone in real space, $\mathrm{FLC}\equiv\{x\in \mathbb{R}^{d-1,1}|~x^2\leq0,t\geq0\}$. From the definition of the Fourier transform
\begin{equation}\label{Fourierddims}
    {G}(k^\mu)=\int_{x\in{\rm FLC}}\dd[d]{x}\,e^{-ik\cdot x}G(x)\,,
\end{equation}
 it is possible to see that $G(k^\mu)$ is analytic in the region $\mathbb{R}^{d-1,1}+i \mathring{\rm{FLC}}$, i.e.~when the imaginary part of the complex momentum $k^{\mu}=k_R^{\mu}+ik_I^{\mu}$ is timelike and future directed. The overring symbol, $\mathring{ }\,$, emphasizes that the analyticity domain is the {\em interior} of the FLC, i.e. $\mathring{\rm{FLC}}=\{k\in \mathbb{R}^{d-1,1}|~k^2<0,k^0>0\}$. The exponential factor in \cref{Fourierddims} improves the convergence of the integral only when $k^{\mu}$ acquires an imaginary part which lies strictly inside the $\rm{FLC}$, provided that $G(x)$ is polynomially bounded (which is guaranteed by the temperedness assumption). Analyticity then follows directly from the existence of the complex derivative or the Cauchy–Riemann equations.\footnote{When time translations are broken, it is helpful to work in a mixed representation in which only the spatial coordinates are Fourier transformed. In this case it is possible to prove \cite{Hui:2025aja} that the mixed transformed ${G}(t,\vec{k})$ will be analytic for any $\vec{k}\in \mathbb{C}^{d-1}$.} 

Conversely, suppose that ${G}(k^\mu)$ is analytic in the region $\mathbb{R}^{d-1,1}+i\,\mathring{\rm FLC} \subset \mathbb{C}^{d}$ and is polynomially bounded at large momenta. Then its inverse Fourier transform is necessarily a tempered distribution supported on the forward lightcone. The argument goes as follows: given $x=(t,\vx)$, the inverse Fourier transform reads
\eq{
G(x)=\int\frac{\dd{\omega}\dd{k_\|}}{(2\pi)^2}\frac{\dd[d-2]\vk_{\perp}}{(2\pi)^{d-2}}\,e^{-i\omega t+ik_\| r}\,G(k^\mu)\,,
}
where we have decomposed the spatial momenta as $\vk=\vk_\|+\vk_\perp$, with $\vk_\perp\cdot\vx=0$, and $r=|\vx|$. When $t<0$ it is enough to deform the contour in the upper half $\omega$-plane (as in $d=1$ case) to conclude that it vanishes. For $0<t<r$, it is convenient to change variables to $z_+=\omega+k_\|$ and $z_-=\omega-k_\|$, leaving $\vk_\perp$ untouched. The integral becomes
\eq{
G(x)=\int\frac{2\dd{z_+}\dd{z_-}}{(2\pi)^2}\frac{\dd[d-2]\vk_{\perp}}{(2\pi)^{d-2}}\,e^{-iz_+(t-r)/2-iz_-(t+r)/2}\,G(z_+,z_-,\vk_\perp)\,.
}
As a function of complex $z_{+}$, $G(z_+,z_-,\vec{k}_{\perp})$ is analytic in the upper half-plane. Hence, we may stretch the $z_+$ contour to infinity and conclude that $G(x)=0$ for $0<t<r$, since the factor $e^{-i z_+ (t-r)/2}$ provides an exponential suppression. One gets the same result for $t < -r$ using $z_-$. Altogether, we conclude that $G(x)$ is supported only within the FLC, i.e. $t \geq |\vec{x}| \geq 0$. 

More rigorously, the necessary and sufficient conditions that characterize a retarded and micro-causal tempered distribution are summarized by the following theorem:
\vspace{-0.5cm}
\begin{center}
\rule{\textwidth}{0.4mm}
\end{center}
\vspace{-0.7cm}
\paragraph{Microcausality-Analyticity Theorem} \emph{A tempered distribution $G(x)\in S'(\mathbb{R}^{d-1,1})$ is retarded and micro-causal if and only if its Fourier transform $G(k^\mu)$, with $k^\mu=k^\mu_R+ik^\mu_I$, is analytic in the domain $\mathbb{R}^{d-1,1}+i\mathring{\rm FLC}$ and it admits the bound
\begin{equation}\label{estimateddims}
|{G}(k^\mu)|\leq A (1+|k^{\mu}|)^n\left( 1+\frac{1}{(k^0_I-|\vec{k}_I|)^m}\right)\,,\quad k^\mu\in \mathbb{R}^{d-1,1}+i\mathring{\rm FLC}\,,
\end{equation}
where the positive constant $A$ and the powers $m,n\in \mathbb{N}$ depend on $G$.}
\vspace{-0.5cm}
\begin{center}
\rule{\textwidth}{0.4mm}
\end{center}

This can be viewed as a generalization of the theorem stated in the previous section. A detailed proof is provided in \cref{AppA}, following \cite{bogolubov2012general}.\footnote{In \cite{bogolubov2012general}, the statement appears as a compilation of several intermediate results, summarized in Corollary B.9. More explicitly: Theorems B.2 and B.4 apply to general (not necessarily tempered) distributions, while Theorems B.7 and B.8 specialize to tempered distributions supported in a pointed cone.} Similarly to the previous case, the Fourier transform $G(k^\mu)$ is an analytic function which is polynomially bounded for large momenta. The asymptotic polynomial growth is controlled by the power $n$ in \cref{estimateddims}, where $|k^{\mu}|\equiv \sqrt{|k^0|^2+...+|k^{d-1}|^2}$ is the Euclidean norm of the complex vector $k^{\mu}\in \mathbb{C}^d$. This is related to the short distance behaviour of $G(x)$ (see the discussion around \cref{u1u2}).

Conformal field theory (CFT) provides an example in which the Green’s function exhibits power-law growth at large momenta. For an operator with scaling dimension $\Delta > 0$, the two-point function is fixed (up to an overall normalization) to be\footnote{Correlation functions in a CFT are typically expressed in Euclidean signature. The corresponding Lorentzian correlators follow by an appropriate Wick rotation; see \cite{Hartman:2015lfa,Bautista:2019qxj}.} $G(x)=x^{-2\Delta}$. This corresponds to 
\eq{
G(k^\mu)\propto \Big(-(\omega+i\epsilon)^2+\vk^2\Big)^{\Delta-d/2}\,,
}
in Fourier space, where the $i\epsilon$ prescription is chosen to ensure analyticity in the FLC, as appropriate for the retarded Green’s function. For $\Delta > d/2$, this expression grows as a power for complex momenta.

The denominator in \cref{estimateddims} controls the growth of $|G(k^\mu)|$ when we approach the boundary of the domain of analyticity, i.e. $k^0_I-|\vec{k}_I|\to 0^{+}$: the Fourier transform can diverge at most as $(k^0_I-|\vec{k}_I|)^{-m}$ ($|\vec{k}_I|$ being the Euclidean norm of the $d-1$ dimensional vector $\vec{k}_I$). Notice that, in more than one dimension, this limit does not necessarily require $k_I^\mu \to 0$: $G(k)$ can develop singularities also for complex values of $k^{\mu}$, on the edge of the $\rm{FLC}$.

Similar to the $d=1$ case, the bound in \cref{estimateddims} does not exclude essential singularities on the boundary of the FLC. Consider the following function in $1+1$-dimensions:
\begin{equation}
    G(\omega,k)= e^{-i\left(\frac{1}{\omega+k}+\frac{1}{\omega-k}\right)}\,.
\end{equation}
It is evident that $G$ possesses singularities along the null directions $\omega_I=|k_I|$, even though its anti-Fourier transform is compatible with retardedness and microcausality when mapped back to real space. 

Let us comment on the region of analyticity of the retarded Green’s function. A notable feature of several complex variables, compared to the single-variable case, is the notion of domain of holomorphy. For one complex variable, any open set $D\subset\mathbb{C}$ can be the maximal domain of analyticity for some function: given any such $D$, one can construct a function that is analytic in $D$ and develops a singularity on its boundary. In contrast, for a domain $D\subset \mathbb{C}^d$ with $d>1$, it may happen that \emph{every} function analytic in $D$ can be analytically continued beyond $D$. The biggest domain $H(D)\supseteq D$ to which all analytic functions on $D$ extend, and for which at least one function develops a singularity everywhere on its boundary, is called the envelope of holomorphy of $D$. A domain $D$ is said to be a \emph{natural domain of holomorphy} if it coincides with its envelope of holomorphy. A theorem states that any convex subset of $\mathbb{C}^d$ is necessarily a natural domain of holomorphy; see \cite{Sommer:1970mr,Curry:2024mua} for details. In our setting, the FLC is indeed a convex region in $\mathbb{C}^d$, and hence a natural domain of holomorphy. Notice that the above discussion concerns the space of all analytic functions on the domain; a given Green’s function may well admit analytic continuation beyond the FLC.

\subsection*{Rotational invariance}
Many interesting systems enjoy rotational symmetry. In this case, the dimensionality of the problem is effectively reduced to $\tilde{d}=1+1$ and the Fourier transform will only depend on $\omega$ and $k\equiv|\vec{k}|$:
\begin{equation}\label{Gomegak}
\begin{aligned}
{G}(\omega,\vk)&=\int\dd[d]{x}\,e^{i \omega t - i \vec{k}\cdot \vec{x}}G(t,r)\\
&=\int_0^{+\infty}\dd{t} ~e^{i\omega t}\int_0^t \dd{r}r^{d-2}~ G(t,r)\int \dd{\Omega_{d-2}}~e^{-ik r \cos{\theta}}\\
    &=\frac{2 \pi^{(d-2)/2}}{\Gamma(\frac{d-2}{2})}\frac{i}{k}\int_0^{+\infty}\dd{t} \int_0^t \dd{r}r^{d-3}~ G(t,r)~e^{i\omega t}(e^{-ik r }-e^{ik r })\,,
    \end{aligned}
\end{equation}
where we have used rotational symmetry to perform the angular integration. As expected, we can parametrize the Green's function as a function of $\omega$ and $k$ rather than $\vec{k}$, i.e. ${G}(\omega,k)$. Notice that the limit $k\to0$ is regular. Moreover, eq.~\eqref{Gomegak} implies that we can consider the Green's function as a function of complex $k$. Taking $\omega=\omega_R+i\omega_I$ and $k=k_R+ik_I$, the same argument as before reveals that the region of analyticity is 
\begin{equation}
    \label{eq:2ddomain}
    D=\{ (\omega,k)\in \mathbb{C}^2 \>|\> \omega_I>|k_I|\geq0\}\,.
\end{equation}
Notice that the defining region $k_R>0$ of the Green's function has been extended to the whole real line by using \cref{Gomegak}, which is even in $k$. Moreover, the reality condition implies that 
\eq{\label{eq:realityG}{G}(\omega,k)^*={G}(-\omega^*,-k^*)={G}(-\omega^*,k^*)}
We conclude that, for systems that enjoy rotational symmetry, one can effectively reduce to study complex functions of the two variables $\omega$ and $k$. Notice that this is somewhat non-trivial, since the real and imaginary part of the vector $\vec k$ are in general not aligned.\footnote{\label{rotinv}Let us show in another way that the analyticity of $G(\omega, \vec k)$ in FLC implies the analyticity of $G(\omega, k)$ in the domain \cref{eq:2ddomain} {\em and viceversa}. The direct way is obvious, because one can take $\vec k_R$ and $\vec k_I$ aligned and reduce to the 1+1 case. To prove the converse one has to show that given $\omega = \omega_R + i \omega_I$, $\vec k_R$ and $\vec k_I$ (in general not aligned) in the analyticity region, one can find another point with the same $\vec k^2$ that has the spacial vectors aligned and is inside the analyticity region. Since we are assuming that the function only depends on $\vec k^2$, this will prove the analyticity of $G(\omega, \vec k)$ in the region of interest. Concretely, one has the equation $(\vec k_R+ i \vec k_I)^2 = (\tilde k_R + i \tilde k_I)^2$, which must be solved for $\tilde k_R$ and $\tilde k_I$ with the requirement that $(\omega_I, \tilde k_I)$ is in the domain \cref{eq:2ddomain}. It is straightforward to check that a solution can always be found, provided one starts from a point of analyticity $(\omega_I, \vec k_I) \in$ FLC. See the related discussion in app.~B of \cite{Hui:2025aja}.}

In this paper we focus on scalar operators. The results can be easily generalised to the case of higher-spin (bosonic) operators, most notably conserved currents and the stress-energy tensor. In this case the Green's function for a given $\vec k$ can be decomposed in the various helicity contributions, which transform differently under the rotations that keep $\vec k$ fixed.  Each of these contributions has the same property as a scalar Green's function and the results of this work apply.  The various helicity contributions must be compatible in the limit $k \to 0$: in this limit one has a relation among them, as explained in \cite{Creminelli:2024lhd} for the case of conserved currents.

\section{Multi-variable dispersion relation}\label{muldisp}

It is well-known that the real and the imaginary part of an analytic function of a single variable, assuming suitable decay at infinity, are related through dispersion relations (Kramers-Kronig relations). As we have seen in the previous section, any retarded Green's function is an analytic function of several complex variables. In that case, it is possible to reduce the problem to a family of single variables. This results in a family of Kramers-Kronig relations which we call Leontovich relations, after M. Leontovich, who wrote them first \cite{Leon} (see also \cite{Melrose_1977,Creminelli:2022onn,Creminelli:2024lhd}). In this section, we will discuss another generalization of the standard dispersion relations to multiple variables, which involves an integration over all variables rather than a linear combination of them. Such a dispersion relation, apart from being a more natural generalisation of Kramers-Kronig, is useful to argue the absence of zeros in the region of analyticity guaranteed by microcausality, as we will see in \cref{pos}. In what follows, we mainly focus on the case with rotational symmetry, where, as we have seen, the number of variables reduces to $(\omega,k)\in\mathbb{C}^2$. The case without rotational symmetry will be discussed in \cref{norot}. 

The main idea is to use, instead of $(\omega,k)$, new variables $(z_1,z_2)$ such that the region of analyticity is mapped to the upper half-plane (UHP) in each of the new variables (or at least includes the UHPs). In that case, the problem of writing dispersion relations for multiple variables reduces to writing multiple single-variable dispersions, one variable at a time, keeping the other fixed. To avoid unnecessary complications, we take the transformation to be linear.\footnote{See \cite{Creminelli:2022onn} for a discussion about more generic ones.} Moreover, we demand that keeping, say, $z_2$ fixed and taking $z_1$ to complex infinity will correspond to approach infinity in $(\omega_I,k_I)$ plane via a time-like direction. This is important to have control over the UV behaviour of the function.   

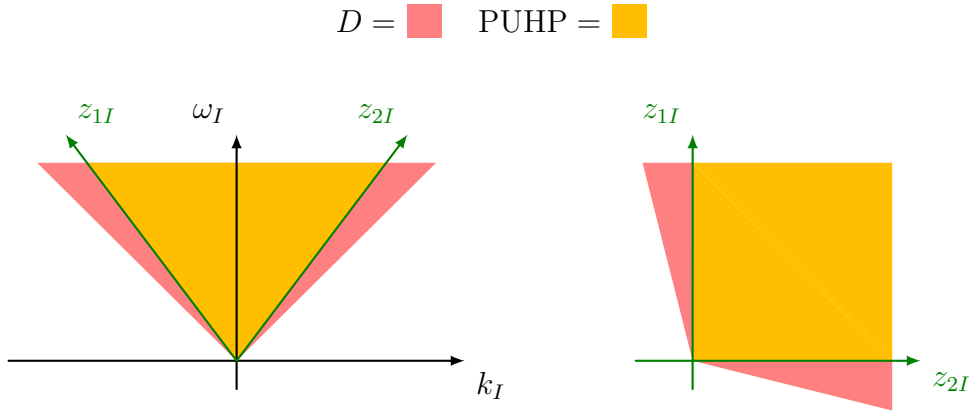
\begin{figure}[hbt!]
\centering
\adjustbox{valign=c}{\tikzset{every picture/.style={line width=0.85pt}}

\definecolor{ultramarine}{rgb}{0.07, 0.04, 0.56}
\definecolor{jasper}{rgb}{0.84, 0.23, 0.24}

\begin{tikzpicture}[>=latex, line join=round, line cap=round, scale=1.5]

\definecolor{royalblue(web)}{rgb}{0.25, 0.41, 0.88}

\coordinate (O) at (0,0);
\coordinate (w) at (0,2);
\coordinate (k) at (2,0);
\coordinate (tk) at (-2,0);
\coordinate (z2) at (1.5,2);
\coordinate (z1) at (-1.5,2);


\coordinate (L1) at (-1.75,1.75);
\coordinate (L2) at (1.75,1.75);
\coordinate (P1) at (1.31,1.75);
\coordinate (P2) at (-1.31,1.75);

\fill[red!50, opacity=1] (O) -- (L1) -- (L2) -- cycle;
\fill[orange!50!yellow, opacity=1] (O) -- (P1) -- (P2) -- cycle;
\draw[->, thick] (tk) -- (k) node[below right] {$k_{I}$};
\draw[->, thick] (0,-0.25) -- (w) node[above left] {$\omega_{I}$};
\draw[->, thick, green!50!black] (O) -- (z1)
node[above right] {$z_{1I}$};
\draw[->, thick, green!50!black] (O) -- (z2)
node[above left] {$z_{2I}$};

\coordinate (Ot) at (4,0);
\coordinate (z1t) at (4,2);
\coordinate (z2t) at (6,0);
\coordinate (z2tt) at (3.5,0);
\coordinate (L1t) at (3.56,1.75);
\coordinate (L2t) at (5.75,-0.44);
\coordinate (P1t) at (4,1.75);
\coordinate (P2t) at (5.75,0);
\coordinate (P3t) at (5.75,1.75);

\fill[orange!50!yellow, opacity=1] (Ot) -- (P1t) -- (P2t) -- cycle;
\fill[orange!50!yellow, opacity=1] (P3t) -- (P1t) -- (P2t) -- cycle;

\fill[red!50, opacity=1] (Ot) -- (L1t) -- (P1t) -- cycle;
\fill[red!50, opacity=1] (Ot) -- (L2t) -- (P2t) -- cycle;
\draw[->, thick,green!50!black] (4,-0.25) -- (z1t)
node[above left] {$z_{1I}$};
\draw[->, thick, green!50!black] (z2tt) -- (z2t)
node[below right] {$z_{2I}$};

\node[anchor=east] (legD) at (1.5,3) {$D=$};
\draw[fill=red!50, opacity=1,draw=none] (1.8,2.85) rectangle (1.5,3.15);

\node[anchor=east] (legD) at (3.3,3) { $~~~~~~\rm PUHP=$};
\draw[fill=orange!50!yellow, opacity=1,draw=none] (3.6,2.85) rectangle (3.3,3.15);
\end{tikzpicture}}
\caption{The region of analyticity guaranteed by microcausality $D$ in red and the poly upper-half plane (PUHP) in the $(z_1,z_2)$ space in yellow. Plot in the $(\omega_I,k_I)$ plane (left) and in the $(z_{1I},z_{2I})$ plane (right). For $\alpha>1$ the region of analyticity $D$ always includes the $\rm{PUHP}$.}\label{PlotRegions}
\end{figure}

An appropriate linear combination is\footnote{\label{genz1z2} The set of linear transformations that maps the region of analyticity to the PUHP is larger than \cref{z1z2}. The most general linear transformation is
\eq{
\nonumber
z_1=a\,\omega+b\,k\,,\qquad z_2=c\,\omega+d\,k\,,
}
assuming $ad-bc\neq0$ to be invertible. Demanding that $\omega_I\to\infty$ corresponds to $z_{1I},z_{2I}\to\infty$ requires that $a,c>0$. We can then re-scale them away setting $a=c=1$, and call $b\to\alpha_1$ and $d\to\alpha_2$. Demanding that the region of analyticity in terms of $z_1$ and $z_2$ includes PUHP implies that $\alpha_1>1$ and $\alpha_2<-1$. Another possibility is that $\alpha_1<-1$ and $\alpha_2>1$, but that exchanges the role of $z_1$ with $z_2$. Indeed the variables we have in \cref{z1z2} correspond to the case $\alpha_1=-\alpha_2$. In terms of the more general variables, the double dispersion relations of \cref{2dcauchy} and \cref{2disp} remain unchanged while \cref{2dispomegak} gets simply modified. This more general transformation will be be useful in \cref{imcons3}.}
\eq{
z_1\equiv\omega+\alpha k\,,\qquad z_2\equiv\omega-\alpha k \,,
\label{z1z2}
}
for some real constant $\alpha>1$. In these new variables, we denote the function as $\tG(z_1,z_2)=G(\omega,k)$. The domain of analyticity, $\omega_I>|k_I|$, in terms of the new variables is written as follows
\begin{equation}
    D=\left\{ (z_1,z_2)\in \mathbb{C}^2 \>|\> z_{2I}>-\frac{\alpha-1}{\alpha+1}z_{1I}\,,\>\text{and}\> z_{2I}>-\frac{\alpha+1}{\alpha-1}z_{1I}\right\}\,.
    \label{z1z2region}
\end{equation}
As shown in \cref{PlotRegions}, for $\alpha>1$, the region of analyticity always contains the UHP in each of the variables. We will call the subset of $\mathbb{C}^2$ with this property the poly-upper half-plane (PUHP)
\begin{equation}
    {\rm PUHP}=\{(z_1,z_2)\in \mathbb{C}^2 \>|\> z_{1I}>0\,,\>\text{and}\> z_{2I}>0\}\,.
\end{equation}
It is clear from \cref{PlotRegions} that $\tG(z_1,z_2)$ is analytic in the closure of the PUHP. In the limit $\alpha\to1$, the PUHP coincides with $D$.

The main advantage of introducing the parameter $\alpha$ is that the asymptotic behaviour in each variable can be related to short-distance physics, as we clarify now. We write the definition of the Fourier transform in the new variables
\eq{
\tG(z_1,z_2)=\frac{\alpha}{2}\int\dd{u_1}\dd{u_2}e^{i(z_1u_1+z_2u_2)/2}\tG(u_1,u_2)\,,
\label{u1u2}
}
where we have defined $u_1\equiv t-x/\alpha$ and $u_2\equiv t+x/\alpha$, with $\tG(u_1,u_2)=G(t,x)$, and the prefactor is the Jacobian of the transformation. Notice that the function $\tG(u_1,u_2)$ has support only when $u_1 > 0$ and $u_2 > 0$.
In the limit $|z_1|\to \infty$ in the UHP, keeping $z_2$ fixed (or vice versa), $z_1$ acquires an arbitrarily large imaginary part, leading to an exponential suppression $e^{-z_{1I}u_1}$ in the integrand. The integral is dominated by the values of $u_1$ for which $e^{-z_{1I}u_1}\sim \order{1}$, namely $z_{1I}u_1\ll 1$. Therefore, the main contribution comes from the neighbourhood of $u_1=0$, which, using causality and the fact that $\alpha>1$, corresponds to $t,x\to 0$. We conclude that the $z_{1I}\to +\infty$ limit probes the short distance regime of $\tG(u_1,u_2)$, or, equivalently, its UV behaviour (see also footnote 11 of \cite{Creminelli:2022onn}). A similar result holds if the role of $z_1$ and $z_2$ is exchanged.

Notice that if one sets $\alpha=1$ this property does not necessarily hold. The neighborhood of $u_1=t-x\to0$ corresponds to the response at null directions but not necessarily at short distances.\footnote{Consider for example $G(\omega,k)=\frac{i\omega}{\omega^2-k^2}$, which is the retarded two point function between a massless field and its conjugate momentum. For $\alpha=1$ we have $\tG(z_1,z_2)=i\frac{z_1+z_2}{4z_1z_2}$. While $G$ goes to zero at high frequencies, $\tG$ goes to a constant for $z_1\to\infty$ for fixed $z_2$. On the other hand, for $\alpha>1$, the asymptotic behavior of $\tG$ is the same as $G$.}

\fg{
\adjustbox{width=.75\textwidth,valign=c}{
\tikzset{every picture/.style={line width=0.85pt}}

\definecolor{ultramarine}{rgb}{0.07, 0.04, 0.56}
\definecolor{jasper}{rgb}{0.84, 0.23, 0.24}

\begin{tikzpicture}

\tikzset{ma/.style={decoration={markings,mark=at position 0.35 with {\arrow[scale=0.8]{>}}},postaction={decorate}}}
\tikzset{ma2/.style={decoration={markings,mark=at position 0.5 with {\arrow[scale=0.8]{<}}},postaction={decorate}}}

\draw[gray,thin] (-2.5,0) -- (2.5,0)[->]; 
\draw[gray,thin] (0,0) -- (0,2.5)[->]; 

\draw[gray,thin] (3,0) -- (8,0)[->]; 
\draw[gray,thin] (5.5,0) -- (5.5,2.5)[->]; 

\draw[ma,color=ultramarine,thin] (2,0) arc (0:180:2);
\draw[ma,color=ultramarine,thin] (-2,0) -- (2,0);

\draw[ma,color=ultramarine,thin] (7.5,0) arc (0:180:2);
\draw[ma,color=ultramarine,thin] (3.5,0) -- (7.5,0);

\draw[gray] (2.2,2.2) node{\footnotesize $\zeta_1$}; 
\draw[gray,thin] (2,2.3)|-(2.3,2);
\draw[gray] (7.7,2.2) node{\footnotesize $\zeta_2$}; 
\draw[gray,thin] (7.5,2.3)|-(7.8,2);

\filldraw[gray] (0.5,0.5) circle (0.6pt);
\draw[black] (0.7,0.3) node{\footnotesize $z_1$};  

\filldraw[gray] (4.5,0.7) circle (0.6pt);
\draw[black] (4.7,0.5) node{\footnotesize $z_2$};  

\draw[black] (2.75,1) node{\large$\times$};  

%
%
%

\end{tikzpicture}}
\caption{Integration contour used in \cref{2dcauchy}.}\label{2dcauchyarcs}
}
Since every analytic function of several variables is analytic in every variable keeping the others fixed (see for example \cite{Scheidemann_2023}), we can use Cauchy theorem multiple times, along the contour shown in \cref{2dcauchyarcs}
\eq{
\spl{
\tG(z_1,z_2)&=\frac{1}{2\pi i}\int\dd{\z_1}\frac{\tG(\z_1,z_2)}{(\z_1-z_1-i\epsilon)}\\
&=\frac{1}{(2\pi i)^2}\int\dd{\z_1}\dd{\z_2}\frac{\tG(\z_1,\z_2)}{(\z_1-z_1-i\epsilon)(\z_2-z_2-i\epsilon)}\,,
}
\label{2dcauchy}
}
where we have taken points $(z_1,z_2)$ in PUHP, the integral is along the real values and we have neglected the arc at infinity in each step assuming $\tG\to0$. In \cref{subtract} we will discuss dispersion relations with subtraction when one cannot assume that at infinity $\tG\to0$. The $i\epsilon$ is written explicitly to specify the integration contour when either of the variables becomes real.

Another useful relation can be derived by noticing that analyticity constrains the real part of the function, on the real plane, in terms of its imaginary part. As a result, one should be able to write a dispersion relation in terms of ${\rm Im}\,{\tG}(\z_1,\z_2)$ only. The easiest way to see this is to note that, by Cauchy theorem, if we put the poles in the lower half-plane we simply get zero. More precisely for $(z_1,z_2)$ in PUHP we obtain
\eq{
\frac{1}{(2\pi i)^2}\int\dd{\z_1}\dd{\z_2}\frac{\tG(\z_1,\z_2)}{(\z_1-z_1^*+i\epsilon)(\z_2-z_2^*+i\epsilon)}=0\,.
\label{2dcauchy2}
}
Subtracting the complex conjugate of \cref{2dcauchy2} from \cref{2dcauchy} gives the desired result\footnote{The argument leading to \cref{2disp} is analogous to the usual argument for a single variable. In that case, we put the point on the real line and deform the contour to circle around it from above, i.e.~equivalently moving the pole to the lower half plane. Calculating the contribution from the infinitesimal circle leads to a relation between real and imaginary part.}
\begin{equation}
   \boxed{\displaystyle\tG(z_1,z_2)=\frac{-i}{2\pi^2}\int\dd{\z_1}\dd{\z_2}\frac{{\rm Im}\,\tG(\z_1,\z_2)}{(\z_1-z_1-i\epsilon)(\z_2-z_2-i\epsilon)}}
\label{2disp} 
\end{equation}

This is the generalisation of the standard dispersion relation for a function of a single variable, assuming decay at infinity. Notice that in \cref{2disp}, the dependence on the parameter $\alpha$ is implicit through the definition of $\tG$ in terms of $G$. 

Let us make few comments about \cref{2disp}. First of all, we can re-write \cref{2disp} in terms of the original Fourier variables, namely in frequency and momentum. After trivial algebra, one obtains
\eq{
G(\omega,k)=\frac{-i\alpha}{\pi^2}\int\dd{\omega'}\dd{k'}\frac{{\rm Im}\,G(\omega',k')}{(\omega'-\omega-i\epsilon)^2-\alpha^2(k'-k)^2}\,,
\label{2dispomegak}
}
where we have changed the integration variables $\z_1=\omega'+\alpha k'$ and $\z_2=\omega'-\alpha k'$. Notice that it would have been  difficult to guess this relation from the beginning without using $z_1$ and $z_2$ variables.

Moreover, the multi-variable dispersion relation is equivalent to the Leontovich relations, both being direct consequence of analyticity (and decay at infinity). As mentioned briefly in the beginning of this section, to obtain Leontovich relation we use the fact that $G(\omega,k+\omega\xi)$, considered as a function of complex $\omega$, is analytic in the UHP as long as $|\xi|<1$.\footnote{We are imposing rotational invariance here. More generally, the second argument is $\vec{k}+\omega\vec{\xi}$ for $|\vec{\xi}|<1$. See \cite{Creminelli:2024lhd}.} We can then write a dispersion relation
\eq{
G(\omega,k)=\frac{1}{2\pi i}\int\frac{\dd{\z}}{(\z-\omega-i\epsilon)}G\Big(\z,k+(\z-\omega)\xi\Big)\,,
\label{leontovich}
}
where we have simply re-defined $k\to k-\omega\xi$. Indeed, it is not difficult to see that this simply corresponds to single dispersion relation with respect to $z_1$ (or equivalently $z_2$) with the parameter $\xi=1/\alpha$. More explicitly, starting from the first line of \cref{2dcauchy}, for fixed real $z_2$, we obtain \cref{leontovich}, after a suitable change of variables. Conversely, the double dispersion relation can be obtained from repeated application of \cref{leontovich} with two different parameters, i.e.
\eq{
G(\omega,k)=\frac{1}{(2\pi i)^2}\int\frac{\dd{\z}}{(\z-\omega-i\epsilon)}\int\frac{\dd{\z'}}{(\z'-\z-i\epsilon)}G\Big(\z',k+(\z-\omega)\xi+(\z'-\z)\xi'\Big)\,.
}
Selecting $\alpha=1/\xi=-1/\xi'$, after a change of variable recovers \cref{2disp}.\footnote{For $\xi\neq-\xi'$, this corresponds to \cref{2disp} with the more general transformation discussed in \cref{genz1z2} for $\alpha_1=1/\xi$ and $\alpha_2=1/\xi'$.}

Finally, single variable dispersion relations, evaluated on the real axis, relate the real and the imaginary part of the analytic function on the real axis. More precisely, it is easy to see that for real $z_1$ and $z_{2I}\geq0$
\eqa{
&{\rm Re}\,\tG(z_1,z_2)=\frac{1}{\pi}{\rm PV}_{z_1}\int\frac{\dd{\z_1}}{\z_1-z_1}{\rm Im}\,\tG(\z_1,z_2)\,,\label{realimag}\\
&{\rm Im}\,\tG(z_1,z_2)=-\frac{1}{\pi}{\rm PV}_{z_1}\int\frac{\dd{\z_1}}{\z_1-z_1}{\rm Re}\,\tG(\z_1,z_2)\,,
}
and similarly with the role of $z_1$ and $z_2$ exchanged. Double dispersion relations, such as \cref{2disp}, evaluated on the real plane give a consistency relation for the imaginary and the real part. From \cref{2disp}, using Sokhotski–Plemelj formula, $(x\pm i\epsilon)^{-1}={\rm PV}x^{-1}\mp i\pi\delta(x)$, and \cref{realimag} for real $z_1$ and $z_2$ we obtain 
\eqa{
&{\rm Im}\tG(z_1,z_2)=-\frac{1}{\pi^2}{\rm PV}_{z_1,z_2}\int\frac{\dd{\z_1}\dd{\z_2}}{(\z_1-z_1)(\z_2-z_2)}{\rm Im}\tG(\z_1,\z_2)\,,\label{imcons}\\
&{\rm Re}\tG(z_1,z_2)=-\frac{1}{\pi^2}{\rm PV}_{z_1,z_2}\int\frac{\dd{\z_1}\dd{\z_2}}{(\z_1-z_1)(\z_2-z_2)}{\rm Re}\tG(\z_1,\z_2)\,.\label{recons}
}
In the above relations the Cauchy principle value is imposed for both integrals. In other words, not only the real and imaginary part of analytic functions of several variables are related, as in the single variable case, but also they cannot be arbitrary functions as they must satisfy \cref{imcons} and \cref{recons}. We will discuss this fact in more detail in the following section.

\section{\label{sec:imconst}Microcausality constraints on the imaginary part}

The previous section showed that, assuming $G$ decays at infinity, knowledge of its imaginary part on the real plane is sufficient to determine the whole analytic function, via \cref{2disp}. (In \cref{subtract} we will relax the assumption of decay at infinity, introducing dispersion relations with subtractions.) In this section, we want to study the following question: are there constraints that ${\rm Im}\,G$, for real $\omega$ and $k$, must satisfy? (We could ask the same question about ${\rm Re}\,G$ with similar conclusions. We concentrate on the imaginary part, since this, as we will study in \cref{pos}, must also satisfy positivity properties. Moreover it is connected to fluctuations via the fluctuation-dissipation theorem, see \cref{sec:FD}.) In the single-variable case, the answer is negative: any function (or more generally tempered distribution) with suitable integrability conditions is viable and gives an analytic function in the UHP. 

This is, in general, no longer true for response functions of several variables. The easiest way to see this is to start from the position-space response function. A few lines of algebra show that 
\eq{
{\rm Im}\,G(k^\mu)=\int\dd[d]{x}\,e^{-ik\cdot x}\left[\frac{G(x)-G(-x)^*}{2i}\right]\,,
\label{imcomut}
}
where to obtain the second term, we have changed $x\to-x$ after complex conjugation of the Fourier transform, and we have used $d-$vector notation. Therefore, ${\rm Im}\,G(k^\mu)$ is the Fourier transform of the combination appearing in the brackets. Since the response function is retarded and micro-causal, the object in brackets is only micro-causal, i.e.~it vanishes outside the (complete) light cone. This implies that ${\rm Im}\,G(k^\mu)$ cannot be a completely generic function, as it is the Fourier transform of functions that are zero outside the light cone. Notice that this logic does not work for $d=1$: in that case, the function in brackets is neither retarded nor advanced, and therefore completely generic. This is consistent with the fact that there are no constraints for the single-variable case. Moreover, the argument above is independent of having rotational symmetry or satisfying un-subtracted dispersion relations.

In the rest of this section, $(i)$ we show that this implies that the imaginary part cannot have compact support in the $(\omega,k)$ plane, and $(ii)$ we write constraints on the functional form of the imaginary part, which we check explicitly in several examples. In the last two subsections, we comment on the fluctuation-dissipation theorem and on the extension of these arguments to higher-point functions.

\subsection{Support of the imaginary part}\label{imsupsec}

\fg{
\centering
\adjustbox{valign=c}{\begin{tikzpicture}[>=latex, line join=round, line cap=round, scale=1]

\coordinate (O1) at (0,0);
\coordinate (w1) at (0,2);
\coordinate (k1) at (2,0);
\coordinate (O2) at (4,0);
\coordinate (w2) at (4,2);
\coordinate (k2) at (6,0);
\coordinate (O3) at (8,0);
\coordinate (w3) at (8,2);
\coordinate (k3) at (10,0);


\draw[->, thick] (-1,0) -- (k1) node[below right] {$k$};
\draw[->, thick] (0,-1) -- (w1) node[above left] {$\omega$};
\draw[->, thick] (3,0) -- (k2) node[below right] {$k$};
\draw[->, thick] (4,-1) -- (w2) node[above left] {$\omega$};
\draw[->, thick] (7,0) -- (k3) node[below right] {$k$};
\draw[->, thick] (8,-1) -- (w3) node[above left] {$\omega$};

\path[
  pattern=north east lines,
  pattern color=red!60
]
(4.5,-1) rectangle (5.5,2);

\draw[thick, red!90!black] (4.5,-1) -- (4.5,2);
\draw[thick, red!90!black] (5.5,-1) -- (5.5,2);

\path[
  pattern=north east lines,
  pattern color=red!60
]
(7,0.5) rectangle (10,1.5);

\draw[thick, red!90!black] (7,0.5) -- (10,0.5);
\draw[thick, red!90!black] (7,1.5) -- (10,1.5);

\path[
  pattern=north east lines,
  pattern color=red
] plot [smooth cycle, tension=0.9]
  coordinates {(0.7,0.6)(.9,.3)(0.3,-0.3)(1.5,0)(1.7,0.6)(1.2,0.9)};

\draw[red!90!black,thick]
  plot [smooth cycle, tension=0.9]
  coordinates {(0.7,0.6)(.9,.3)(0.3,-0.3)(1.5,0)(1.7,0.6)(1.2,0.9)};

\end{tikzpicture}}
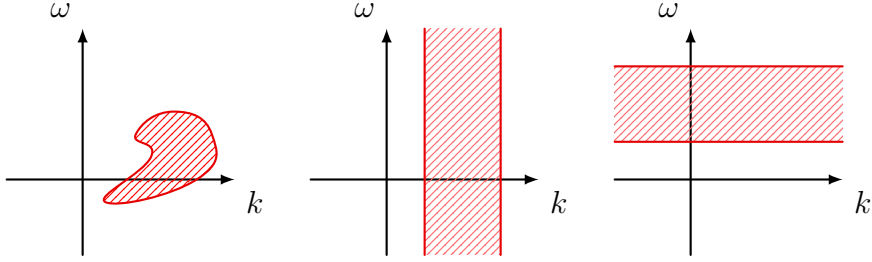
\caption{The three cases of compact support of ${\rm Im}\,G(\omega, \vec k)$ discussed in the text.}\label{Imsupport}
}

Weakly coupled systems can be described in terms of quasi-particles with given ``dispersion relations'' $\omega(k)$ (not to be confused with the dispersive integrals we study in this paper) that extend to infinity, both in $\omega$ and in $k$. In these cases, ${\rm Im}\,G(k^\mu)$ is not of compact support. We will now prove that this property is fully general and follows from Eq.~(\ref{imcomut}). To see this, imagine that ${\rm Im}\,G(k^\mu)$ is nonzero only in a compact subset, $K$, of the real $(\omega,\vec{k})$ space,\footnote{Since the discussion does not rely on the assumption of spherical symmetry, we have reinstated the vector notation on the spatial part. Moreover, we have chosen $d=4$ spacetime dimensions only for concreteness, while it is clear that the result holds for any $d>1$.} and zero elsewhere, see Fig.~\ref{Imsupport} left. Consider the inverse Fourier transform of the imaginary part\footnote{One can show that $G_c(x)$ corresponds to the expectation value of the commutator of the local operator divided by 2, hence the index $c$.}
\eq{
G_c(x)\equiv\int\frac{\dd[4]{k}}{(2\pi)^4}\,e^{ik\cdot x}\,{\rm Im}\,G(k^\mu)\,,
\label{eq:invFT}
}
where $G_c(x)$ is the combination in brackets on the right-hand side of \cref{imcomut} for real $x^\mu$. Paley–Wiener-Schwartz theorem (see for example \cite{stein1971introduction}, see also \cite{Hui:2025aja}) ensures that $G_c(x)$, being Fourier transform of a function with compact support, is an entire function of complex $x^\mu=x^\mu_R+ix^\mu_I$, i.e.~analytic in the whole $\mathbb{C}^4$.\footnote{Moreover, the exponential growth of the function is controlled by the size of $K$, but this will not be needed in the proof.} Intuitively this is because, whatever $x^\mu$ is, the integral in eq.~\eqref{eq:invFT} converges since the domain is restricted to $K$.  Micro-causality, on the other hand, implies that $G_c(x)$ vanishes for $x^\mu$ real with $x^2>0$. But then $G_c(x)$ must vanish \emph{everywhere} in $\mathbb{C}^4$, simply because all of the coefficients in its (convergent) power series expansion must be zero.\footnote{
More explicitly, consider a power series expansion around $x_*$ on the real plane, with $x_*^2>0$. All the real partial derivatives $\p G_c(x_*)/\p x_R^\mu$ vanish simply because $G_c(x)=0$ in the neighborhood of $x_*$. In addition, all the imaginary derivatives vanish due to analyticity, i.e. $\p G_c(x_*)/\p x_I^\mu=i\p G_c(x_*)/\p x_R^\mu$. Therefore, all complex derivatives vanish $\p G_c(x_*)/\p x^\mu=0$. The logic extends to all higher derivatives. This means that all of the coefficients in the series expansion around $x_*$ vanish. Since the function is everywhere analytic, it coincides with its series expansion and it vanishes.} We conclude that ${\rm Im}\,G(k^\mu)$ cannot be of compact support, unless it identically vanishes.

We can also rule out the case of compact support in $\vec{k}$ but otherwise arbitrary for $\omega$, see Fig.~\ref{Imsupport} center. One can do a Fourier transform in $\omega$ and get $G_c(t,\vec k)$, which is still of compact support in $\vec k$. This implies that $G_c(t, \vec x)$ is analytic in $\vec x$ for fixed $t$ \footnote{It is of course sloppy to fix $t$ since we are dealing with distributions. We should convolve with a test function in $t$ with a compact support, so that the response must still vanish for large enough $|\vec x|$.}, but this is not compatible with microcausality which, requires the function to vanish exactly for $|\vec x| > t$.

The case in which the support is compact in $\omega$ and arbitrary in $\vec{k}$, Fig.~\ref{Imsupport} right, is slightly different. In this case, following the same logic, $G_c(t,\vec{x})$ is an entire function of complex $t$ for fixed $\vec{x}$. $G_c(t,\vec{x})$ must vanish for $|t|>|\vec{x}|$ and this set exists only for nonzero $\vec{x}$. Therefore, any distribution supported in $\vec x =0$ is allowed. This is a linear combination of derivatives of $\delta(\vec x)$ in real space or polynomial in $\vec k$ in Fourier space. We conclude that, in this case, ${\rm Im}\,G(\omega,\vec{k})$ is a polynomial in $\vec k$ with coefficients which are functions of $\omega$ with compact support.

It is easy to verify these non-compactness properties in the examples we will study below.
%
%
%
\subsection{Functional form of the imaginary part }\label{imconssec}
Micro-causality restricts the functional form of the imaginary part as well as its support. We have already pointed out in the previous section that consistent imaginary parts must satisfy \cref{imcons}. This is a non-trivial consistency condition that a generic function of two variables does not satisfy. It is useful to rewrite \cref{imcons} in a more suggestive way. Notice that the integral operator acting on the right-hand side of \cref{imcons} is a double Hilbert transform with respect to each of the variables, namely \cref{imcons} can be written as 
\eq{
{\rm Im}\tG(z_1,z_2)=-\mathbb{H}_{z_1}\mathbb{H}_{z_2}{\rm Im}\tG(\z_1,\z_2)\,,
\label{HH1}
}
where it is implicit that the first operator acts on $\z_1$ and the second one acts on $\z_2$, and the Hilbert transform is defined as
\eq{\label{hilbertt}
\mathbb{H}_{z}\big(\dots\big)=\frac{1}{\pi}{\rm PV}\int\frac{\dd{\z}}{\z-z}\big(\dots\big)\,.
}
Remembering that $\mathbb{H}_{z}^{-1}=-\mathbb{H}_{z}$, we can re-write \cref{HH1} as $\mathbb{H}_{z_1}{\rm Im}\tG(\z_1,z_2)=\mathbb{H}_{z_2}{\rm Im}\tG(z_1,\z_2)$. More explicitly, a consistent imaginary part must satisfy
\eq{
\frac{1}{\pi}{\rm PV}\int\dd{\z_1}\frac{{\rm Im}\tG(\z_1,z_2)}{\z_1-z_1}=\frac{1}{\pi}{\rm PV}\int\dd{\z_2}\frac{{\rm Im}\tG(z_1,\z_2)}{\z_2-z_2}\,.
\label{imcons2}
}
Indeed, this relation is satisfied by the imaginary part of any function which is analytic in the PHUP and decays at infinity. The left-hand side, using a single dispersion relation for fixed $z_2$ as in \cref{realimag}, gives ${\rm Re}\tG(z_1,z_2)$, and the same holds for the right-hand side. However, it is very easy to violate this relation. Consider, as an example, a function of two variables that is factorized ${\rm Im}f_1(\z_1){\rm Im}f_2(\z_2)$ for some analytic functions $f_1$ and $f_2$ of single variables. Then the left-hand side of \cref{imcons2}, using the analyticity of $f_1$, is ${\rm Re}f_1(z_1){\rm Im}f_2(z_2)$, while the right-hand side, using the analyticity of $f_2$, is ${\rm Im}f_1(z_1){\rm Re}f_2(z_2)$, which are not equal in general.\footnote{Of course we know that $f_1(z_1)f_2(z_2)$ is analytic in the PUHP, and therefore, for instance, one combination that satisfies \cref{imcons2} is ${\rm Im}(f_1(\z_1)f_2(\z_2))={\rm Im}f_1(\z_1){\rm Re}f_2(\z_2)+{\rm Re}f_1(\z_1){\rm Im}f_2(\z_2)$.}   

Following the same argument leading to the Leontovich relation in \cref{leontovich}, we can re-write \cref{imcons2} in terms of ${\rm Im}G(\omega,k)$ as follows
\eq{
\frac{1}{\pi}{\rm PV}\int\dd{\z}\frac{{\rm Im}G(\z,k+\xi_1(\z-\omega))}{\z-\omega}=\frac{1}{\pi}{\rm PV}\int\dd{\z}\frac{{\rm Im}G(\z,k+\xi_2(\z-\omega))}{\z-\omega}
\label{imcons3}
}
for any $\omega$, $k$, $|\xi_1|<1$ and $|\xi_2|<1$. To derive this relation from \cref{imcons2}, one has to use the more general linear transformation discussed in \cref{genz1z2}.\footnote{Using \cref{z1z2}, one obtains for the right-hand side of \cref{imcons3}, $\xi_2=-\xi_1$.} An equivalent way to derive \cref{imcons3} is to start from the Leontovich relation, \cref{leontovich}, and demand consistency of the result by changing $\xi$. This argument has been discussed previously in \cite{Creminelli:2024lhd}. 

We should note that \cref{imcons3} is in fact a direct consequence of micro-causality and the decay assumption. More precisely, it is easy to see that
\eq{\label{commxi}
G_c(x)=G_c(x)\big[\theta(t-\vec{x}\cdot\vec{\xi_1})+\theta(-t-\vec{x}\cdot\vec{\xi_2})\big]\,,
}
where $G_c(x)$ is the combination in brackets in \cref{imcomut} which vanishes outside the light-cone, and $\vec{\xi}_1$ and $\vec{\xi}_2$ are generic vectors with norm less than unity, as in the Leontovich relation. Taking the Fourier transform of both sides, using the convolution theorem together with the decay assumption gives \cref{imcons3}. We also conclude that the constraints on the functional form of the imaginary part exist as a consequence of micro-causality only, without the need of rotational symmetry and decay at infinity. We will see their explicit form in \cref{subtract} and \cref{norot}.

The consistency condition of \cref{imcons3} implies that the integral on the left-hand side is independent of $\xi_1$. Therefore, equivalently, after taking the derivative with respect to $\xi_1$, we must have
\eq{
\int\dd{\z}\p_k{\rm Im}\,G(\z,k+\xi(\z-\omega))=0\,,
\label{imcons4}
}
for any $\omega$, $k$, and $|\xi|<1$, and there is no need for the Cauchy principal value as there is no pole left after differentiation. Since this must be true for arbitrary $k$, we can drop $\omega$ dependence by setting it to zero. In case we know that the integral over the imaginary part (without derivative) is convergent, then \cref{imcons4} implies its $k$ independence. More explicitly, we must have 
\eq{
\int\dd{\z}{\rm Im}\,G(\z,k+\xi \z)=\int\dd{\z}{\rm Im}\,G(\z,\xi \z)\,,
\label{imconsmid}
}
where on the right-hand side we have chosen $k=0$. The right-hand side is particularly simple for physical systems: reality of the response function in real space implies that its imaginary part in Fourier space must be an odd function of $\omega$, while, assuming rotational invariance, it is an even function of $k$. As a result, the integral on the right-hand side is vanishes. This implies
\begin{equation}
\boxed{\displaystyle\int \dd{\z}\,{\rm Im}\,G(\z,k+\xi \z)=0}
\label{imcons5}
\end{equation}
We should stress that \cref{imcons5} holds only if we have reality condition, as is the case for physical systems, and rotational symmetry (or at least parity), together with the assumption of convergence of the integral. The latter is stronger than the convergence of the dispersion relation that we started with. On the other hand, \cref{imcons4}, does not rely on neither the reality condition nor any stronger integrability assumptions on the imaginary part.

It is important to stress that, assuming proper decay at infinity, \cref{imcons4} (or \cref{imcons5}) is a necessary {\em and sufficient} condition for microcausality. Indeed, \cref{imcons4} implies \cref{imcons3} and the latter, once ${\rm Re}\,G$ is suitably chosen to be compatible with the imaginary part, implies the Leontovich relations, which are equivalent to microcausality. Eq.~(\ref{imcons4}) and \cref{imcons5} represent the main result of this work.

We should note that dispersion relations, given some input on the right-hand side, generate analytic functions. In other words, in \cref{2disp}, whatever function we choose for the imaginary part in the integrand, the result of the integration, if convergent, turns out to be an analytic function of $z_1$ and $z_2$ in the PUHP. This is because the kernel $(\z_1-z_1-i\epsilon)^{-1}(\z_2-z_2-i\epsilon)^{-1}$ is analytic. This is regardless of the fact that our chosen function for the imaginary part satisfies the integral constraint of \cref{imcons2} or not. If \cref{imcons2} is not satisfied, however, the imaginary part of the generated analytic function will not coincide with the function we started with. Once again, consider the example discussed above. Taking the imaginary part to be ${\rm Im}f_1(\z_1){\rm Im}f_2(\z_2)$, the integration of \cref{2disp} simply gives $-if_1(z_1)f_2(z_2)/2$ which is indeed analytic in the PUHP. On the other hand, its imaginary part is $({\rm Im}f_1(\z_1){\rm Im}f_2(\z_2)-{\rm Re}f_1(\z_1){\rm Re}f_2(\z_2))/2$ which is not the original function.
 

\subsection{Illustrative examples for the integral constraints}\label{examp}
In this subsection, we check the integral constraints \cref{imcons5} (or \cref{imcons4}) on the imaginary part of the response function in several physical systems. These examples show that the integral condition, which is nothing but constraints imposed by microcausality, acts as a nontrivial consistency requirement, which constrain the admissible parameter space of a given model.

\subsubsection{Lorentz invariant case}\label{lorinv}
The integral constraints on the imaginary are direct consequence of microcausality, and therefore must be satisfied in the special case of Lorentz invariant systems, where the response function is essentially function of a single variable $\omega^2-k^2$. In general, the imaginary part of the response function can be written as follows
\eq{
{\rm Im} \,G(\omega,k)={\rm Sgn}(\omega)\theta(\omega^2-k^2)\rho(\omega^2-k^2)\,,
}
where ${\rm sgn}$ is the sign function to ensure that the imaginary part is odd in $\omega$, a consequence of reality. The spectral function $\rho$ is only a function of the Lorentz-invariant combination $\omega^2-k^2$. The step function is to ensure that the support is only on $\omega^2-k^2 \ge 0$, i.e.~positive mass squared.\footnote{Although this is not needed here, the spectral function must be positive, see Sec.~\ref{pos}.} Assuming suitable convergence, we can check whether \cref{imcons5} is satisfied. For the left-hand side, we get
\eq{
\int_0^{+\infty}\dd{\z}\Big[\theta(\z^2-(k+\z\xi)^2)\rho(\z^2-(k+\z\xi)^2)-\{\z\to-\z\}\Big]\,.
}
Changing the integration variable $u^2=\z^2-(k+\z\xi)^2$, for the first term, we obtain
\eq{
\int_0^{+\infty}\frac{u\dd{u}}{\sqrt{k^2+u^2(1-\xi^2)}}\rho(u^2)\,,
\label{ztou}
}
while for the second term, we choose $v^2=\z^2-(k-\z\xi)^2$, with the same result as \cref{ztou}.  (Notice that the presence of the step function forces to select only one of the square-root branches.) The two contributions thus cancel and \cref{imcons5} is satisfied.

It is easy to see that adding a speed of sound, i.e. $\rho(\omega^2-c_s^2k^2)$, does not change much in the above analysis, as long as $c_s \le 1$. This proves the validity of the integral constraints, at least for a subset of systems with broken Lorentz symmetry.

\subsubsection{Response by a single mode}
There are cases in which the response of the system is dominated by a single mode with finite lifetime and a dispersion with $c_s \le 1$. After suitable normalization, the response function takes the following form
\eq{
G(\omega,k)=\frac{1}{-\omega^2-i\gamma\omega+c_s^2k^2+m^2}\,,
\label{drude}
}
with $\gamma>0$, $0 \le c_s^2 \le 1$ and $m^2 \ge 0$ (see \cite{Creminelli:2024lhd}, Appendix E, for a discussion). The corresponding imaginary part is 
\eq{
{\rm Im}\,G(\omega,k)=\frac{\gamma\omega}{(-\omega^2+c_s^2k^2+m^2)^2+\gamma^2\omega^2}\,.
}
Since the imaginary part is a rational function, we can explicitly perform the integral in \cref{imcons5} using Cauchy theorem. The function has four simple poles that appear in complex conjugate pairs. Deforming the contour to either the upper or the lower half plane, we pick up two poles while the contribution from the arc at infinity vanishes. It is not difficult to see that the residue of the two poles are equal in magnitude but opposite in sign, and therefore their sum cancels, consistent with \cref{imcons5}. This is a direct consequence of causality; if we violate one of the conditions $\gamma>0$, $0 \le c_s^2 \le 1$ and $m^2 \ge 0$, then there exist $k$ and $\xi$ such that \cref{imcons5} is no longer satisfied.\footnote{In the case of \cref{drude}, for $c_s<1$ the region of analyticity is larger than the generic case: the constraints remain valid for $|\xi| \le c_s^{-1}$. On the other hand for $c_s>1$, \cref{imcons5} is violated for $c_s^{-1}>\xi>1$.}  

\subsubsection{Kinetic theory}\label{kin}
For more complicated systems, a description in terms of quasi-particles, such as the one studied above, may not be possible. Alternatively, it has been shown that the dynamics of the collective modes of the system is captured qualitatively (and even quantitatively) by (relativistic) Boltzmann equation \cite{Romatschke:2015gic,Kurkela:2017xis}. In this approach the system is described by a phase space distribution function $f(x^i,p^i;t)$ with time evolution described by
\eq{
\Big(p^\mu\p_{x^\mu}+F^\mu\p_{p^\mu}\Big)f=\frac{p^\alpha u_\alpha}{\tau}(f-f_{eq})\,,
\label{BoltRTA}
}
in which $x^\mu=(t,x^i)$, $p^{\mu}=(E(p^i),p^i)$ with $E(p^i)$ being the (relativistic) single-particle dispersion relation, $F^\mu$ is any external force applied to the system, $u^\mu(x)$ is the macroscopic local $4-$velocity, $\tau$ is a time scale associated to the relaxation of the system after perturbing near an equilibrium state, and $f_{eq}(x^i,p^i)$ is an equilibrium distribution function which depends on $u^\mu(x)$ as well as the local temperature and chemical potential, $T(x)$ and $\mu(x)$. In \cref{BoltRTA} the collision term is approximated by a term which tends to recover the equilibrium state in time $\sim\tau$. This is know as relaxation time approximation (RTA) and the term on the right-hand side is called Bhatnagar–Gross–Krook (BGK) operator. By looking at different external forces one can obtain the corresponding response function at linear order. For instance, by putting the system in an external electromagnetic field, for which $F^\mu=F^{\mu\nu}p_\nu$ with $F_{\mu\nu}=\p_\mu A_\nu-\p_\nu A_\mu$, we can obtain the two point function, $G^{\mu\nu}(x)=i\theta(t)\ev{[J^\mu(x),J^\nu(0)]}$, of the associated conserved current
\eq{
J^\mu(x)=\int\frac{\dd[3]{p}}{(2\pi)^3}\frac{p^\mu}{E(p^i)}f(x^i,p^i;t)\,,
}
which describes diffusion of the conserved charge. In Fourier space, we can have only two structures\footnote{This definition differs from \cite{Creminelli:2024lhd} in the transverse part: $G_T^{\rm here}=k^2G_T^{\rm there}/\omega^2$. With the new definition, $G_L\propto \varepsilon_L-1$ and $G_T\propto \varepsilon_T-1$. Moreover, there is a minus sign difference between our definition of the retarded two-point function and the one in \cite{Romatschke:2015gic}.}
\eq{
G^{ij}(\omega,k^i)=\omega^2G_L(\omega,k)\frac{k^ik^j}{k^2}+\omega^2G_T(\omega,k)\left(\delta^{ij}-\frac{k^ik^j}{k^2}\right)\,,
}
where the two functions $G_L$ and $G_T$ specify the response to a longitudinal and transverse perturbation. Following \cite{Romatschke:2015gic}, one can consider external perturbation with $k^i=(0,0,k)$. Then
\eqa{
&G_L=\frac{1}{k^2}G^{00}=\frac{\chi}{k^2}\frac{2k\tau+(1-i\omega\tau)i\log\left(\frac{\omega-k+i/\tau}{\omega+k+i/\tau}\right)}{2k\tau+i\log\left(\frac{\omega-k+i/\tau}{\omega+k+i/\tau}\right)}\,,\\
&G_T=\frac{1}{\omega^2}G^{11}=\frac{-i\tau\chi}{4\omega}\left[\frac{2(1-i\omega\tau)}{(k\tau)^2}+\frac{(1-i\omega\tau)^2+(k\tau)^2}{(ik\tau)^3}\log\left(\frac{\omega-k+i/\tau}{\omega+k+i/\tau}\right)\right]\,,
}
in which $\chi>0$ is called static charge susceptibility. It is easy to check that for $\omega\to\infty$ with fixed $k$, both functions decay as 
\eq{
G_L\sim G_T\sim-\frac{\chi/3}{\omega^2}\,,
\label{plasma}
}
known as plasma behavior. The same holds for $\omega\to\infty$ with the parametrization $k=q+\omega\xi$. One can explicitly check that, as a consequence of causality, these functions are analytic in the FLC and all the singularities lie outside of this region, for $\tau>0$. This shows that the response obtained in \cite{Romatschke:2015gic} is indeed microcausal. Moreover, in the small momentum limit, the difference $G_T-G_L$ goes to zero at least as fast as $\sim k^2$ which is required by analyticity of the full matrix $G^{ij}$ \cite{Creminelli:2024lhd}. 

Equivalently, we can check the integral constraints on the imaginary part\cref{imcons5}. Because of the complicated structure of the response function, we could only check \cref{imcons5} numerically: the constraints are satisfied for ${\rm Im}\,G_L$ for a wide range of parameters $k$ and $\xi$ as expected. For the transverse response one has to be careful about convergence of the integrals. Both $G_T$ and ${\rm Im}\,G_T$ diverge as $\sim 1/\omega$ at small frequencies and therefore the dispersion relations and consequently \cref{imcons5} do not converge. Instead, we can look at $G^{11}=\omega^2G_T$ which is finite at low frequencies, but does not decay at high frequencies according to \cref{plasma}. Therefore, we must write dispersion relations with subtractions and slightly modified integral constraints \cref{eq:CLSZsub}. We have verified numerically that the modified constraints are in fact satisfied.

Finally, we can repeat this exercise for the two-point function of the stress-tensor, corresponding to the response of the system to a background metric, which describes transport of energy and momentum in the system. There will be three response functions associated to scalar (sound), vector (shear), and tensor modes. The expressions are given explicitly in \cite{Romatschke:2015gic}, for which we checked numerically that \cref{imcons5} is satisfied.  

\subsubsection{\label{sec:chemical}Finite chemical potential}
Starting from a Lorentz invariant field theory with a $U(1)$ symmetry, we can study states at finite charge density or chemical potential. Such a state minimizes the modified Hamiltonian $H-\mu Q$, with $Q$ the conserved charge and $\mu$ the chemical potential, and breaks Lorentz boosts spontaneously. In practice, we can access this state by introducing a background gauge field $A^\mu=(\mu,\vec{0})$ with $\p_\mu\to\p_\mu-iA_\mu$. Equivalently, we can redefine the complex fields as $\Phi\to e^{i\mu t}\Phi$.   

\paragraph{Renormalizable $U(1)$ model} The simplest example arises from a renormalizable theory of a conformal complex scalar field (see for example \cite{Creminelli:2022onn,Creminelli:2023kze,Hui:2023pxc}). Following the notation of \cite{Creminelli:2023kze} one has
\begin{equation}
\mathcal{L}=-\partial \Phi^{\dagger}\partial \Phi-\lambda (\Phi^{\dagger}\Phi)^2
=-\frac{(\partial \rho)^2}{2}-\frac{\rho^2}{2 v^2} (\partial \theta)^2-\frac{\lambda}{4}\rho^4\,,
\end{equation}
where we have written $\Phi=\rho e^{i\theta/v}/\sqrt{2}$. A finite chemical potential can be introduced by expanding the phase around a time-dependent background, $\theta=\hat\mu^2 t/2+\pi$, leading to $v=\hat\mu/(4\lambda)^{1/4}$.\footnote{What is indicated with $\mu$ in ref.~\cite{Creminelli:2023kze} is not the chemical potential: for this reason we call it $\hat\mu$ here. The chemical potential can be written as $\mu = \hat\mu^2/v$.} This state spontaneously breaks boosts; also the global $U(1)$ symmetry and time translations are broken to a diagonal subgroup. Nonzero $\hat{\mu}$ induces an energy-dependent mixing between the phase and radial modes, such that the quadratic Lagrangian cannot be diagonalized by a local field redefinition. Consequently, both the phase and radial fields, $\pi$ and $\rho$, interpolate the two propagating degrees of freedom, whose dispersion relations are
\begin{equation}
\omega^2_{\pm}(k)=k^2+\frac{3}{4}\frac{\hat\mu^4}{v^2}\pm \sqrt{\frac{\hat\mu^4}{v^2} k^2+\left(\frac{3}{4}\frac{\hat\mu^4}{v^2}\right)^2}\, .
\end{equation}
Because of the mixing, the retarded Green’s function of $\pi$ (and similarly for $\rho$) receives contributions from both $\pm$ states:
\begin{equation}
G^{\pi}(\omega,k)= \frac{|Z^{\pi}_{-}|^2}{(\omega+i \epsilon)^2-\omega_{-}^2(k)}
+\frac{|Z^{\pi}_{+}|^2}{(\omega+i \epsilon)^2-\omega_{+}^2(k)}\,,
\end{equation}
where $Z^{\pi}_{\pm}(k)$ are the interpolating factors defined by $Z^{\pi}_{\pm}\equiv \langle 0|\pi(0)|\pm,\vk\rangle$; their explicit expression is given in \cite{Creminelli:2023kze}. The imaginary part is then
\begin{equation}
{\rm Im}\,G^{\pi}(\omega,k)
=\pi\,{\rm Sgn}(\omega)\left[|Z^{\pi}_{-}|^2\delta(\omega^2-\omega^2_{-})
+|Z^{\pi}_{+}|^2\delta\big(\omega^2-\omega^2_{+}\big)\right]\,.
\end{equation}
When evaluating the integral in \cref{imcons5}, it is important that \emph{both} gapped and gapless modes are included; the integral does not vanish if one considers only the gapless mode contribution. Indeed the separation between the two modes is non-local and manifests in the interpolating factors $Z^{\pi}_{\pm}$ being nontrivial functions of $k$.  The integral vanishes in \cref{imcons5} due to a detailed cancellation between the two modes, for arbitrary $k$ and $0 \le \xi<1$. 

\bigskip

It is important to stress that the microcausality condition \cref{imcons5}, even though it integrates over all frequencies, it effectively receives contributions for energies $\zeta \sim k$ (we assume $\xi$ of order unity). Indeed, the discussion above \cref{imcons5} implies that the integral does not receive contribution both for $\zeta \ll k$, when one can neglect $\xi \zeta$ in the second argument of $G$, and for $\zeta \gg k$ when $k$ can be neglected. This implies, as expected, that a violation of microcausality cannot be ``cured'' adding new states at arbitrarily high mass. Let us, for example, consider the theory of a Goldstone, like the one just studied, with a dispersion relation characterised by a scale $M$: $\omega^2 = c_s^2 k^2 +k^4/M^2 + \ldots$ This dispersion relation is not micro-causal on its own and one gets a contribution to  \cref{imcons5} going as $k^2/M^2$ for small $k$. In this regime, the addition of a heavy mode with mass $M_h$ will be suppressed in \cref{imcons5} by $k^2/M_h^2$, so that one cannot have a parametric separation $M^2 \ll M_h^2$. This says that micro-causality is something that can be checked at low energy without the full knowledge of the UV. Indeed, in the example above, the radial mode appears at the same scale at which the Goldstone would start showing problems with micro-causality.\footnote{For fixed $\mu$ and $\xi$, the value of the integral constraint, considering only the contribution from the gapless mode, reaches a maximum at $k\sim \sqrt{\mu^2/v}$.}

\paragraph{Degenerate Fermions (the Lindhard function)}
The electromagnetic response of a degenerate Fermi gas, i.e.~at finite chemical potential, at zero temperature has been calculated by Lindhard in the non-relativistic limit. A derivation, including relativistic effects, is presented in \cite{Creminelli:2024lhd}. Here, we focus on the longitudinal component of the retarded current–current correlator, corresponding to the $(\mu=0,\nu=0)$ part of the tensor---see eq.~(C.10) of \cite{Creminelli:2024lhd}. The imaginary part of the retarded correlator arises from the logarithmic term, which is supported only when its argument becomes negative. The imaginary part is given by\footnote{Note that in eq.~(C.10) of \cite{Creminelli:2024lhd}, the $\omega\to-\omega$ term can be translated to $E_q\to-E_q$ because the building blocks  $(2E_p+\omega)^2$ and $\omega(\omega+E_p)$ goes to $(-2E_p+\omega)^2$ and $\omega(\omega-E_p)$ respectively.}
\eq{
\begin{aligned}
\mathrm{Im}\,G(\omega,k) = \sum_{\eta = \pm 1}\int_{0}^{k_F}\frac{q^2\dd{q}}{2\pi E_q} 
&\frac{(2 \eta  E_q+\omega)^2-k^2}{4 q k^3}~\theta\Bigg(\frac{2 q k-k^2+\omega \left(2 \eta  E_q+\omega\right)}{2 q k+k^2-\omega \left(2 \eta  E_q+\omega\right)}\Bigg)
{\rm Sgn}\Big(k( \eta  E_q+\omega)\Big)\,,
\end{aligned}
\label{LindH}
}
where $E_q=\sqrt{q^2+m^2}$, $k_F$ is the Fermi momentum and $m$ is the mass of the fermion. We want to check that this function satisfies the integral constraints \cref{imcons5}, over ${\rm Im}\,G(\z,k+\z\xi)$. Before performing the $q$ integral in \eqref{LindH}, we perform the $\zeta$ integral first and show that the $\zeta$ integral vanishes for any value of $q$. For fixed $q$ satisfying $E_q<k/3$, the step function is nonzero in two disjoint intervals,
\eq{
\zeta^{-}(q) \le \zeta \le \zeta^{-}(-q)\,,\quad\text{and}\,\quad \zeta^{+}(-q) \le \zeta \le \zeta^{+}(q)\,,
}
where the sign function is positive over the first and negative over the second, and the boundaries are defined
\eq{
\z^{\pm}(q)=\frac{k \xi-\eta E_q+\xi  q\mp \sqrt{2 k \left(q-\xi  \eta E_q\right)+k^2+\left(\eta E_q-\xi  q\right){}^2}}{1-\xi ^2}\,.
}
The integral over $\z$ can be performed analytically,
\eq{
\sum_{\eta=\pm1}\Bigg[\frac{(\xi ^2-1)}{4\xi^3q}\log(k+\xi\zeta)-\frac{(k+2 \xi\eta E_q)(3 k-2 \xi  \eta E_q+4 \xi \zeta)}{8 \xi ^3 q\left(k+\xi  \zeta\right){}^2}\Bigg]_{\zeta^-(q)}^{\zeta^-(-q)}-\Bigg[\cdots\Bigg]_{\zeta^+(-q)}^{\zeta^+(q)}\,.
}
It is not difficult to check that the logarithmic terms give $\log (1) =0$ independent of $q$, while the rest cancel out. The algebra for fixed $q$ in $E_q>k/3$ is similar: it involves multiple intervals which integrate to zero. This confirms the integral constraints in \cref{imcons5}.

\subsubsection{Finite temperature}\label{finT}
Another way to break Lorentz boosts spontaneously is to consider a Lorentz invariant theory at finite temperature. The state can be described by the density matrix $\rho=e^{-\beta H}/Z$, with $H$ the Hamiltonian, $\beta$ the inverse temperature, and $Z=\Tr(e^{-\beta H})$ the partition function. Calculations at finite temperature are usually done using the imaginary time formalism. The real time observables, such as two-point correlation functions, can be obtained by suitable analytic continuation. One can show that the retarded Green's function is
\eq{
G(\omega,k)=G_E(i\omega_n\to\omega+i\epsilon,k)\,,
}
in which $G_E(\tau,x)=\ev{T_E\oo(\tau,x)\oo(0)}$ is the Euclidean time-ordered two point function. Since the two point function is periodic in Euclidean time, the Fourier transform is nonzero only at discrete Matsubara frequencies $\omega_n=2\pi n/\beta$, with $n>0$. 

The simplest example (perhaps too simple) to consider is a free theory at finite temperature. In this case, the retarded two-point function is the same as in the Lorentz invariant vacuum, since $[\phi(x),\phi(0))]$ is a $c-$number in free theory. Therefore, the discussion of \cref{lorinv} applies. 

A more interesting example is provided by a conformal field theory (CFT) at finite temperature. The calculation is in particular simple for $d=2$ in which the two-point function is fixed (up to normalization) by symmetries (see for example \cite{Iliesiu:2018fao}). The imaginary part of the retarded Green's function has a simple closed form \cite{Son:2002sd,Becker:2014jla}
\eq{
\mathrm{Im}\,G(\omega,k)=\sinh(\frac{\beta\omega}{2}) \left| \Gamma\left(\frac{\Delta}{2}+\frac{i\beta(\omega +k)}{4\pi}\right) \Gamma\left(\frac{\Delta}{2}+\frac{i\beta(\omega -k)}{4\pi}\right)\right| ^2\,,
\label{imcft}
}
up to a normalization constant, in which $\Delta>0$ is the scaling dimension of the operator. By conformal symmetry, the asymptotic behavior of the two point function is $\sim \omega^{2\Delta-2}$ which is the same also for the imaginary part in \cref{imcft}. Therefore, we can write un-subtracted dispersion relation for $\Delta<1$ and the integral over the imaginary part is convergent for $\Delta<1/2$. One can verify that for $\Delta<1$ the integral constraints in the form of \cref{imcons4}, or \cref{imcons5} for $\Delta<1/2$, are indeed satisfied. For higher values of $\Delta$, one has to write subtracted dispersion relations, to be discussed in \cref{subtract}: the constraints involve additional $k-$derivatives compared with \cref{imcons4}, see \cref{eq:CLSZsub}. One can check that, as expected, these are satisfied for \cref{imcft}.

\subsection{\label{sec:FD}Causality constraints on fluctuations}
The integral constraints \cref{imcons5} on the imaginary part of the retarded two-point function must be satisfied for any state as long as the underlying theory is micro-causal. For the specific case of thermal states, different types of two-point functions, i.e.~retarded, advanced etc., are simply related; they are all different analytic continuations of the Euclidean two-point function. As a result, one can derive constraints on the other correlation functions, for which the conditions of micro-causality are not manifest (e.g.~they do not contain a commutator).

An example of such relation is the well-known fluctuation-dissipation theorem, which relates the fluctuation (also known as power spectrum) in Fourier space to the imaginary part of the linear response, i.e.~dissipation. 
We begin by defining the basic two-point correlation function
\begin{equation}
S(t,\vec{x}) \equiv 
\langle \mathcal{O}(t,\vec{x}) \mathcal{O}(0,0) \rangle \,,
\end{equation}
where time translation and spatial homogeneity allow us to evaluate one operator at the origin.  
The Fourier transform is then
\begin{equation}
S(\omega,\vec{k}) = 
\int \dd t\, \dd^{d-1}x\; e^{i(\omega t - \vec{k}\cdot \vec{x})}
S(t,\vec{x}) \,.
\end{equation}
Using the definition of $G$ in terms of commutators, \cref{retG}, the dissipation (imaginary part) of the response function in position-space \eqref{imcomut} can be written as
\begin{align}
-\frac{i}{2}\Big[ G(t,\vec{x}) - G(-t,-\vec{x}) \Big]
&= 
\frac{1}{2}\Big[
   \langle \mathcal{O}(t,\vec{x}) \mathcal{O}(0,0) \rangle
 - \langle \mathcal{O}(0,0) \mathcal{O}(t,\vec{x}) \rangle
   \Big]\;.
\end{align}
One can connect this with the above two-point correlator $S$ using the Kubo-Martin-Schwinger (KMS) condition (work in the canonical ensemble, where the thermal average is taken with trace over $\rho = e^{-\beta H}$)
\eq{
 \ev{\oo(0)\oo(t,\vec{x})}=\ev{\oo(t-i\beta,\vec{x})\oo(0)}\,.
 \label{kms}
 }
Hence,
\begin{equation}
-\frac{i}{2}\Big[ G(t,\vec{x}) - G(-t,-\vec{x}) \Big]
= 
\frac{1}{2}\Big[
   \langle \mathcal{O}(t,\vec{x}) \mathcal{O}(0,{0}) \rangle
 - \langle \mathcal{O}(t - i\beta,\vec{x}) \mathcal{O}(0,{0}) \rangle
   \Big].
\end{equation}
Finally, Fourier transforming to frequency--momentum space gives
\begin{equation}\label{eq:FDtheorem}
\Im G(\omega,\vec{k})
= 
\frac{1}{2}\Big[ 1 - e^{-\beta\omega} \Big] S(\omega,\vec{k})\,.
\end{equation}
This is the celebrated fluctuation-dissipation theorem \cite{Kubo:1966fyg}.\footnote{The KMS condition implies that $S(-\omega,\vec k)=e^{-\beta \omega}S(\omega, \vec k)$: this is consistent in \cref{eq:FDtheorem} with the fact that $\Im G(\omega,\vec{k})$ is odd in $\omega$.} 
Until now, we kept the notation without assuming rotational invariance, since a modified version of \cref{imcons5} will be applicable in such cases as well (see \cref{norot}). 
 Assuming rotational invariance, we can rewrite the integral constraints of \cref{imcons5} in terms of fluctuations 
 \eq{\boxed{
 \int\dd{\z}\Big[ 1 - e^{-\beta\z} \Big]S(\z,k+\z\xi)=0}
 }

 It is remarkable that, via the fluctuation-dissipation theorem, the spectrum of fluctuations is sensitive to the microcausality constraints. In particular, also the discussion above about the support of the imaginary part can be rephrased in terms of the power spectrum $S(\omega,k)$. This remains true at $T=0$, i.e.~$\beta\to \infty$ when the power spectrum $S$ coincides with $\Im G$.


\subsection{Non-linear response}
The focus of this article is on the linear response. On the other hand, the mathematics that led to the constraints of \cref{imcons5}, which simply was analyticity in the PUHP, can be used in different contexts. One example is the non-linear response function. Let us ignore spatial dependence for simplicity, then the first non-linear response $G_{NL}(t_1,t_2)$ corresponds to 
\eq{
\oo_{J}(t)=\oo_{J=0}(t)+\int\dd{t'}G(t-t')J(t')+\frac{1}{2}\int\dd{t_1}\dd{t_2}G_{NL}(t-t_1,t-t_2)J(t_1)J(t_2)+\dots\,,
}
which relates the observable to the external source beyond linear order. Causality, i.e. retardation in this case, ensures that $G_{NL}(t_1,t_2)=0$ if either $t_1<0$ or $t_2<0$. Then the same logic of \cref{Bogo} implies analyticity of
\eq{
G_{NL}(\omega_1,\omega_2)=\int\dd{t_1}\dd{t_2}e^{i\omega_1t_1+i\omega_2t_2}G_{NL}(t_1,t_2)\,,
}
in the PUHP of complex $\omega_1$ and $\omega_2$. This argument is not new and it has been discussed and used especially in the non-linear optics literature (see for example \cite{PhysRevB.44.8446,Hutchings1992}). 

Here we want to stress that one can write a multi-variable dispersion relation of the form \cref{2disp}, and that the imaginary part ${\rm Im}G(\omega_1,\omega_2)$ must satisfy the analog of the integral constraint in \cref{imcons2}, i.e.~$\mathbb{H}_{\omega_1}{\rm Im}\,G_{NL}(\z_1,\omega_2)=\mathbb{H}_{\omega_2}{\rm Im}\,G_{NL}(\omega_1,\z_2)$. The same logic can be applied to the real part of the response. In a similar way one can rephrase the above discussion about the support of the imaginary part. These are general constraints on the non-linear response of the system, as a consequence of retardation.  

As an example, consider the simple Drude model, i.e.~the one-dimensional analog of \cref{drude}, with linear response $D(\omega)=(-\omega^2-i\gamma\omega+m^2)^{-1}$, and add an an-harmonic (quadratic in the field) term to the dynamical equation. Perturbation theory gives for the non-linear response
\eq{
G_{NL}(\omega_1,\omega_2)=D(\omega_1)D(\omega_2)D(\omega_1+\omega_2)\,.
}
This is a simple model for second-harmonic generation in nonlinear optics (see for instance \cite{10.5555/1817101}). Analyticity in the PUHP is obvious from the analyticity of the Drude model. It is also straightforward to verify (numerically) that the integral constraints are indeed satisfied.

\section{Positivity in the complex plane}\label{pos}

The imaginary part of the response function in Fourier space specifies the dissipation of energy of an external source coupled to the system, as in \cref{diss}.\footnote{This is true only if we are studying the two point function of a single observable. In cases that there are many observables, the dissipative part of the linear response (matrix) will be $(G-G^\dagger)/2i$. This coincides with the imaginary part if the system has spacetime parity symmetry. See \cite{Creminelli:2024lhd} for a discussion. For multiple operators the positivity property turns into the positive-definitness of the matrix $(G-G^\dagger)/2i$. In the following discussion we will concentrate on the single-operator case.} See for example \cite{Creminelli:2024lhd,Hartnoll:2009sz} for more details. For passive systems, which by definition can only absorb energy from the source, the imaginary part is necessarily always positive for $\omega>0$, and by the reality condition \cref{eq:realityG}, negative for $\omega<0$, for all $k$.\footnote{On the contrary, the imaginary part of the non-linear response functions does not have a definite sign. See \cite{PhysRevB.44.8446}.} This is true for stable systems in their ground state and also for thermal states, among others. More generally, this positivity property holds for a system described by a stationary density matrix for which the occupancy of the energy eigenstates is monotonically decreasing with energy \cite{Creminelli:2024lhd}. . 

In this section we are going to show that this positivity condition of the imaginary part extends to complex frequencies and momenta, by using dispersion relations. Consider first the single-variable case (see \cite{lan84,Creminelli:2024lhd}). By taking the imaginary part of the dispersion relation (see \cref{2dispomegak} with $\xi=0$), and assuming $G \to 0$ at infinity, we obtain
\eq{
{\rm Im}\,G(\omega)=\frac{{\rm Im}\,\omega^2}{\pi}\int_0^{+\infty}\frac{\dd{\z^2}}{|\z^2-\omega^2|^2}{\rm Im}\,G(\z)\,,
\label{impos11}
}
where we have re-written the kernel $(\z-\omega)^{-1}=(\z-\omega^*)|\z-\omega|^{-2}$, suppressing the $i\epsilon$ as $\omega$ is taken to be in the UHP, and then used the reality condition to integrate over positive frequencies. The positivity condition implies that the integral in \cref{impos11} is strictly positive, unless $G(\omega) \equiv 0$, and therefore the sign of ${\rm Im}\,G(\omega)$ is the same as the sign of ${\rm Im}\,\omega^2=2{\rm Im}\,\omega\,{\rm Re}\,\omega$ in the UHP, i.e.~positive for ${\rm Re}\,\omega>0$, negative for ${\rm Re}\,\omega<0$, and zero on the imaginary axis.

In this section we will generalize the argument above when $G$ also depends on $k$.  It is useful to define
\eq{\label{defF}
F(\omega,k)\equiv\omega \, G(\omega,k)\,,
}
or equivalently, $\tF(z_1,z_2)=(z_1+z_2)\tG(z_1,z_2)/2$. The imaginary part ${\rm Im}\,F(\omega,k)=\omega\,{\rm Im}\,G(\omega,k)$ is positive for all real $\omega$ and $k$.\footnote{It is natural to define $F$, as ${\rm Im}\,F(\omega,k)$ appears in the energy dissipation formula, \cref{diss}.} The new function is analytic in the same region, therefore, we can write dispersion relations for $\tF(z_1,z_2)$. Notice, however, that $F$ decays at infinity slower than $G$ by one factor of $\omega$. Let us assume momentarily that $\tF(z_1,z_2) \to 0$ at infinity. Fixing $z_2$ to be some real point $\z_2$, we can write
\eq{
\tF(z_1,\z_2)=\frac{1}{\pi}\int\frac{\dd{\z_1}}{\z_1-z_1-i\epsilon}{\rm Im}\,\tF(\z_1,\z_2)\,.
}
Taking the imaginary part this gives a manifestly positive result ($i\epsilon$ is suppressed for simplicity)
\eq{
{\rm Im}\,\tF(z_1,\z_2)=\frac{{\rm Im}\,z_1}{\pi}\int\frac{\dd{\z_1}}{|\z_1-z_1|^2}{\rm Im}\,\tF(\z_1,\z_2)>0\,.
\label{imFpos1}
}
Let us now fix $z_1$ in the UHP and write a dispersion relation for $z_2$; after taking the imaginary part one gets
\eq{
{\rm Im}\,\tF(z_1,z_2)=\frac{{\rm Im}z_2}{\pi}\int\frac{\dd{\z_2}}{|\z_2-z_2|^2}{\rm Im}\,\tF(z_1,\z_2)>0\,,
\label{imFpos2}
}
where the right-hand side is strictly positive, because of \cref{imFpos1}. This proves positivity of ${\rm Im}\,\tF(z_1,z_2)$ in the PUHP. Remember that from \cref{z1z2region} we see that the PUHP is a subset of the whole region of analyticity for $\alpha>1$ (see \cref{PlotRegions}). However, for every point in the region of analyticity, there exists a suitable $\alpha>1$ such that it is included in the PUHP in terms of $z_1$ and $z_2$ variables. We concludes that $F(\omega,k)$ has positive imaginary part in the whole region of analyticity guaranteed by microcausality.  

The derivation presented above assumes $\tF(z_1,z_2)\to0$ at complex infinity, which is a stronger assumption compared to $\tG(z_1,z_2)\to0$. This assumption is in fact not needed.\footnote{The same argument that leads to \cref{imFpos2} can be followed with one subtraction for $\tF$. But the details are slightly more involved. So we decided to present the argument in a different way.} We start from the double Cauchy integral, \cref{2dcauchy}, for $\tG(z_1,z_2)$, replacing $\tG(z_1,z_2)=2\tF(z_1,z_2)/(z_1+z_2)$, 
\eq{
\tF(z_1,z_2)=\frac{1}{(2\pi i)^2}\int\dd{\z_1}\dd{\z_2}\frac{(z_1+z_2)}{(\z_1-z_1)(\z_2-z_2)(\z_1+\z_2)}\tF(\z_1,\z_2)\,,
\label{2dcauchy3}
}
and suppressing the $i\epsilon$ to avoid clutter. The integral can be simplified using the analyticity of $\tF$. Using the trick of putting a pole at $z_1^*$ (or $z_2^*$), in analogy with \cref{2dcauchy2}, we obtain
\eq{
0=\int\dd{\z_1}\frac{1}{(\z_1-z_1^*)(\z_1+\z_2)}\tF(\z_1,\z_2)=\int\dd{\z_1}\frac{\z_1-z_1}{|\z_1-z_1|^2(\z_1+\z_2)}\tF(\z_1,\z_2)\,.
\label{2dcauchy4}
}
A similar expression holds with $z_1\leftrightarrow z_2$ and $\z_1\leftrightarrow \z_2$. This means that we can simplify the integral kernel of \cref{2dcauchy3} in multiple steps as follows
\eq{
\spl{
\frac{(z_1+z_2)}{(\z_1-z_1)(\z_2-z_2)(\z_1+\z_2)}&=\frac{(z_1+z_2)(\z_1-z_1^*)(\z_2-z_2^*)}{|\z_1-z_1|^2|\z_2-z_2|^2(\z_1+\z_2)}\\
&=\frac{(z_1+z_2)(\z_1-z_1+2i{\rm Im}\,z_1)(\z_2-z_2+2i{\rm Im}\,z_2)}{|\z_1-z_1|^2|\z_2-z_2|^2(\z_1+\z_2)}\\
&\to(2i{\rm Im}\,z_1)(2i{\rm Im}\,z_2)\frac{(z_1+z_2)}{|\z_1-z_1|^2|\z_2-z_2|^2(\z_1+\z_2)}\\
&=(2i{\rm Im}\,z_1)(2i{\rm Im}\,z_2)\frac{(\z_1+\z_2)-(\z_1-z_1)-(\z_2-z_2)}{|\z_1-z_1|^2|\z_2-z_2|^2(\z_1+\z_2)}\\
&\to(2i{\rm Im}\,z_1)(2i{\rm Im}\,z_2)\frac{1}{|\z_1-z_1|^2|\z_2-z_2|^2}\,.
}\label{kernel}}
The first equality follows from $(\z_1-z_1)^{-1}=(\z_1-z_1^*)|\z_1-z_1|^{-2}$ and similarly for the other variable. In the next step we re-write $\z_1-z_1^*=(\z_1-z_1)+(z_1-z_1^*)=(\z_1-z_1)+2i{\rm Im}\,z_1$ and similarly for the other one. For each variable, fixing the other one, the integral over the first parenthesis, i.e. $(\z_1-z_1)$, vanishes due to analyticity, as in \cref{2dcauchy4}, which reduces the kernel to the third line. In the fourth line, we re-write $z_1+z_2=(\z_1+\z_2)-(\z_1-z_1)-(\z_2-z_2)$, in which, following the same logic, the last two terms drop out and simplifies the kernel further in the last line of \cref{kernel}. This last form is particularly useful since, after taking the imaginary part of \cref{2dcauchy3}, it gives
\eq{\boxed{
{\rm Im}\,\tF(z_1,z_2)=\frac{{\rm Im\,}z_1{\rm Im}\,z_2}{\pi^2}\int\dd{\z_1}\dd{\z_2}\frac{{\rm Im}\,\tF(\z_1,\z_2)}{|\z_1-z_1|^2|\z_2-z_2|^2}>0}
\label{imrep}
}
which is manifestly positive in the PUHP. Notice that this is the same result as above (the combination of \cref{imFpos1} and \cref{imFpos2}) but with the weaker assumption $G \to 0$ at infinity.

Analytic functions of one or several variables that have a sign-definite imaginary part (or real part) in the region of analyticity, are well-studied in the mathematics literature: they are called Herglotz-Nevanlinna functions. We will briefly discuss this class of functions in \cref{herg}, after discussing subtracted dispersion relations. Let us mention two properties of ${\rm Im}\,\tF(\zeta_1,\zeta_2)$ which follows from the Herglotz-Nevanlinna condition (see \cite{Luger_2017} for the proof). The first is that ${\rm Im}\,\tF(\zeta_1,\zeta_2)$ cannot be integrable
\eq{
\int\dd{\z_1}\dd{\z_2}{\rm Im}\,\tF(\z_1,\z_2) = + \infty \;.
}
This statement is much stronger than the non-compactness discussed above, but it requires positivity (and decay at infinity). The second property is that one cannot have contributions of the form ${\rm Im}\,\tF(\z_1,\z_2) \supset \delta^{(2)}(\z_1- \z_1^0, \z_2-\z_2^0)$, i.e.~a single point $(\z_1^0, \z_2^0)$ must have zero measure. For more related results we refer to \cite{Luger_2020,Nedic_2024}.

Let us verify the positivity of ${\rm Im}\,F(\omega,k)$ in the PUHP for an example: the case of response mediated by a single mode given in \eqref{drude}. In this case the functional form is simple enough that the positivity can be checked analytically. For $\omega=\omega_R+i ~\omega_I$ and $k=k_R+i ~k_I$, we obtain  
\eq{
{\rm Im}\,F(\omega,k)=\frac{\left(\gamma +\omega _I\right) \left(\omega _I^2+\omega _R^2\right)+c_s^2\big(\omega _I (k_R^2-k_I^2)-2 k_I k_R \omega _R\big)+\omega _I m^2}{|-\omega^2-i\gamma\omega+c_s^2k^2+m^2|^2}\,.
}
The denominator is always positive. The numerator can be rewritten in a manifestly positive form
\eq{
{\rm Im}\,F(\omega,k)\propto\Bigg[\gamma\left(\omega _I^2+\omega _R^2\right)+\omega _I m^2+\omega_Ic_s^2\left(k_R-\frac{k_I\omega_R}{\omega_I}\right)^2+\left(\omega_I+\frac{\omega_R^2}{\omega_I}\right)(\omega_I^2-c_s^2k_I^2)\Bigg]\,,
}
where each term is non-negative for $0 \le c_s^2 \le 1$, $\gamma\ge0$ and $m^2 \ge0$, in the region of analyticity $\omega_I>|k_I| \ge 0$.

In the next section we will discuss a simple consequence of positivity of the imaginary part in PUHP, with relevant physics applications.

\subsection{The absence of zeros in the PUHP}

The retarded Green's function is free of singularities in the complex PUHP, as required by micro-causality. Furthermore, the positivity condition on its imaginary part guarantees that it admits no zeros in the PUHP. Indeed, any zero of $G$ would necessarily imply a zero of $F=\omega \,G$, which is excluded by the strict positivity of ${\rm Im}\,F$ in this region, as argued above. This, in turn, ensures the analyticity of $G^{-1}$ in the same region: $G^{-1}$ cannot have poles and other kind of singularities are forbidden by the analyticity of $G$.

Let us see what are the implications of this property, focusing on the concrete example of electrodynamics in a medium, assumed to be translationally invariant and passive. Assuming that the effect of the medium can be studied in linear-response theory, the Maxwell equations for the average field in the medium are (see \cite{Creminelli:2024lhd} for details)
\eq{\label{eq:Maxmedia}
\partial_\alpha F^{\alpha \mu}+\int \mathrm{d}^4 y \,\Pi^{\mu \nu}(x, y) A_\nu(y)=-J_{\mathrm{ext}}^\mu(x)\,.
}
$J_{\mathrm{ext}}^\mu(x)$ is the external current, while the effect of the medium is captured by the self-energy tensor 
\eq{
\Pi^{\mu \nu}(x, y)=i \theta\left(x^0-y^0\right)\left\langle\left[J^\mu(x), J^\nu(y)\right]\right\rangle_{1 \mathrm{PI}}+ \text{contact terms}\,,
}
where $J^\mu$ is the conserved current operator, and contact terms involve $\delta(x-y)$ or its derivatives. $\Pi^{\mu\nu}$ gives the induced current as response to the the {\em total} field (the 1PI prescription means one only considers graphs which are one-particle irreducible with respect to the photon line). In contrast, the retarded current two-point function, $i \theta\left(x^0-y^0\right)\left\langle\left[J^\mu(x), J^\nu(y)\right]\right\rangle$, gives the induced current as response to the {\em external} field. Analyticity of the latter easily follows from causality, while analyticity of the former is not obvious because of the 1PI prescription. Here however is where the absence of zeros comes to a rescue.

The equation of motion \cref{eq:Maxmedia} implies that the full photon propagator $G_\gamma$ satifies (we neglect the tensor structure, see \cite{Creminelli:2024lhd} for details)
\eq{
G_\gamma^{-1}=\Delta_\gamma^{-1}-\Pi\,,
}
in which $\Delta_\gamma$ is the \emph{free} retarded Green's function of the photon. The analyticity of $G_\gamma^{-1}$ (and of $\Delta_\gamma^{-1}$) discussed above guarantees that $\Pi$ is analytic as well, in the usual domain $\Im (\omega,k) \in $ FLC. Notice that the analyticity of $\Pi$ is not a consequence of micro-causality only: positivity of the imaginary part is necessary as well, together with the assumption of decay at infinity.\footnote{Ref.~\cite{Creminelli:2024lhd} showed the analyticity of $\Pi$ in a smaller region parametrized as 
$\Pi(\omega,\vec{k}+\vec{\xi}\,\omega)$ with $\vec{k}\cdot\vec{\xi}=0$ and $\omega \in$ UHP.}

The tensor $\Pi^{\mu\nu}$ that appears in the Maxwell equations in a medium, \cref{eq:Maxmedia}, is usally written in terms of the electric permittivity $\epsilon(\omega,k)$ and the magnetic permeability $\mu^{-1}(\omega,k)$. In terms of these quantities, Maxwell equations take the well-known form 
\eq{
\epsilon \, \vec \nabla \cdot\vec E=\rho_{\rm ext}\,,\qquad\qquad \frac{1}{{\mu}}\vec{\nabla}\times \vec B-{\epsilon}\, \p_t\vec E=\vec J_{{\rm ext}}\,.
}
(Since $\epsilon$ and $\mu^{-1}$ are not constant in general, one should understand the products above as convolutions in real space.) The discussion above proves the analyticity of $\epsilon(\omega,k)$ and $\mu^{-1}(\omega,k)$ for $\Im (\omega,k) \in $ FLC. To the best of our knowledge, this is the first proof of this property.

The discussion above is not limited to electromagnetism and the same result applies in other cases. Notice also that the zeros of Green's functions have physical information: they appear in the Luttinger's theorem in condensed matter \cite{PhysRev.118.1417} and in the recent bounds on chaos, see e.g.~\cite{Blake:2018leo}.



\section{Subtracted dispersion relations}\label{subtract}

So far we have mainly restricted ourselves to cases where the asymptotic behavior of the Green's function is such that we can neglect the contribution from the infinity arc. We have already seen examples (in \cref{examp}) that violate this assumption. In case when the Green's function does not vanish at complex infinity, we have to take into account the contribution from the large semi-circle. Equivalently, we can write dispersion relations for a modified version of the function which goes to zero at infinity, perhaps using division by a polynomial. This result is known as dispersion relation with subtractions. 

Consider $G(z)$, analytic in the UHP, that does not vanish at infinity. Instead we look at $G(z)/P(z)$, where $P(z)$ is a polynomial with degree $n$ such that $G(z)/P(z)\to0$ at infinity. It is more economic to choose the degree as minimum as possible to avoid unnecessary complications. The polynomial is otherwise arbitrarily and can have zeros both in the upper half-plane (UHP) or lower half-plane (LHP). In case there is a real zero, we move it away from the real axis, i.e. by an $i\epsilon$ prescription. Following the standard argument we obtain
\eq{
G(z)=Q(z)+\frac{P(z)}{\pi}\int\dd{\z}\frac{{\rm Im}\,G(\z)}{(\z-z-i\epsilon)P(\z)}\,.
\label{1dispsub}
}
More details are given in \cref{subapp}. Here, $Q(z)$ is a polynomial of degree $n-1$, given by
\eq{
Q(z)=P(z)\left[\sum_a\frac{G(z_a)}{P'(z_a)(z-z_a)}+\sum_b\frac{G(z_b^*)^*}{P'(z_b)(z-z_b)}\right]\,,
\label{Q}
}
where the first sum is over the zeros of $P(z_a)=0$ that lie in the UHP, thereby picked by the contour integration, while the second sum is over the zeros $P(z_b)=0$ in the LHP, and prime means differentiation. The zeros in the LHP appear when we use the complex conjugation trick, as in \cref{2dcauchy2}, to get ${\rm Im}\,G$ in the integrand. It is easy to see that $Q(z_a)=G(z_a)$ and $Q(z_b)=G(z_b^*)^*$. As a special case, we can evaluate \cref{1dispsub} for real $z$ to obtain
\eq{
{\rm Re}\,G(z)=Q(z)+\frac{P(z)}{\pi}{\rm PV}_z\int\dd{\z}\frac{{\rm Im}\,G(\z)}{(\z-z)P(\z)}\,,\qquad\text{(real $z$)}
\label{reim1dispsub}
}
where the principle value is with respect to the pole at real $z$. Subtracted dispersion relation for functions of two variables is discussed in \cref{subapp}. It is obtained by repeating the steps leading to \cref{1dispsub} for the other variable. 

Let us discuss the integral constraints on the imaginary part when subtractions are needed. One expects a set of constraint equations from \cref{imcomut}, which holds regardless of the asymptotic behavior of the imaginary part. As in \cref{imcons5}, it is sufficient to work with single-variable dispersion relations. Therefore, we write subtracted Leontovich relation for $G(\omega,k+\omega\xi)$ as was described in \cref{1dispsub}. In particular we use \cref{reim1dispsub} choosing a polynomial with all the zeros $z_a$ on the real line
\eq{
{\rm Re}\,G(\omega,k+\omega\xi)=P(\omega)\sum_a\frac{{\rm Re}\,G(z_a,k+z_a\xi)}{P'(z_a)(\omega-z_a)}+\frac{P(\omega)}{\pi}{\rm PV}_{\omega,\{z_a\}}\int\dd{\z}\frac{{\rm Im}\,G(\z,k+\z\xi)}{(\z-\omega)P(\z)}\,,
}
where the principle value is for all the real zeros $z_a$ and for the pole at real $\omega$. Redefining $k\to k-\omega\xi$, makes the left-hand side independent of $\xi$, which then gives zero after differentiation with respect to $\xi$
\eq{
0=-P(\omega)\sum_a\frac{\p_k{\rm Re}\,G(z_a,k+(z_a-\omega)\xi)}{P'(z_a)}+\frac{P(\omega)}{\pi}{\rm PV}_{\{z_a\}}\int\dd{\z}\frac{\p_k{\rm Im}\,G(\z,k+(\z-\omega)\xi)}{P(\z)}\,.
}
By an appropriate redefinition of $k$, it is possible to make every term in the above summation independent of $\xi$, one at a time. For instance, fixing a term $z_{a_0}$, we redefine $k\to k-(z_{a_0}-\omega)\xi$, and then taking $\xi-$derivative we obtain
\eq{
0=-\sum_{a\neq a_0}\frac{(z_a-z_{a_0})}{P'(z_a)}\p_k^2{\rm Re}\,G(z_a,k+(z_a-z_{a_0})\xi)+\frac{1}{\pi}{\rm PV}_{\{z_a\neq z_{a_0}\}}\int\dd{\z}\frac{\p_k^2{\rm Im}\,G(\z,k+(\z-z_{a_0})\xi)}{P(\z)/(\z-z_{a_0})}\,.
}
Continuing this process, we get rid of the summation over zeros, and denominator in the second term. A final redefinition $k$ then gives
\eq{\label{eq:CLSZsub}\boxed{
\int\dd{\z}\p_k^{n+1}{\rm Im}G(\z,k+\xi \z)=0}
}
for $n$ subtractions. Indeed, it is possible that the convergence behavior of the imaginary part is better, which means that we could take a few, or perhaps all, of the derivatives outside of the integral.\footnote{Consider, as an example, a Green's function that decays at infinity and with imaginary part decays faster than $\omega^{-1}$ such that \cref{imcons5} is true. Then add to it a generic polynomial in $\omega$ and $k^2$. The new function is analytic in the correct region, but satisfies the subtracted dispersion relation. On the other hand, if the polynomial is chosen to be real, its imaginary part is equal to the original function which satisfies \cref{imcons5}.} We used \cref{eq:CLSZsub} in a few examples discussed in \cref{kin} and \cref{kin}.

We remark that positivity in the PUHP cannot be achieved for generic asymptotic behavior. This point will be discussed further in \cref{herg}.

\section{Broken rotational symmetry}\label{norot}

Until now, we have assumed rotational invariance, but many results of this paper simply generalize to cases when rotational symmetry is spontaneously broken. A notable example is a system placed in a background electric or magnetic field (see \cite{Hattori:2012je,Karbstein:2013ufa}). Another example will be studied at the end of this section. The theorem of \cref{Bogo}, concerning the relation between microcausality and analyticity, does not rely on rotational symmetry as presented. The same holds for the discussion in \cref{imsupsec} regarding the support of the imaginary part. In what follows, we first formulate a multi-variable dispersion relation—analogous to \cref{2dcauchy2}—without assuming rotational invariance. We then derive the analogue of \cref{imcons5}, which provides integral constraints on the imaginary part, and we also discuss the positivity of the imaginary part within the region of analyticity. 

It should be noted that in many physical systems only a subgroup of the rotational symmetry is broken. If the unbroken group is $SO(d_r)$, rather than $SO(d-1)$, then the effective dimension of the system is $\tilde{d}\equiv d-d_r+1$, composed of the number of broken directions plus one for the unbroken ones, and one for time. We have seen in \cref{Gomegak} for the rotational invariant case that $\tilde{d}=2$, corresponding to $G(\omega,k)$. When breaking is due to a background vector, we have $G(\omega,k_{\|},k_\perp)$ as a function of momentum along or orthogonal to the direction of the breaking, and consistently $\tilde{d}=3$. For generic $\tilde{d}$, the Green's function $G(\omega,\vk)$, with $\vk\in\mathbb{R}^{\tilde{d}-1}$, is analytic in the ($\td-$dimensional) FLC: $\omega_I>|\vk_I|\ge0$. The argument is similar to what was given for the rotational invariant case in \cref{Gomegak}, and also in \cref{rotinv}. 

To write a multi-variable dispersion relation, as in \cref{2disp}, it is less straightforward to find the analog of the mapping \cref{z1z2}. We need an invertible map between the (now $\td$-dimensional) poly-upper half-plane $\mathrm{PUHP}\equiv \{ (z_1,\dots,z_{\td})\in \mathbb{C}^{\td}|z_{i,I}>0,\forall i=1,\dots,\td\}$ and a subset of the region of analyticity, $\mathbb{R}^{\td-1,1} + i P\>,\>P\subset \mathring{\rm{FLC}}$. As before, we can restrict our attention to linear transformations of the form 
\begin{equation}
\omega=\frac{1}{\td}\sum_{i=1}^{\td}z_i\>,\qquad\vec{k}=\frac{1}{\alpha\td}\sum_{i=1}^{\td} \vec{n}_i z_i\in\mathbb{R}^{\td-1}\>,
\label{norotmap}
\end{equation}
with $\alpha>1$, and a set of unit vectors ($\vec{n}_i^2=1$), satisfying $\vec{n}_i\cdot \vec{n}_j=-1/(\td-1)$ for $i\neq j$. More details on constructing the map is given in \cref{AppMapsNonRot}. It is easy to verify that with complete rotational symmetry ($\td=2$), this is identical to \cref{z1z2}, with $\vec{n}_1=-\vec{n}_2=1$. We see that $z_{i,I}>0$ is strictly contained inside $\mathring{\rm{FLC}}$
\eq{
\omega_I=\frac{1}{\td}\sum_{i} |\vec{n}_i| z_{i,I}\ge\frac{1}{\td}\Big|\sum_{i} \vec{n}_i z_{i,I}\Big|>|\vk_I|\,,
}
where we have used the Cauchy–Schwarz inequality. The inclusion of $\alpha>1$ is for the same reason as before: to have control over the high energy behavior. Intuitively, $P$ is a hyperpyramid emanating from the tip of the $\rm{FLC}$ along future-directed timelike directions, whose constant-$\omega_I$ slices correspond to regular simplexes inscribed in $S^{\td-2}$. This convenient choice ensures that only $1+(\td-1)(\td-2)/2$ parameters are necessary in order to cover the whole interior of $\rm{FLC}$. As shown in \cref{PyramidsMain}, the value of the parameter $\alpha>1$ controls the spread of the pyramid (i.e. in the limit $\alpha \to 1^{+}$, the generators of the pyramid become null directions), while the remaining freedom corresponds to the possibility of rotating the pyramid about the $\omega_I$-axis. 
\fg{
\centering
\adjustbox{valign=c}{\begin{tikzpicture}[>=latex, line join=round, line cap=round, scale=1.6]

\coordinate (O) at (0,0,0);
\coordinate (k2) at (2,0,0);
\coordinate (w) at (0,2,0);
\coordinate (k1)  at (0,0,2);

\draw[->, thick] (O) -- (k1) node[below right] {$k^{\parallel}_{I}$};
\draw[->, thick] (O) -- (k2) node[above left] {$k^{\perp}_{I}$};
\draw[->, thick] (O) -- (w)  node[left] {$\omega_{I}$};

\coordinate (A) at (0,1.5,0.75);
\coordinate (B) at (0.65,1.5,-0.38);
\coordinate (C) at (-0.65,1.5,-0.38);

\coordinate (D) at (0,1.5,1.37);
\coordinate (E) at (1.19,1.5,-0.68);
\coordinate (F) at (-1.19,1.5,-0.68);

\draw[thick] (O) -- (A);
\draw[thick] (O) -- (B);
\draw[thick] (O) -- (C);
\draw[dashed,thick] (A) -- (0,2,1);
\draw[dashed,thick] (B) -- (0.83,2,-0.51);
\draw[dashed,thick] (C) -- (-0.83,2,-0.51);
\draw[thick] (O) -- (D);
\draw[thick] (O) -- (E);
\draw[thick] (O) -- (F);

\draw[dashed,thick] (D) -- (0,2,1.82);
\draw[dashed,thick] (E) -- (1.57,2,-0.91);
\draw[dashed,thick] (F) -- (-1.57,2,-0.91);

\draw[thick] (A) -- (B) -- (C) -- cycle;
\draw[thick] (D) -- (E) -- (F) -- cycle;

\fill[green!20, opacity=0.25] (A) -- (B) -- (C) -- cycle;
\fill[yellow!20, opacity=0.25] (D) -- (E) -- (F) -- cycle;

\def\hcone{2}
\def\rcone{2}

\foreach \ang in {0,5,...,355}{
  \path[fill=gray!30, draw=none, opacity=0.05]
    (O)
    -- ({\rcone*cos(\ang)},   {\hcone}, {\rcone*sin(\ang)})
    -- ({\rcone*cos(\ang+5)}, {\hcone}, {\rcone*sin(\ang+5)})
    -- cycle;
}

\draw[gray!70!black, dashed]
  plot[domain=0:360, samples=120]
    ({1.5*cos(\x)}, {1.5}, {1.5*sin(\x)});
\draw[gray!70!black, dashed]
  plot[domain=0:360, samples=120]
    ({cos(\x)}, {1}, {sin(\x)});
\draw[gray!70!black, dashed]
  plot[domain=0:360, samples=120]
    ({0.5*cos(\x)}, {0.5}, {0.5*sin(\x)});

\node[gray!70!black] at (1.6,1.2,0.2)
  {$\partial{\rm FLC}$};

\end{tikzpicture}}
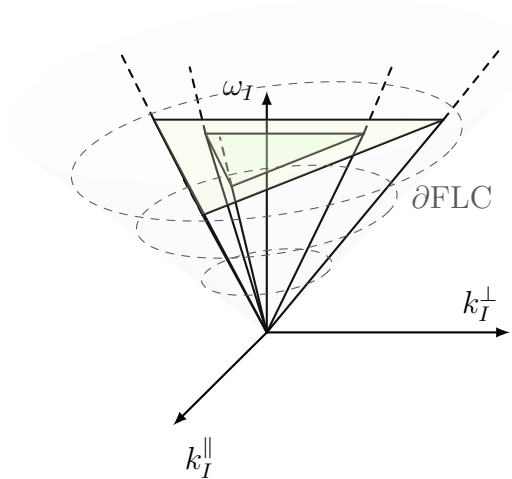
\caption{Comparison between constant-$\omega_I$ sections in $\td=3$. Two pyramids inside the ${\rm FLC}$ in $\td=3$. Varying $\alpha$ changes the area of their constant-$\omega_I$ sections. Setting $\omega_I=3/2$, the green and yellow sections are obtained by choosing $\alpha=2$ (green) or $\alpha=11/10$ (yellow).}
\label{PyramidsMain}
}

Once we have \cref{norotmap} then it is straightforward to write multi-variable dispersion relation for $\hat{G}(z_1,\dots,z_{\td})\equiv G(\omega,\vk)$; we exploit analyticity in the $\td-$dimensional $\rm{PUHP}$ and use the Cauchy theorem multiple times, as described in \cref{muldisp}. Assuming $G\to0$ at infinity, we obtain\footnote{See \cite{Vladimirov_1977} for a different construction using spinor representation of 4-vectors.} 
\begin{equation}\label{dispGd+1}
\tG(z_1,\dots,z_{\tilde{d}})=\frac{2i}{(2\pi i)^{\td}}\int \dd\z_1 \cdots \dd\z_{\td}\frac{ {\rm Im}\>\tG(\z_1,\dots,\z_{\td})}{(\z_1-z_1-i\epsilon)\cdots(\z_{\td}-z_{\td}-i\epsilon)}\,,
\end{equation}
which is the generalization of \cref{2disp} when rotational symmetry is (partially) broken. Notice that one can derive this relation by using multiple times the Leontovich relation, analog of \cref{leontovich},
\eq{
G(\omega,\vk)=\frac{1}{\pi}\int\frac{\dd{\z}}{(\z-\omega-i\epsilon)}{\rm Im}\,G\Big(\z,\vk+(\z-\omega)\vxi\Big)\,,
\label{leovec}
}
with $\vk,\vxi\in\mathbb{R}^{\td-1}$ and $|\vxi|<1$. Finally, the generalization to dispersion relations with subtractions is straightforward. 

By using \cref{dispGd+1}, following the argument in \cref{pos}, we can argue positivity of the imaginary part in the PUHP. Once again we look at $F(\omega,\vk)=\omega G(\omega,\vk)$, which, for passive systems, has positive imaginary part for real frequency and momentum. Assuming $G\to0$ at infinity, we can re-write \cref{dispGd+1} for $F$,
\begin{equation}
     \tF(z_1,\dots,z_{\td})=\frac{1}{ (2\pi i)^{\td}}\int \dd\z_1 \dots \dd\z_{\td}\frac{ (z_1+\dots+z_{d})\tF(\z_1,\dots,\z_{\td})}{(\z_1+\dots+\z_{\td})(\z_1-z_1)\cdots(\z_{\td}-z_{\td})}\,.
\end{equation}
Following the very same logic that led from \cref{2dcauchy3} to \cref{imrep}, we can simplify the integral kernel and obtain the more familiar dispersion relation for $\tF$
\begin{equation}\label{dis:F_norot}
     \tF(z_1,\dots,z_{\td})=\frac{\prod_{i=1}^{\td} {\rm Im}\,z_{i}}{ \pi^{\td}}\int \dd\z_1 \dots \dd\z_{\td}\frac{ \tF(\z_1,\dots,\z_{\td})}{|\z_1-z_1|^2\cdots|\z_{\td}-z_{\td}|^2}
\end{equation}
Finally, taking the imaginary part on both sides yields
\begin{equation}
\boxed{   {\rm Im}\,\tF(z_1,\dots,z_{\td})=\frac{\prod_{i=1}^{\td} {\rm Im}\,z_{i}}{ \pi^{\td}}\int \dd\z_1 \dots \dd\z_{\td}\frac{{\rm Im}\tF(\z_1,\dots,\z_{\td})}{|\z_1-z_1|^2\cdots|\z_{\td}-z_{\td}|^2}>0
   }
\end{equation}
which is valid on the $\td$-dimensional $\rm{PUHP}$ upper-half plane. As a consequence, we can map the result back to $(\omega,\vec{k})$-space and conclude that $\tF$ has positive imaginary part in the region $\mathbb{R}^{\td-1,1}+i P$, where the shape of $P$ depends on the specific choice of parameters $(\alpha,\vec{n}_i)$ of the linear transformation we started with. Varying these parameters enables to cover all the interior of $\rm{FLC}$, extending the positivity property throughout the whole forward tube. 

Microcausality constraints on the functional form of the imaginary part of $G(\omega,\vec{k})$, discussed in \cref{imconssec}, survive also when rotations are spontaneously broken. The logic is the same as before. Starting from \cref{leovec} we conclude that
\begin{equation}
    \frac{1}{\pi}{\rm PV}\int \dd{\z}\frac{{\rm Im}\,G(\z,\vec{k}+\vec{\xi}_1(\z-\omega))}{\z-\omega}=\frac{1}{\pi}{\rm PV}\int \dd{\z}\frac{{\rm Im}\,G(\z,\vec{k}+\vec{\xi}_2(\z-\omega))}{\z-\omega},
\end{equation}
Taking a derivative with respect to $\vec{\xi}$ reduces to
\begin{equation}\label{CLSZnorot}
    \int \dd \z \partial_{k^i}\mathrm{Im}\>G(\z, \vec{k}+\vec{\xi}(\z-\omega))=0\>,
\end{equation}
for any $(\omega,\vec{k})\in \mathbb{R}^{1,\td-1}, |\vec{\xi}|<1$. Notice that \cref{CLSZnorot} consists of $\td-1$ equations, one for each spatial component of $\vec{k}$. As before, when $\mathrm{Im}\>G$ decays faster than $\omega^{-1}$ we can push the derivatives outside and hence claim that the integral of ${\rm Im}\,G(\z,\vec{k}+\vec{\xi}\z)$ is a function of $\vec{\xi}$ only. In particular, specializing to the case $\vec{k}=0$ allows to conclude that
\begin{equation}\label{Intconstnorot}
     \int \dd\z \>\mathrm{Im}\>G(\z,\vec{k}+\vec{\xi}\z)=0\>,\quad \forall \vec{k}\in \mathbb{R}^{\td-1},|\vec{\xi}|<1\>,
\end{equation}
where we used ${\rm Im}\,G(-\z,-\vec{\xi}\z)=-{\rm Im}\,G(\z,\vec{\xi}\z)$ (which follows from reality of $G(x)$).
As a final remark, when the decay assumptions are not satisfied, one can write subtracted Leontovich relation for $G(\omega,\vec{k}+\vec{\xi}\omega)$. Following the same logic leading to \cref{eq:CLSZsub}, it is possible to obtain the general form of the integral constraints for the imaginary part
\begin{equation}
    \int \dd\z\>\partial_{k^{i_1}}...\partial_{k^{i_{n+1}}} \mathrm{Im}\>G(\z,\vec{k}+\vec{\xi}\z)=0
\end{equation}
where $n$ is the number of subtractions.

\paragraph{ $U(1)$ with space-dependent background} A nontrivial check of the integral constraint \cref{Intconstnorot} is given by the massive version of the previously discussed $U(1)$. In this case, the starting point is the Lagrangian 
\begin{equation}
\mathcal{L}=-\partial \Phi^{\dagger}\partial \Phi+\frac{m^2}{2}\rho-\lambda (\Phi^{\dagger}\Phi)^2
=-\frac{(\partial \rho)^2}{2}+\frac{m^2}{2}\rho^2-\frac{\rho^2}{2 v^2} (\partial \theta)^2-\frac{\lambda}{4}\rho^4\,,
\end{equation}
where we have written the polar decomposition of $\Phi$. This time, we expand the phase about a space-dependent background $\theta=\vec{C}\cdot \vec{x}+\pi$, leading to the $SO(3)\to SO(2)$ breaking pattern for rotations, as well as the breaking of boost-invariance along the direction of $\vec{C}$. Notice that a diagonal combination of the global $U(1)$ and translations along the $\vec{C}$ direction is preserved, yielding effectively full spacetime translations unbroken. Similarly to the previous scenario, the breaking induces a mixing between the fluctuations of the phase and of the radial mode, as it can be seen from the quadratic Lagrangian\footnote{Where $M^2=\lambda v^2/2$ and $\lambda v^2=m^2-\vec{C}^2/v^2$}
\begin{equation}\label{eq:quadraticL}
    \mathcal L_{(2)} = \frac{1}{2} (\partial h)^{2} -\frac{M^2}{2}h^2+\frac{1}{2} (\partial \pi)^{2} -\frac{2}vh\vec{C}\cdot \vec{\nabla}\pi\>
\end{equation} 
As a consequence, the dispersion relations of d.o.f.s of the model acquire the form
\begin{equation} \label{eq:disprelation}
    \omega^2_{\pm} =\vec{k}^2 + \frac{M^2}{2}  \pm \sqrt{ \frac{(\vec{C}\cdot \vec{k})^2}{v^2}  + \frac{M^4}{4}  } \,.
\end{equation}
Notice that transverse fluctuations (whose $\vec{k}$ is orthogonal to $\vec{C}$) do not see the breaking, hence their dispersion relation is Lorentz invariant. We can then focus on the retarded Green's function of $\pi$ and extract its imaginary part, which receives contributions from both $\pm$ states
\begin{equation}
{\rm Im}\,G^{\pi}(\omega,k)
=\pi\,{\rm Sgn}(\omega)\left[|Z^{\pi}_{-}|^2\delta(\omega^2-\omega^2_{-})
+|Z^{\pi}_{+}|^2\delta\big(\omega^2-\omega^2_{+}\big)\right]\,,
\end{equation}
where the interpolating functions are given by
\begin{equation}
    |Z^{\pi}_l(\vec{k})|^2=l\frac{\omega^2_l(\vec{k})-\vec{k}^2-M^2}{\omega^2_{+}(\vec{k})-\omega^2_{-}(\vec{k})}.
\end{equation}
We stress that transverse fluctuations are blind to the breaking. Indeed, when $\vec{k}\cdot \vec{C}=0$, the interpolating functions of $\pi$ collapse to 
\begin{equation}
    |Z^{\pi}_l(\vec{k})|^2=\begin{cases}
        1& {\rm if}\>l=-\\
        0& {\rm if}\>l=+
    \end{cases}
\end{equation}
In this limit, $G^{\pi}(\omega,\vec{k})$ (and its imaginary part) acquire a Lorentz invariant form, leading the integral constraint to be satisfied identically (by the argument presented in \cref{lorinv}). The interesting situation, therefore, is given by considering $\vec{k}\cdot \vec{C}\neq0$. In this case one can numerically check the validity of \cref{Intconstnorot}. In particular, the integral constrain is satisfied for arbitrary $k$ and $0 \le \xi<1$, provided both gapped and gapless contributions are included in ${\rm Im}\,G$. As before, indeed, the integral does not vanish if one ignores the gapped mode.

\section{Herglotz-Nevanlinna functions}\label{herg}
In \cref{pos} we have established that, assuming $\tG(z_1,z_2)$ goes to zero at infinity, the function $\tF(z_1,z_2)$ has a non-negative imaginary part in the PUHP, i.e.\ for $\Im z_1 > 0$ and $\Im z_2 > 0$. This property is precisely the defining attribute of a \emph{Herglotz--Nevanlinna} function in two variables. By definition, such a function is a holomorphic map $q(z_1,z_2):{\rm PUHP} \to \mathbb{C}$ with ${\rm Im}\,q(z_1,z_2)\geq0$. We showed that $\tF(z_1,z_2)$ is a Herglotz--Nevanlinna function starting from the multi-variable dispersion relation in \cref{2disp}.

In this section we want to connect with the mathematical literature on Herglotz--Nevanlinna functions. In particular we will discuss a theorem that characterizes this class of functions in terms of a positive measure (which corresponds to $\Im\hat F(\omega,k)$ in our discussion). This measure, when one has more than one variable, must satisfy a set of constraints, which we will prove to be equivalent to the integral constraints studied above. Moreover the theorem gives a complete characterization of Herglotz--Nevanlinna functions, dropping our assumption that $G \to 0$ at infinity.

The proof of the theorem for several variables is a generalisation of the classical Riesz--Herglotz theorem by Korányi--Pukánszky~\cite{koranyi1963holomorphic}, and an independent derivation was given by Vladimirov~\cite{vladimirov1974holonomic, vladimirov1969holomorphic}, which characterises holomorphic functions with positive imaginary part on the unit polydisk in $\mathbb{C}^n$. In the following we will follow the presentation of \cite{Luger_2017,LUGER20191189}. The theorem holds for an arbitrary number of variables, although we will mostly comment only on the one and two-variable case. Notice that the case with more than two variables is relevant for systems without rotational symmetry, in \cref{norot}.

\begin{center}
\rule{\textwidth}{0.4mm}
\end{center}
\paragraph{Theorem (Luger, Nedic \cite{Luger_2017,LUGER20191189})} 
\emph{
A holomorphic function $q : {\rm PUHP}^{n} \to \mathbb{C}$ is of Herglotz--Nevanlinna type 
if and only if it admits a representation of the form
\begin{equation}
q(z) = a +\sum_{\ell=1}^nb_\ell z_\ell + \frac{1}{\pi^n} \int_{\mathbb{R}^n} 
K_n(z,\z) \, \dd{\mu}(\z)\,,
\label{intrepherg}
\end{equation}
in which the kernel is 
\eq{
K_n(z,\z)=\frac{i}{(2i)^n}\left[2\prod_{\ell=1}^n\left(\frac{1}{\zeta_\ell-z_\ell}-\frac{1}{\zeta_\ell+i}\right)-\prod_{\ell=1}^n\left(\frac{1}{\z_\ell-i}-\frac{1}{\z_\ell+i}\right)\right]\,,
\label{kern}
}
and $a \in \mathbb{R}$, $b_\ell \geq 0$, and $\mu$ is a positive measure on $\mathbb{R}^n$ satisfying the growth condition
\begin{equation}
\int_{\mathbb{R}^n}\dd{\mu}(\zeta) \prod_{\ell=1}^n\frac{1}{(1+\zeta_\ell^2)} \,  < \infty\,,
\label{grow}
\end{equation}
together (for $n>1$) with the Nevanlinna condition
\begin{equation}
\sum_{\rho\in\{-1,0,1\},-1\in\rho\wedge1\in\rho}\int_{\mathbb{R}^n} N_{\rho_1,1}N_{\rho_2,2}\cdots N_{\rho_n,n} \, \dd\mu(\zeta) = 0\,,
\label{nevan}
\end{equation}
for all $z\in {\rm PUHP}^{n}$, with 
\eq{
N_{-1,\ell}=\frac{1}{\zeta_\ell-z_\ell}-\frac{1}{\zeta_\ell-i}\,,\quad N_{0,\ell}=\frac{1}{\zeta_\ell-i}-\frac{1}{\zeta_\ell+i}\,,\quad N_{+1,\ell}=\frac{1}{\zeta_\ell+i}-\frac{1}{\zeta_\ell-z_\ell^*}\,.
}
}
\begin{center}
\rule{\textwidth}{0.4mm}
\end{center}

\subsubsection*{Single variable}\label{singlevar}

Let us first clarify the theorem in the simplest case of single variable functions, with $n=1$. Notice that in this case there is no Nevanlinna condition on the measure, in line with our discussion above. A simple calculation gives
\eq{
K_1(z,\z)=\frac{z\z+1}{(\z-z)(\z+i)(\z-i)}\,.
}
The denominator suggests that this corresponds to a dispersion relation with two subtraction points, $\z=\pm i$, and the measure corresponds to the imaginary part on the real axis. On the other hand the growth condition in \cref{grow}, ensures that the measure decays fast enough such that the integral with only one subtraction converges. See the discussion around \cref{ImGimprov}. We choose $P(z)=z+i$ with only a zero in the LHP. The kernel of such dispersion relation, as given in \cref{1dispsub}, is
\eq{
\frac{(z+i)}{(\z-z)(\z+i)}=K_1(z,\z)+\frac{i}{\z^2+1}\,.
\label{2dispker}
}
By the growth condition \cref{grow}, each term in \cref{2dispker} is separately convergent. Using \cref{1dispsub}, \cref{Q}, and \cref{2dispker} we see that (since we want to connect with the rest of the paper, we now call the function $F$)
\eq{
\spl{
F(z)&=F(i)^*+\frac{1}{\pi}\int\dd{\z}\frac{(z+i)}{(\z-z)(\z+i)}{\rm Im}\,F(\z)\\
&={\rm Re}\,F(i)+\frac{1}{\pi}\int\dd{\z}K_1(z,\z){\rm Im}\,F(\z)
}\label{expl2dispker2}}
where to obtain the second line we use the fact that
\eq{
{\rm Im}\,F(i)=\frac{1}{\pi}\int\dd{\z}\frac{{\rm Im}\,F(\z)}{(\z^2+1)}\,,
\label{n1Imi}
}
which is the first line in \cref{expl2dispker2} evaluated at $z=i$. Hence, the integral representation in \cref{intrepherg} is a one-subtracted dispersion relation of \cref{1dispsub}, with $P(z)=z+i$, plus a linear term. The measure in \cref{intrepherg} corresponds to $\dd{\mu(\z)}={\rm Im}\,F(\z)\dd{\z}$. This is indeed proven in \cite{Luger_2017,LUGER20191189} for any number of variables.  The constant piece $a={\rm Re}\,F(i)$, which is always valid for this choice of subtraction points. Moreover, the coefficient of the linear piece is the given by $b=\lim_{z\to\infty}F(z)/z$ \cite{Luger_2017}. Notice that addition of the linear term $bz$ with real $b$ does not change ${\rm Re}\,F(i)$, and the measure ${\rm Im}\,F(\z)$. 

The fact that \cref{intrepherg} gives a non-negative imaginary part in the UHP, given $b\geq0$ and a positive measure, simply follows from the observation that
\eq{
{\rm Im}\,K_1(z,\z)=\frac{{\rm Im}\,z}{|\z-z|^2}>0\,,
}
which one can verify easily from the definition. The non-trivial statement of the theorem is that {\em all}  Herglotz-Nevanlinna functions can be put in this form. For instance it is not obvious that one cannot write expressions with more subtractions.

As a side remark, it should be noted that, while for generic subtractions the function is not Herglotz-Nevanlinna, some ``positivity property'' of this type is still present. Indeed, one can always choose the subtraction polynomial $P(z)$ to be real and positive. Hence, repeating \cref{1dispsub} for $F$, we have
\eq{
F(z)=Q(z)+\frac{P(z)}{\pi}\int\frac{1}{(\z-z)}\left(\frac{{\rm Im}\,F(\z)}{P(\z)}\dd{\z}\right)\,.
}
We can interpret the term in parentheses as a new positive measure which satisfies the growth condition. The integral, therefore, defines a Herglotz-Nevanlinna function. As a result, $F(z)$ has the structure of a polynomial plus a polynomial times a Herglotz-Nevanlinna function. This representation may deserve further study.

\subsubsection*{Two-variable case}

Let us move on to $n=2$. Following the discussion of the single variable case, we expect that this corresponds to a single subtraction for each variable, plus linear terms. We choose $P_1(z_1)=z_1+i$ and $P_2(z_2)=z_2+i$, with zeros in the LHP to avoid complications discussed in the previous section for \cref{2dispsub}. A few lines of algebra shows that the kernel in \cref{2dispsub} can be written as\footnote{The algebra is quite straightforward and can be easily extended beyond $n=2$. The kernel on the left-hand side of \cref{kerdisp2} is a product of terms as in \cref{2dispker}. Moreover, the structure the kernel in \cref{kern} can actually be written as
$K_n(z,\z)=\frac{i}{(2i)^n}\left[2\prod_{\ell=1}^n(N_{-1,\ell}+N_{0,\ell})-\prod_{\ell=1}^nN_{0,\ell}\right]
$ which simply gives a recursive relation for $K_n$ in terms of lower orders.
}
\eq{
\frac{-i(z_1+i)(z_2+i)}{2(\z_1-z_1)(\z_2-z_2)(\z_1+i)(\z_2+i)}=K_2(z,\z)+\frac{i}{(1+\z_1^2)(1+\z_2^2)}\,.
\label{kerdisp2}
}
Here $K_2$ is the kernel appeared in statement of the theorem in \cref{kern} for $n=2$. Assuming the growth condition \cref{grow} on the measure, each term is separately convergent. Then \cref{2dispsub} gives
\eq{
\tF(z_1,z_2)={\rm Re}\,\tF(i,i)+\frac{1}{\pi^2}\int\dd{\z_1}\dd{\z_2}K_2(z,\z){\rm Im}\,\tF(\z_1,\z_2)\,,
}
where we have used, as in \cref{n1Imi}, the dispersion relation at $(z_1,z_2)=(i,i)$
\eq{
{\rm Im}\,\tF(i,i)=\frac{1}{\pi^2}\int\dd{\z}_1\dd{\z}_2\frac{{\rm Im}\,\tF(\z_1,\z_2)}{(\z_1^2+1)(\z_2^2+1)}\,.
}
Similar to the single variable case, the most that could be added to this is the linear structure $b_1z_1+b_2z_2$ with $b_\ell\geq0$, as in \cref{intrepherg}. 

Given \cref{intrepherg}, the positivity of the imaginary part for the linear terms is clear, provided that $b_\ell\geq0$. For the integral term, one can show that \cite{Luger_2017,LUGER20191189}
\eq{
{\rm Im}\,K_2(z,\z)=\frac{{\rm Im}\,z_1{\rm Im}\,z_2}{|\z_1-z_1|^2|\z_2-z_2|^2}-\frac{1}{(2i)^2}\sum_{\rho\in\{-1,0,1\},-1\in\rho\wedge1\in\rho} N_{\rho_1,1}N_{\rho_2,2}\,,
\label{imK}
}
in which the second term is zero due to the Nevanlinna condition \cref{nevan}. The first term is manifestly positive and therefore ensures positivity in the PUHP. Notice that the first term is the same result we have obtained in \cref{imrep}. The structure of \cref{imK} actually holds for any number of variables. This is relevant for cases without spherical symmetry, as discussed in \cref{norot}. 

\subsubsection*{Nevanlinna condition}

We now want to verify that the Nevanlinna condition is satisfied by the measure given by ${\rm Im}\,\tF(\z_1,\z_2)$.
One can start from \cref{nevan} and show that
\begin{equation}
\label{nevoth}
\sum_{\rho\in\{-1,0,1\},-1\in\rho\wedge1\in\rho} \int_{\mathbb{R}^2}N_{\rho_1,1}N_{\rho_2,2}\dd{\mu}(\zeta_1,\zeta_2)=\int_{\mathbb{R}^2} 
\mathrm{Re}\big( N_{+1,1} N_{-1,2}\big) \, \dd{\mu}(\zeta_1,\zeta_2)
\end{equation}
By analyticity of $\tF$, and the assumed asymptotic behavior, we see that 
\begin{equation}\label{eq:Fnevo}
0=\int\dd{\z_1} \;\frac{(z_1^*-i)}{(\zeta_1 - z_1^*)(\zeta_1+i \, )}\,\tF(\zeta_1,\zeta_2) \,=\int \dd{\zeta_1}\,N_{+1,1}\,\tF(\zeta_1,\zeta_2)\,,
\end{equation}
for $z_1$ in the UHP. Similar argument for $z_2$ gives
\eq{\label{eq:Fnevo2}
\int \dd{\zeta_2}\,N_{-1,2}^*\,\tF(\zeta_1,\zeta_2)=0\,.
}
Then the integral in \cref{nevoth}, with $\dd{\mu}(\zeta_1,\zeta_2)={\rm Im}\,\tF(\z_1,\z_2)\dd{\z}_1\dd{\z}_2$, can be expanded as
\begin{equation}\label{eq:schematic}
\begin{split}
\frac{-i}{4}\int \dd{\zeta_1}\,\dd{\zeta_2}\;\Big(
& N_{1,1}\,N_{-1,2}\,\tF(\zeta_1,\zeta_2)
- N_{1,1}\,N_{-1,2}\,\tF(\zeta_1,\zeta_2)^* \\
&+ N_{1,1}^*\,N_{-1,2}^*\,\tF(\zeta_1,\zeta_2)
- N_{1,1}^*\,N_{-1,2}^*\,\tF^*(\zeta_1,\zeta_2)\Big)=0\,,
\end{split}
\end{equation}
in which every term is zero by \cref{eq:Fnevo} and \cref{eq:Fnevo2}, as a consequence of analyticity.


\section{\label{sec:conc}Conclusions and future directions}

In this paper, we have investigated the analyticity and positivity properties of the retarded Green’s function. It is well known that retardation and micro-causality imply analyticity in the complex frequency–momentum domain. We discussed a theorem that provides a necessary and sufficient condition for such analyticity, following \cite{bogolubov2012general}. As a consequence, we examined multi-variable dispersion relations and their generalizations, including cases without rotational invariance and with additional subtractions.  

The imaginary part of the Green’s function plays a central role, as it encodes both the analytic structure of the function and its physical interpretation in terms of dissipation. We have shown that micro-causality imposes strong constraints on this imaginary part—for instance, it cannot possess compact support. Moreover, it must satisfy certain integral constraints, \cref{imcons5}, which we have verified through several examples. When further restricting to passive systems that only dissipate energy, the imaginary part must obey certain positivity requirements. This, in turn, ensures the absence of zeros within the domain of analyticity and naturally links the Green’s function to the class of Herglotz–Nevanlinna functions, which are well studied in mathematics literature. 

We conclude by highlighting a few remarks and presenting open questions that suggest directions for future studies.

\paragraph{Solving the constraints} As discussed in \cref{imconssec}, for \emph{any} $\rho(\z_1,\z_2)$, the integral
\eq{
G_{\rho}(z_1,z_2)=\frac{-i}{2\pi^2}\int\dd{\z}_1\dd{\z}_2\frac{\rho(\z_1,\z_2)}{(\z_1-z_1)(\z_2-z_2)}\,,
\label{Grho}
}
defines an analytic function in the PUHP. Moreover, one can show that distinct functions $\rho$ may lead to the same analytic function in the PUHP. In particular, it can be verified that
\eq{
\rho\to\rho'=\rho+(\delta\rho+\mathbb{H}_{\z_1}\mathbb{H}_{\z_2}\delta\rho)\,,
\label{rhored}
}
yields the same analytic function, $G_\rho = G_{\rho'}$, within the PUHP. One can check this showing that the second term cancels by a repeated use of the Cauchy theorem when $z_1$ and $z_2$ are in the PUHP. (The definition of the Hilbert transfrom $\mathbb{H}$ is given in \cref{hilbertt}.) It should be noted, however, that while $G_\rho$ and $G_{\rho'}$ coincide in the PUHP, they generally differ outside this domain, i.e. $G_{\rho'}(z_1^*, z_2) \neq G_{\rho}(z_1^*, z_2)$ for $z_1, z_2$ in the PUHP. Notice that the functions defined by \cref{Grho} outside the PUHP do not in general coincide with the analytic extension, if this exists. See \cite{LUGER20191189}. For a generic $\rho$, the imaginary part ${\rm Im}\,G_\rho$ is not identical to $\rho$. Requiring ${\rm Im}\,G_\rho=\rho$ coincides with \cref{HH1} and selects, among the equivalent representations, the specific one that satisfies the integral constraints in \cref{imcons5}. Indeed, for any $\rho$, we may write 
\eq{
\rho=\frac{1}{2}\big(\rho-\mathbb{H}_{\z_1}\mathbb{H}_{\z_2}\rho\big)+\frac{1}{2}\big(\rho+\mathbb{H}_{\z_1}\mathbb{H}_{\z_2}\rho\big)\,.
}
The second term has the same structure as in \cref{rhored} and therefore does not contribute within the PUHP. The first term, by contrast, can be shown to satisfy the integral constraints \cref{HH1} for any $\rho$. Hence, we conclude that a function $\rho$ satisfies \cref{HH1} if and only if it can be expressed as 
\eq{
\rho=\rho'-\mathbb{H}_{\z_1}\mathbb{H}_{\z_2}\rho'\,,
}
for some $\rho'$. The more subtle issue concerns ensuring the positivity condition. While the combination $(\rho' - \mathbb{H}_{\z_1}\mathbb{H}_{\z_2}\rho')$ automatically satisfies the micro-causality constraints for any $\rho'$, it is not evident what condition on $\rho'$ guarantees that this combination also exhibits the required positivity. Simply taking $\rho'$ to be a positive function does not suffice. It would be interesting to find an integral representation that makes both micro-causality and positivity manifest simultaneously.

\paragraph{Källén–Lehmann representation} The previous discussion naturally connects with extension of the Källén–Lehmann (KL) representation to situations where Lorentz symmetry is spontaneously broken. In a Lorentz-invariant theory, one can always express any two-point function in a spectral form, derived from symmetry considerations, in terms of the propagator of the free theory (see, for example, \cite{Weinberg:1995mt}). In the absence of Lorentz boosts, however, the standard derivation no longer applies straightforwardly (see, e.g., \cite{Alberte:2020eil}). The relations presented in \cref{2disp}—or more precisely in \cref{2dispomegak}—may thus be viewed as the counterpart of the KL representation in the absence of Lorentz symmetry.

It is important, however, to emphasize a few distinctions. First, as we have repeatedly noted, the dispersion relation \cref{2dispomegak} must be supplemented by the micro-causality constraints. In contrast to the standard Lorentz-invariant case, where any positive spectral function is admissible, here the allowed functions must also satisfy the conditions in \cref{imcons5}. Second, the familiar intuition that a two-point function can be expanded in terms of free-theory propagators does not directly carry over when Lorentz boosts are absent. Relatedly, a localized spectral function of the form ${\rm Im}\,G(\omega,k)\propto\delta(\omega-\omega_0)\delta(k-k_0)$ is forbidden, as discussed in \cref{imsupsec}.\footnote{Same holds with Lorentz invariance. On the other hand, here one may instead have a localized spectral function in $\omega^2 - k^2$ space, i.e. $\mathrm{Im}\,G(\omega,k) \propto \delta(\omega^2 - k^2 - \mu_0^2)$.}

Perhaps one possible approach to address this issue is to expand the Green’s function in terms of a set of physically motivated basis functions for which both the positivity and micro-causality constraints are explicitly satisfied. As an illustrative example, one may consider a family of functions of the form given in \cref{drude}, characterized by varying mass, speed of sound, and damping parameters. Of course, one would still need to verify the completeness of such a basis, as well as the uniqueness of the corresponding expansion.

Finally, note that the standard argument readily generalizes to other types of correlation functions—such as time-ordered or advanced propagators—using the same spectral function. However, the representation discussed above fundamentally relies on micro-causality as it applies to the retarded Green’s function, and it remains unclear how such a construction can be consistently extended beyond the Lorentz-invariant setting. 

\paragraph{Scattering amplitudes} An analogous integral representation has been conjectured for scattering amplitudes, known as the Mandelstam representation \cite{PhysRev.112.1344}. This form follows naturally under the assumption of maximal analyticity in the complex $s$- and $t$-planes, although such analyticity has not been rigorously established nonperturbatively from first principles-- see \cite{osti_4050833}. Apart from minor differences—such as an overall prefactor in \cref{Grho} and the explicit symmetrization among the Mandelstam variables $s$, $t$, and $u$ to make crossing symmetry manifest—the two representations are essentially identical. In the S-matrix literature, the function $\rho$ is referred to as the double discontinuity. Indeed, one can show directly from \cref{Grho} that
\eq{
{\rm Disc}_{z_1}{\rm Disc}_{z_2}G_\rho=2i\rho(z_1,z_2)\,,
}
for real $z_1$ and $z_2$, where ${\rm Disc}_{z_1}G_\rho\equiv G_\rho(z_1+i\epsilon,z_2)-G_\rho(z_1-i\epsilon,z_2)$. It should be noted that this definition depends on the behavior of the function beyond the PUHP, and therefore the redundancy discussed in \cref{rhored} does not apply in this context. It would be interesting to investigate whether the double discontinuity is required to satisfy integral constraints analogous to those obeyed by the imaginary part of the retarded Green’s function. However, as a consequence of unitarity, the double discontinuity is subject to more intricate conditions than mere positivity.

\paragraph{Typically-real functions}
All the properties studied in this paper holds for the {\em advanced} Green's function, replacing the forward light-cone with the backward one and flipping the sign of the imaginary part. If $\Im G(\omega,k)$, with $\omega$ and $k$ real, vanishes in an open set then one can apply Bogolubov's edge-of-the-wedge theorem \cite{bogolubov2012general}: the retarded and advanced Green's function are analytic continuation of each other and the region of analyticity is larger than the union of forward and backward light-cone. In this case one can start studying the sign of $\Im G(\omega,k)$ beyond the PUHP and connect to the notion of typically-real functions.
Typically-real functions \cite{Rogo,Raman:2021pkf} are defined to be functions having positive imaginary part in UHP and negative imaginary part in LHP namely $\Im z \Im[f(z)]\geq0$. These functions enjoy the property that their Taylor coefficients around the origin are two-sidedly bounded. Typically real functions appear in many areas of physics with profound applications \cite{Wigner:1947zz,Bender:1968sa, Jin:1964zza, Mickens:1978qb}. Scattering amplitudes are typically real functions in a crossing symmetric variable, enabling one to bound the Wilson coefficients using these Taylor coefficient bounds\cite{Raman:2021pkf,Zahed:2021fkp}. It will be interesting to investigate such properties in case of two variables and their consequences for retarded Green's functions. 

\paragraph{Thermal state}
In the paper we did not commit to a precise form of the density matrix. It would be interesting to understand how the results are modified when one restricts to thermal states. In this case the correlation functions have important properties {\em in real space}: they must satisfy the KMS condition, \cref{kms}, and they are analytic in the complex $t$-plane in the strip $0 < \Im t < \beta$, see for instance \cite{Calzetta:2008iqa}. The interplay of the Fourier-space properties studied above and these real-space features should be quite powerful.

\paragraph{Wider ``light''-cone and Lieb-Robinson velocity}
Relativity implies that the retarded Green's function must vanish outside the lightcone. In many cases, however, the system is actually much slower, with a response that vanishes with excellent accuracy outside a ``slower'' cone characterised by a velocity $v \ll c$. This corresponds in Fourier space to a correspondingly larger region of analyticity. For instance, the system of degenerate fermions (Lindhardt function) studied in \cref{sec:chemical} is characterised by the Fermi velocity, $v_f$,  and one can check that the Leontovich relations, \cref{leontovich}, are satisfied not only for $|\xi| < 1 $, but even for $1 \le |\xi| \lesssim 1/(3 v_f)$ \cite{Creminelli:2024lhd}. Another situation in which an effective cone appears is in non-relativistic quantum systems, via the celebrated Lieb-Robinson bound \cite{Lieb:1972wy}.  It would be interesting to explore how the results of this work change in these cases, taking into account that the vanishing outside the effective light-cone is not exact, but approximate. Interesting results in this direction already appeared in \cite{Chowdhury:2025qyc}.

\paragraph*{Acknowledgements}
It is a pleasure to thank L.~Di Pietro, M.~Nedic, A.~Nicolis, A.~Podo, L.~Senatore, and A.~Sinha for useful discussions. We also thank A.~Podo for important comments on the manuscript, and M.~Nedic for insightful discussions on Herglotz-Nevanlinna functions. AZ has received support from the European Research Council, Grant No.~101039756.


\appendix
\section{Microcausal tempered distributions}\label{AppA}

In this appendix, we present a concise overview of the main properties of tempered distributions, together with a proof of the main theorems discussed in Section \ref{Bogo}. The treatment will be brief but pedagogical. For a more complete and rigorous exposition, we refer to \cite{Nussenzveig:1972tcd, stein1971introduction}.  

As mentioned in Section \ref{Bogo}, tempered distributions are elements of the dual space of the space of Schwarz functions. Therefore, we begin by briefly recalling the definition and key properties of Schwarz functions.  

\subsection{Schwarz functions}

The space of Schwarz functions \(S(\mathbb{R}^d)\)\footnote{It is also useful to single out a subset of $S(\mathbb{R}^d)$ consisting of smooth functions with compact support in $\mathbb{R}^d$. This space is usually denoted by $\mathcal{D}(\mathbb{R}^d)\subset S(\mathbb{R}^d)$. }, also referred to as Schwarz test functions, consists of smooth functions that, together with all their derivatives, decay faster than any polynomial at infinity. More explicitly, \(\phi\in S(\mathbb{R}^d)\), when $\phi(x) \in C^{\infty}$ and   
\begin{equation}
C_{\alpha,\beta}(\phi)\equiv\sup_{x\in \mathbb{R}^d}\abs{x^\alpha \partial^{\beta}\phi(x)}<\infty\>\,,\qquad\forall \alpha,\beta \in \mathbb{N}^d\,.
\label{Calphabeta}
\end{equation}
Notice that we are using multi-index notation here; for \(\alpha=(\alpha_1,\cdots,\alpha_d)\) and \(\beta=(\beta_1,\cdots,\beta_d)\) with non-negative integer entries, and $x^\alpha\equiv x_1^{\alpha_1}\cdots x_d^{\alpha_d}$ and $\p^\beta\equiv\p_{x_1}^{\beta_1}\cdots\p_{x_d}^{\beta_d}$. One can easily check that, given any polynomial $p(x)$, the function $\phi(x)\equiv p(x)e^{-\vec{x}^2}$ (with $\vec{x}^2\equiv\sum_n (x^n)^2$ being the Euclidean norm of $\vec{x}\in \mathbb{R}^d$) is a Schwarz function. Moreover, $S(\mathbb{R}^d)$ is a vector space. It is also easy to see from the definition that derivatives of a Schwarz function are themselves Schwarz functions.

Elements of $S(\mathbb{R}^d)$ are sometimes referred to as ``rapidly decreasing functions''. The motivation for this nomenclature can be appreciated by noticing that, for any $i=1,\cdots,d$ we have 
\begin{equation}\begin{aligned}
    \lim_{x^i\to \infty}|x^{\alpha}\partial^{\beta}\phi|&=\lim_{x^i\to \infty}\frac{|x^ix^{\alpha}\partial^{\beta}\phi|}{|x^i|}<\lim_{x^i\to \infty}\frac{\sup_{x\in \mathbb{R}^d}|x^ix^{\alpha}\partial^{\beta}\phi|}{|x^i|}=\lim_{x^i\to \infty}\frac{C_{\alpha',\beta}(\phi)}{|x^i|}=0\,,
    \end{aligned}
\end{equation}
where in the first line we have multiplied and divided by $|x^i|$, and in the second line we used the definition \cref{Calphabeta}, together with $\alpha'\equiv(\alpha_1,\cdots,\alpha_i+1,\cdots,\alpha_d)$. The rapid decay property makes \(S(\mathbb{R}^d)\) especially useful: the Fourier transform of any Schwarz function is well-defined, and it is itself a Schwarz function. This can be noticed by the following chain of relations
\begin{equation}
    \begin{aligned}|\mathcal{F}[\phi](k)|&=\bigg|\int\dd[d]{x}\,\phi(x)e^{-ik\cdot x}\bigg|\leq \int\dd[d]{x}\,|\phi(x)|=\int\dd[d]{x}\,\frac{|(1+\vec{x}^2)^{n}\phi(x)|}{(1+\vec{x}^2)^{n}}\\
    &\leq\sup_{x\in\mathbb{R}^d}|(1+\vec{x}^2)^{n}\phi(x)|\int\frac{\dd[d]{x}}{(1+\vec{x}^2)^{n}}<\infty\,,
    \end{aligned}
\end{equation}
where we have used Cauchy–Schwarz inequality, and then multiplied and divided by $(1+\vec{x}^2)^n$ for some $n>d/2$. This implies that the Fourier transform exists. Moreover, using multi-index notation as in \cref{Calphabeta}, we have 
\begin{equation}
|k^{\alpha}\partial_k^{\beta}F[\phi]|=\bigg|\int\dd[d]{x}\,e^{-ik\cdot x}\,\partial^{\alpha}\left(x^{\beta}\phi\right)\bigg|=\big|F\left[\partial^{\alpha}(x^\beta \phi)\right](k)\big|<\infty\,,
\end{equation}
where we have integrated by part multiple times. This implies $C_{\alpha,\beta}(\mathcal{F}[\phi])<\infty$, and therefore \(\mathcal{F}: S(\mathbb{R}^d) \to S(\mathbb{R}^d)\). We will see later that the Fourier transform also maps the dual space \(S'(\mathbb{R}^d)\), space of tempered distributions, into itself. 

One can define a family of norms on $S(\mathbb{R}^d)$ as follows. For non-negative integers $m$ and $n$,
\eq{
\norm{\phi}_{m,n}\equiv\max_{|\alpha|\leq m,|\beta|\leq n} C_{\alpha,\beta}(\phi)\,,
\label{normphi}
}
where the multi-index notation is used, with $|\alpha| = \sum_i \alpha_i$. This norm is used to define the notion of continuity on the space of test functions.

While we have introduced Schwartz functions in Euclidean space $\mathbb{R}^d$, the discussion extends straightforwardly to Minkowski space $\mathbb{R}^{d-1,1}$, or more generally to spaces with arbitrary metric signature. This will be the main setting of the second part of the following discussion.

\subsection{Tempered distributions}

As previously anticipated, tempered distributions are elements of the dual space to Schwarz functions, denoted by \(S'(\mathbb{R}^d)\), i.e. continuous linear functionals from \(S(\mathbb{R}^d)\) to $\mathbb{R}$. The action of a tempered distribution $T$ on a test function $\phi$ is denoted by $T[\phi]$. Alternative notations that are sometimes we will use interchangeably are
\eq{
T[\phi]\equiv\bk{T(x)}{\phi(x)}\equiv\int\dd[d]{x}\,T(x)\phi(x)\,,
}
where the last two expressions must be understood symbolically, since it generally makes no sense to speak about the value of a tempered distribution at a given point.\footnote{Indeed, the definition of the support of a distribution requires some care. To be precise, the support of a distribution $T$ is identified as the complement of the largest open set in which $T$ vanishes. In particular, $T$ is said to vanish on an open set $O$ if $T[\phi]=0$ for any compactly supported test function $\phi$ such that ${\rm supp}(\phi)\subset O$. Then, the support of $T$ can be defined as the complement of the largest open set in which $T=0$.} Nonetheless, this notation is particularly useful since many operations on tempered distributions are defined as if they were ordinary functions acting on test functions by integration. 

The key property that characterizes any tempered distribution is to have its action on any test function to be bounded by the norm of the test function itself. To be more precise, one can show that a linear functional \(T\) on $S(\mathbb{R}^d)$ is a tempered distribution if and only if there exist \(K>0\) and non-negative integers \(m\) and $n$ such that
\begin{equation}
    |T[\phi]|\leq K \norm{\phi}_{m,n}\,,\quad\forall \phi \in S(\mathbb{R}^d)\,.
    \label{Tbound}
\end{equation} 
Operations such as linear combination are defined straightforwardly for tempered distributions. Generally speaking, for any diffeomorphism $f$ of $\mathbb{R}^d$, we define the transformation of a tempered distribution, inspired by integration of ordinary functions, as
\eq{
\bk{T\circ f^{-1}(x)}{\phi(x)}\equiv\bk{T(x)}{|\p f|\phi\circ f(x)}\,,
\label{diff}
}
in which $|\p f|$ is the Jacobian. For example, we can define the translation of a distribution as $\tau_{x}T[\phi]\equiv T[\tau_{-x}\phi]$. By using \cref{Tbound} one can show that tempered distributions grow at most polynomially, as follows
\begin{equation}
\begin{aligned}
|\tau_xT[\phi]&|\equiv |\bk{T(x+y)}{\phi(y)}=|\bk{T(y)}{\phi(y-x)}|=|T[\tau_{-x}\phi]|\leq K||\tau_{-x}\phi||_{m,n}=\\
&=K\max_{|\alpha|\leq m,|\beta|\leq n}\sup_{y\in \mathbb{R}^d}|y^{\alpha}\partial^{\beta}\phi(y-x)|=K\max_{|\alpha|\leq m,|\beta|\leq n}\sup_{y\in \mathbb{R}^d}|(y+x)^{\alpha}\partial^{\beta}\phi(y)|\leq\\
&\leq K |x|^m \max_{|\alpha|\leq m,|\beta|\leq n}\sum_{k=0}^{\alpha}\binom{\alpha}{k} C_{k,\beta}(\phi) \leq K'|x|^m\,,
\end{aligned}
\end{equation}
where in the first line we have used the definitions of a tempered distribution and of the norms of Schwarz functions, while in the second line we performed a change of variables and used binomial expansion, together with the definition of Schwarz functions. An inspection at the \(x \to \infty\) limit unveils the characteristic polynomial boundedness of tempered distributions. In this sense, elements of $S'(\mathbb{R}^d)$ are sometimes said to be \emph{slowly increasing }.\footnote{Distributions which grow faster than a polynomial are elements of $\mathcal{D}'(\mathbb{R}^d)$. They naturally act on the space of smooth compactly supported functions. Notice that $\mathcal{S}'(\mathbb{R}^d)\subset\mathcal{D}'(\mathbb{R}^d)$. In fact, one can show that any element of $\mathcal{D}'(\mathbb{R}^d)$ which satisfies \cref{Tbound} can be extended to a unique element of $\mathcal{S}'(\mathbb{R}^d)$.\label{dprimSprim}}

Similarly, one can define the derivative of a tempered distribution as 
\begin{equation}
    \partial^\beta T[\phi]=\bk{\partial_x^\beta T(x)}{\phi(x)}\equiv(-)^{|\beta|}\bk{T(x)}{\partial_x^{\beta}\phi(x)}=(-)^{|\beta|}T[\partial^{\beta}\phi]\,,
    \label{der}
\end{equation}
which is consistent with the intuition from integration by parts. It is therefore clear that the derivatives of a tempered distribution are elements of $S'(\mathbb{R}^d)$. Additionally, if a function $g$ is such that for any test function $\phi$, the product $g\phi$ is a test function, it is called a \emph{multiplicator}, and then its product with a tempered distribution is defined as
\eq{
gT[\phi]\equiv T[g\phi]\,.
\label{mul}
}
One can show that any smooth function of polynomial growth is a multiplicator. In particular, all test functions are multiplicators. On the other hand, products of two tempered distributions are not generally defined.

A crucial object we will frequently use is the Fourier transform of a tempered distribution, which is defined as\footnote{A factor $(2\pi)^d$ in the $k$-space measure is implicitly suppressed to keep the notation light.} 
\eq{
\mathcal{F}[T][\phi]\equiv\int \dd^d k\, \mathcal{F}[T](k)\phi(k)=\int \dd^d k \int \dd^d x\, e^{-ik\cdot x} T(x)\phi(k) 
    = \int \dd^d x\, T(x) \mathcal{F}[\phi](x) = T[\mathcal{F}[\phi]]\,.
\label{Four}
}
Notice that the above object is well defined since \(\mathcal{F}[\phi] \in S(\mathbb{R}^d)\). Thus, it follows that \(\mathcal{F}[T] \in S'(\mathbb{R}^d)\). This property is fundamental: it ensures that the Fourier transform of a tempered distribution (together with all its derivatives) is polynomially bounded for real arguments.   We see that similar to \(S(\mathbb{R}^d)\), the space \(S'(\mathbb{R}^d)\) is preserved under the Fourier transform. 

Another operation we can consider is the  
convolution between a tempered distribution and a test function:
\eq{
[T\ast\phi](x)\equiv \bk{T(y)}{\phi(x-y)}\,,
\label{conv}
}
where $\phi(x-y)$ is regarded as a test function in $y$ for fixed $x$. Notice that \cref{conv} defines, as a function of $x$, a smooth \emph{function} of polynomial growth. Since the convolution is polynomially bounded, one can take its Fourier transform in a distribution sense. Moreover, one can show that
\eq{
\mathcal{F}[T\ast\phi](k)=\mathcal{F}[T](k)\mathcal{F}[\phi](k)\,.
\label{convFou}
}
To see this, we act on some other test function
\eq{
\spl{
\bk{\mathcal{F}[T\ast\phi](k)}{\mathcal{F}[\psi](-k)}&=\bk{T\ast\phi(x)}{\psi(x)}=\bk{T(x)}{\phi(-x)\ast\psi(x)}\\
&=\bk{\mathcal{F}[T](k)}{\mathcal{F}[\phi(-x)\ast\psi(x)](-k)}=\bk{\mathcal{F}[T](k)}{\mathcal{F}[\phi](k)\mathcal{F}[\psi](-k)}\\
&=\bk{\mathcal{F}[T](k)\mathcal{F}[\phi](k)}{\mathcal{F}[\psi
](-k)}\,,
}
}
where in the first line we have used \cref{Four} and properties of convolution of test functions, in the second line we have again used \cref{Four} and the convolution theorem, and in the last line we have used the fact that test functions are multiplicators. It is useful to re-write \cref{conv} in terms of the Fourier transforms, using \cref{Four}, as
\eq{
[T\ast\phi](x)=\bk{\mathcal{F}[T](k)}{e^{ik\cdot x}\mathcal{F}[\phi](k)}=\mathcal{F}[T][e^{ik\cdot x}\mathcal{F}[\phi]]\,,
\label{convFou2}
}
in terms of the action of $\mathcal{F}[T]$ on the test function $e^{ik\cdot x}\mathcal{F}[\phi]$, which is a smooth function of $x$.


\subsubsection*{Laplace transform}

As discussed in the main text, it is often natural to analytically continue the Fourier transform of tempered distributions to complex momenta. This is tantamount to study the properties of the Laplace transform of distributions. We begin by defining, for a generic distribution $T$ (not necessarily tempered), the set
\begin{equation}
   \Gamma(T) \equiv \{q \in \mathbb{R}^d \mid e^{-q \cdot x} T \in S'(\mathbb{R}^d)\}
   \label{GammaT}
\end{equation}
In words, $\Gamma(T)$ is the set of all real vectors $q$ such that multiplication by the exponential weight $e^{-q \cdot x}$ makes $T$ a tempered distribution.\footnote{If \(T\) is tempered, it is clear that \(\{0\} \in \Gamma(T)\). However, not all tempered distributions admit a nontrivial \(\Gamma(T)\).  The issue is that the factor \(e^{-q\cdot x}\) can spoil polynomial boundedness if \(T\) has support over all of \(\mathbb{R}^d\). In that case, the action of \(e^{-q\cdot x} T\) on \(S(\mathbb{R}^d)\) becomes ill-defined. In contrast, if \(\mathrm{supp}(T)\) is such that there exists at least one \(q \neq 0\) with \(q\cdot x \geq 0\) for all \(x \in \mathrm{supp}(T)\), then \(\Gamma(T)\) is nontrivial.} The rationale behind this definition is that, for suitable choices of $q$, the exponential factor may control the growth of $T$ at infinity, thereby ensuring the existence of its Fourier transform. In this way, $\Gamma(T)$ specifies the domain where the Laplace transform—or equivalently the Fourier transform at complex momenta—is well defined. More precisely, for a complex vector $k=p+iq$, the Laplace transform of distribution $T$ is defined as
\eq{
\mathcal{L}[T]\equiv \mathcal{F}[e^{-q \cdot x}T]\,,\qquad q\in \Gamma(T)\,,
\label{laplace}
}
namely the Fourier transform (with respect to $x \mapsto p$) of the tempered distribution $e^{-q \cdot x} T$. Thus, for every fixed $q \in \Gamma(T)$, the Laplace transform is itself a tempered distribution in $p$. Concretely, one can use \cref{laplace} and \cref{Four} to show that $\mathcal{L}T$ acts on a test function as
\eq{
\mathcal{L}[T][\phi]\equiv T[e^{-q\cdot x}\mathcal{F}[\phi]]\,.
}
A key fact is that $\mathcal{L}[T]$ can be shown to be an \emph{analytic function} of $k\equiv p+iq $ within the domain\footnote{An open subset of $\mathbb{C}^d$ of the form $\mathbb{R}^d+i\,K\subset\mathbb{C}^d$, with $K\subset\mathbb{R}^d$, is also called \textit{tube with base $K$}. Thus, the region of analyticity of $\mathcal{L}[T]$ is a tube with base $\mathring{\Gamma}(T)$.} $\mathbb{R}^d+i\,\mathring{\Gamma}(T)$, where $\mathring{\Gamma}(T)$ denotes the (open) interior of $\Gamma(T)$. 

From now on, we will work in Minkowski spacetime, $\mathbb{R}^d\to \mathbb{R}^{d-1,1}$, and focus our attention to retarded and micro-causal tempered distributions. More precisely, we focus on tempered distributions supported on
\begin{equation}
\mathrm{supp}(T)=\mathrm{FLC}\equiv\{x \in \mathbb{R}^{d-1,1} \mid x^2 \leq 0, \ x^0 \geq 0\}\,,
\end{equation}
where we have defined the forward lightcone (FLC) for the vector $x^\mu$ with respect to the Minkowski metric. It is not difficult to see that \({\rm FLC}\subset \Gamma(T)\) simply because for both $x^\mu$ and $q^\mu$ in FLC, we will have $x\cdot q\leq0$\footnote{Notice that we are using a mostly plus signature, hence the exponent in the definition of $\Gamma(T)$ must be sign-flipped.} so there is always an exponential suppression in \cref{laplace}.\footnote{Generally speaking, for a tempered distribution $T$ with support in a convex cone $K$, the domain $\Gamma(T)$ always contains the dual cone $K^*$, consisting of points with $x\cdot q\leq0$.}



\subsection{Microcausality-analyticity theorem}

We now state and prove the main theorem of \cref{Bogo} for retarded and micro-causal tempered distributions (see \cite{bogolubov2012general} for a more comprehensive treatment).

\begin{center}
\rule{\textwidth}{0.4mm}
\end{center}

\paragraph{Theorem}
A tempered distribution $T \in S'(\mathbb{R}^{d-1,1})$ is retarded and micro-causal if and only if its Laplace transform, $\mathcal{L}[T]$ is analytic in the tube
\[
\mathcal{T}= \mathbb{R}^{d-1,1} + i \, \mathring{\rm FLC}\,,
\]
and satisfies the estimate
\begin{equation}
   |\mathcal{L}[T](k)| \leq A \, (1+|k^{\mu}|)^n\left( 1+\frac{1}{(k^0_I-|\vec{k}_I|)^m}\right),
   \qquad \forall\, k_R \in \mathbb{R}^{d-1,1}, \quad k_I \in \mathring{\rm FLC}\,,
   \label{est}
\end{equation}
where $A > 0$ and $m, n \in \mathbb{N}$ are constants that depend on $T$. Furthermore, the Fourier transform of the distribution is the boundary value of the Laplace transform.
\begin{center}
\rule{\textwidth}{0.4mm}
\end{center}

To prove the $\implies$ direction, we study the properties of the Laplace transform on the tube $\mathcal{T}$. Notice that the interior of ${\rm FLC}$ consists exclusively of future-directed timelike vectors:
\[
\mathring{\rm FLC}=\{k_I\in \mathbb{R}^{d-1,1} \,|\, k_I^0>0, \, k_I^2<0\}\,.
\]
By definition, the Laplace transform is a tempered distribution in $k_R$ for every $k_I\in\mathring{\rm FLC}$. In fact, more than that, it is a smooth function. To see this, we note that, since $e^{k_I\cdot x}$ decays exponentially for $x\in{\rm FLC}$, there exists a test function $u_{k_I}(x)\in S(\mathbb{R}^{d-1,1})$ that coincides with $e^{k_I\cdot x}$ on ${\rm FLC}$. Because $T$ is supported only in the ${\rm FLC}$, we may therefore rewrite the product $e^{k_I\cdot x}=u_{k_I}(x)T$. We then exploit the definition in \cref{laplace} to claim that $\mathcal{L}[T]$ is nothing but the Fourier transform of $u_{k_I}(x)T$. According to the analogue of \cref{convFou}, this is the convolution of the tempered distribution $\mathcal{F}[T]$ with the test function $\mathcal{F}[u_{k_I}]$. As recalled in \cref{conv}, convolution of a tempered distribution with a test function gives a smooth function. Concretely, from \cref{conv} and \cref{convFou2}, we obtain
\eq{
\mathcal{L}[T](k)=\mathcal{F}[T]\ast\mathcal{F}[u_{k_I}]=\bk{T(x)}{e^{-ik_R\cdot x}u_{k_I}(x)}=\bk{T(x)}{e^{-ik\cdot x}}=T[e^{-ik\cdot x}]\,,
\label{laplace2}
}
where in the last step we replaced $u_{k_I}(x)$ with $e^{k_I\cdot x}$, which is legitimate because $T$ has support in ${\rm FLC}$. Thus, the Laplace transform $\mathcal{L}[T]$ is indeed a smooth function of $k$ within the tube $\mathcal{T}$, since tempered distributions are continuous functionals. It is easy to see that it is in fact an analytic function of $k$, simply because the exponential is analytic in $k$. In particular, the Cauchy Riemann equations are satisfied for any $k\in \mathcal{T}$.

Next, we must verify that the stated estimate holds on $\mathcal{T}$. From \cref{laplace2}, fixing $k \in \mathcal{T}$, we compute the modulus of $\mathcal{L}T(k)$:
\begin{equation}
\begin{aligned}
|\mathcal{L}T(k)| &= \big| T\big[ e^{-ik\cdot x} \big] \big| 
    \leq K \norm{e^{-ik\cdot x}}_{m,n}= K \max_{|\alpha|\leq m,\,|\beta|\leq n} \,|k|^{\beta}\,\sup_{x\in {\rm FLC}}\, |x^{\alpha} e^{k_I\cdot x}| 
\end{aligned}
 \end{equation}
where we have used bounds on tempered distributions in \cref{Tbound} and taken the derivatives. We remind that we are using multi-index notation $|k|^\beta=|k^0|^{\beta_0}\cdots|k^{d-1}|^{\beta_{d-1}}$, and $x^\alpha=(x^0)^{\alpha_0}\cdots(x^{d-1})^{\alpha_{d-1}}$. First of all, let us try to maximize with respect to $x\in{\rm FLC}$. To do so, we use the fact that $|x^i|\leq|\vx|\>,\forall i=1,...,d-1$, and $\vk_I\cdot\vx\leq|\vk_I||\vx|$. Moreover, since we are looking for the sup in the FLC, we can implement the constraint $|\vec{x}|\leq x^0$ to obtain
\eq{
\sup_{x\in {\rm FLC}}\, |x^{\alpha} e^{k_I\cdot x}|\leq \sup_{x\in {\rm FLC}} |x^0|^{\alpha_0}|\vec{x}|^{\alpha_1+...+\alpha_d}e^{-k_I x^0+|\vec{k}_I||\vec{x}|}\leq\sup_{x^0\geq 0} |x^0|^{|\alpha|}|e^{-(k_I^0-|\vec{k}_I|)x^0}|=\left( \frac{|\alpha|}{k^0_I-|\vec{k}_I|}\right)^{|\alpha|}
}
where we evaluate the sup over $x^0$ just by computing the maximum of the function. Finally, we need to maximize with respect to the d-tuples,  $\alpha$ and $\beta$. Dealing with $\beta$ is more straightforward, as we can notice that the following chain of inequalities holds 
\begin{equation}
    |k|^{\beta}=|k^0|^{\beta_0}...|k^{d-1}|^{\beta_{d-1}}\leq |k|^{|\beta|}\leq (1+|k|)^{|\beta|}\>,
\end{equation}
where $|k|\equiv \sqrt{|k^0|^2+...+|k^{d-1}|^2}$ is the Euclidean norm of the complex vector $k$. It is immediate to argue that the maximum is attained at $|\beta|=n$. On the other hand, maximizing wrt $\alpha$ requires some care, since we can notice that 
\begin{equation}
    \max_{|\alpha|\leq m}\left( \frac{|\alpha|}{k^0_I-|\vec{k}_I|}\right)^{|\alpha|}=\begin{cases}
        1\>,\>{\rm if}\>k^0_I-|\vec{k}_I|>m\\
        \left( \frac{m}{k^0_I-|\vec{k}_I|}\right)^{m}\>,\>{\rm if}\>k^0_I-|\vec{k}_I|<m
    \end{cases}\>.
\end{equation}
In particular, in the large $k^0_I-|\vec{k}_I|$ limit, the maximum is attained at $|\alpha|=0$. Therefore, the bound acquires the form 
\eq{
|\mathcal{L}[T](k)|\leq A  (1+|k|)^n\left( 1+ \frac{1}{(k_I^0-|\vec{k}_I|)^m}\right) \,.
}
It is clear that the two integers $n,m$ control the growth in two different regimes. Being more explicit, $m$ captures the behavior of the l.h.s. in the limit in which the boundary of $\mathring{{\rm FLC}}$ is reached (i.e. which is the limit $k_I^0-|\vec{k}_I|\to 0$). To make contact with \cite{bogolubov2012general} we can recast the denominator into a more Lorentz-invariant-looking form using the following chain of inequalities:
\begin{equation}
    \frac{1}{(k_I^0-|\vec{k}_I|)^m}=\frac{(k^0_I+|\vec{k}_I|)^m}{(-k_I^2)^m}\leq\frac{|k_I|^m}{\left(\sqrt{-k_I^2}\right)^{2m}}\leq \frac{|k|^m}{\left(\sqrt{-k_I^2}\right)^{2m}}\leq \frac{(1+|k|)^m}{\left(\sqrt{-k_I^2}\right)^{2m}}
\end{equation}
As a consequence, the bound can be recast into 
\eq{
|\mathcal{L}T(k)|\leq A (1+|k|)^n\left( 1+ \frac{(1+|k|)^m}{\left(\sqrt{-k_I^2}\right)^{2m}}\right)\,,
}
or, more compactly, as
\begin{equation}
    |\mathcal{L}T(k)|\leq A  \frac{(1+|k|)^{n+m}}{\left(\sqrt{-k_I^2}\right)^{2m}}.
\end{equation}
Redefining the integers $m'\equiv n+m\>,\>l'\equiv 2m$ yields the same form used in Corollary B.9 in \cite{bogolubov2012general} for the case $n=1$ (since the domain of analyticity is just a single tube rather than a product of forward tubes).

For the $\impliedby$ direction, we start from an analytic function $L(k)$ in the tube $\mathcal{T}=\mathbb{R}^{d-1,1}+i\mathring{\rm FLC}$, that satisfies the estimate in \cref{est}. Since, for fixed $k_I$, the estimate guarantees polynomial boundedness in $k_R$, we can regard $L(k_R+ik_I)$ as a tempered distribution in $k_R$. Analyticity implies that the inverse Fourier transform of $L(k)$ satisfies
\begin{equation}
\big(x_{\mu} - \partial_{k_I^{\mu}}\big)\mathcal{F}^{-1}[L(k_R+ik_I)](x) = 0\,,
\label{fouana}
\end{equation}
where we simply took inverse Fourier transform (with respect to $k_R$) of Cauchy-Riemann equation, i.e. $( \partial_{k_R^{\mu}} - i \partial_{k_I^{\mu}}) L(k) = 0$. Then, let us define
\begin{equation}
    T(x,k_I) \equiv e^{-k_I\cdot x} \, \mathcal{F}^{-1}[L](x) \in \mathcal{D}'(\mathbb{R}^{d-1,1})\,,
    \label{Txki}
\end{equation}
which is in general an element of $\mathcal{D}'(\mathbb{R}^{d-1,1})$, not $S'(\mathbb{R}^{d-1,1})$ due to the exponential factor. From \cref{fouana}, we see that 
\begin{equation}
    \partial_{k_I^{\mu}} T(x,k_I) = 0\,,
    \label{pki}
\end{equation}
therefore, we can drop the dependence on $k_I$, i.e. $T(x)$. We conclude that $L(k)$ is the Laplace transform a distribution $T(x)\in \mathcal{D}'(\mathbb{R}^{d-1,1})$, with $\mathring{\rm FLC}\subset D(T)$. 

Next, we prove that $T$ is in fact a tempered distribution. Notice that we cannot simply set $k_I=0$ in \cref{Txki}, since this is on the boundary of the tube. Instead, we show that the $T$ satisfy the bound in \cref{Tbound} for a generic test function $\phi\in\mathcal{D}(\mathbb{R}^{d-1,1})$ with compact support. Such distribution will necessarily be tempered (see \cref{dprimSprim}). We simply have
\eq{
\bk{T(x)}{\phi(x)}=\bk{e^{-k_I\cdot x} \, \mathcal{F}^{-1}[L](x)}{\phi(x)}=\bk{L(k)}{\mathcal{F}[e^{-k_I\cdot x}\phi](-k_R)}\,,
}
where we have used the definition in \cref{Txki}, and properties \cref{mul} and \cref{Four}, noting that the exponential is a multiplicator for compactly supported $\phi(x)$. Since, $L(k)$ is a tempered distribution, it satisfies the bound \cref{Tbound}. So we have
\begin{equation}
    |T[\phi]| \leq K \, \norm{\mathcal{F}[e^{-k_I\cdot x}\phi](-k_R)}_{m,n}\,.
\end{equation}
We will show that the right-hand side can be written as $K'\,\norm{\phi}_{m',n'}$, and therefore $T(x)$ is in fact a tempered distribution. To see this, we use the definition of the norm, in \cref{normphi}, and use properties of the Fourier transform to obtain
\eq{
\begin{aligned}
|T[\phi]|&\leq K\max_{|\alpha|\leq m,\beta\leq n}\sup_{k_R\in \mathbb{R}^{d-1,1}}\big|\mathcal{F}[\p_x^{\alpha}(x^\beta\,e^{-k_I\cdot x}\phi(x))](-k_R)\big|\\
&\leq K\max_{|\alpha|\leq m,\beta\leq n}\int\dd[d]{x}\big|\p_x^{\alpha}(x^\beta\,e^{-k_I\cdot x}\phi(x))\big|\\
&\leq K\, |{\rm supp}(\phi)|\max_{|\alpha|\leq m,\beta\leq n}\sup_{x\in {\rm supp}(\phi)}\big|\p_x^{\alpha}(x^\beta\,e^{-k_I\cdot x}\phi(x))\big|\,,
\end{aligned}\label{Tphib}
}
where in the second step we have used Cauchy–Schwarz inequality, while the last line is an upper bound on the integration over ${\rm supp}(\phi)$. Clearly, the last part in \cref{Tphib}, can be written as a sum over terms of the form $|k_I^a e^{-k_I\cdot x}||x^b \p^c\phi|$, with $|a|,|c|\leq|\alpha|$ and $|b|\leq|\beta|$. This is almost what we need to prove, except for the factors $|k_I^a e^{-k_I\cdot x}|$ and the dependence on the size of the support of the test function, $|{\rm supp}(\phi)|$. The strategy is, given ${\rm supp}(\phi)$, to decompose it into small boxes, i.e. ${\rm supp}(\phi)=\sum_i B_i$ where $B_i$ is a box in $\mathbb{R}^{d-1,1}$ centered at $x_i\in\mathbb{R}^{d-1,1}$ with unit volume. Then we use arbitrariness of $k_I$, due to \cref{pki}, to put an upper bound on $|k_I^a e^{-k_I\cdot x}|$, by choosing $k_I^\mu\cdot x_i^\mu=1$ for all $B_i$. Then it is easy to see that $|k_I^a e^{-k_I\cdot x}|\leq e$ in every $B_i$. As a result, we conclude our desired resul\footnote{More precisely, first, we restrict our attention to test functions $\phi$ such that $\mathrm{supp}(\phi)=\{x\in \mathbb{R}^{d-1,1}|X^{\mu}-2\leq |x^{\mu}|\leq X^{\mu}\>,\forall \mu\}$ for some $X^{\mu} \in \mathbb{Z}_0\times \mathbb{Z}_0\times ...\times \mathbb{Z}_0$. Thus, we can set $k_I^0=1/(X^0+|\vec{X}|),|\vec{k}_I|=1/(2X^0+2|\vec{X}|)\>,\forall \mu$. Then we see that 
\begin{equation}\nonumber\begin{aligned}
    |k_I^a e^{-k_I\cdot x}|&\leq |(k_I^0)^{a_0+a_1+...+a_{d-1}}e^{k_I^0(x^0+|\vec{x}|)}|\leq \bigg| \frac{1}{(X^0+|\vec{X}|)^{a_0+a_1+...+a_{d-1}}}e^{(x^0+|\vec{x}|)/(X^0+|\vec{X}|)}\bigg| \leq e\,.
\end{aligned}\end{equation} 
We can then extend the estimate to any compactly supported test function. To do this we consider the partition of unity $\{e_{i_0...i_d}\}$ subordinated to a covering of $\mathbb{R}^{d,1}$ by sets of the kind $\{x\in \mathbb{R}^{1,d}|k^{\mu}-2< |x^{\mu}|<k^{\mu}\>,\forall \mu\}$ and decompose any test function $\phi\in \mathcal{D}(\mathbb{R}^{1,d})$ wrt $\{e_{i_0...i_d}\}$. In particular, for any compactly supported test function, only a finite number of elements of $\{e_{i_0...i_d}\}$ will be nonzero on $\mathrm{supp}[\phi]$, let's call this number $N_{\phi}$. We can find the largest $N_{\phi}$ scanning over all the set $\mathcal{D}(\mathbb{R}^{1,d})$ of compactly supported test functions and replace $A$ in the previous estimate with $A'\equiv 2* \sup_{\phi \in \mathcal{D}(\mathbb{R}^{1,d})}N_{\phi}$, generalizing the estimate to any test function $\phi \in \mathcal{D}$. But since this holds for any test function $\phi \in \mathcal{D}(\mathbb{R}^{1,d})$, we can extend $g$ to a distribution in $S'(\mathbb{R}^{1,d})$. Hence, $g$ is tempered.}

\begin{equation}
    |T[\phi]|\leq K^{'}\norm{\phi}_{m',n'}\,,
\end{equation}
hence $T$ must be tempered. 

It only remains to show that $T$ is supported in FLC. The strategy is almost identical as what described above. For any test function that has compact support outside FLC, we choose $k_I$ such that the upper estimate \cref{Tphib} is as small as possible. In particular, if the maximum in \cref{Tphib} is attained for $x^0<0$, then we simply choose $k^0\to\infty$. If it is attained at $|\vx|>x^0>0$, we choose $\vk_I$ such that $\vk_I\cdot\vx>0$, and $|\vk_I|\to\infty$ keeping the ratio $1<k^0/|\vk_I|<|\vx|/x^0$ fixed. As a result, the upper bound in \cref{Tphib} can be chosen to be exponentially suppressed due to the factor $e^{-k_I\cdot x}$. Thus it must vanish for any $x\notin {\rm FLC}$.   

\section{More on subtractions}\label{subapp}

Since $P(z)$ is arbitrary, apart from its degree which must be greater than a minimum value, we can choose it to be real on the real axis, i.e. choose its coefficients to be real. In this case, it is easy to see that $Q(z)$ will be real as well; zeros of $P(z)$ appear in complex conjugate pairs, and therefore, in \cref{Q} for real $z$, the second term is simply the complex conjugate of the first.\footnote{As mentioned above, for the special case that a zero lies on the real line, we push it to the UHP or LHP by $i\epsilon$. Then we simply use Sokhotski–Plemelj identity to simplify. We note that, then, the integral in \cref{1dispsub} will  be the Cauchy principal value with respect to such points.} This is also evident from \cref{reim1dispsub}. As a result, in addition to the imaginary part on the real axis, we require $n$ real numbers to reconstruct the function in the UHP. As an example, for $n=1$, we choose $P(z)=z-\omega_0-i\epsilon$, with real $\omega_0$, and we get\footnote{Taking $\omega_0\to\infty$, the relation simplifies as $G(z)={\rm Re}\,G(\infty)+\frac{1}{\pi}\int\dd{\z}\frac{{\rm Im}\,G(\z)}{(\z-z-i\epsilon)}$. Here we are assuming that the imaginary part goes to zero at infinity. An equivalent derivation uses the fact that when $G(z)\to{\rm const.}$ at infinity, the arc contribution can be easily calculated directly. See \cite{Creminelli:2024lhd} for an application of this form. 
}
\eq{
G(z)={\rm Re}\,G(\omega_0)+\frac{z-\omega_0}{\pi}{\rm PV}_{z_0}\int\dd{\z}\frac{{\rm Im}\,G(\z)}{(\z-z-i\epsilon)(\z-\omega_0)}\,.
}
Moreover, notice that we can choose the $P(z)$ to be positive on the real line. For instance, we can choose all the zeros on the imaginary axis, i.e. $z_a=iy_0$ and $z_b=-iy_0$ for $y_0>0$, hence $P(z)$ consists of product of terms $(z^2+y_0^2)$. Indeed if the minimum allowed $n$ is odd, we need to choose slightly more complicated $P(z)$ with degree $n+1$ to ensure its positivity. 

It is possible that ${\rm Im}\,G$ goes to zero at infinity faster compared to $G$, and therefore convergence of its integral requires a polynomial with lower degree. Consider as an example a real polynomial plus some analytic function that goes to zero at infinity. In this case, the imaginary part decays at infinity and hence the polynomial in the denominator of \cref{1dispsub} is not needed for convergence. More generally, let us assume that $P(z)=P_0(z)P_1(z)$ such that only $P_0(z)$ is required by the convergence of the integral over the imaginary part in \cref{1dispsub}, while $P_1(z)$ was introduced to ensure $G(z)/P(z)\to0$ at infinity. Starting from \cref{1dispsub}, it is easy to separate the two terms
\eq{
\frac{P_0(z)P_1(z)}{\pi}\int\dd{\z}\frac{{\rm Im}\,G(\z)}{(\z-z)P_0(\z)P_1(\z)}=\frac{P_0(z)}{\pi}\int\dd{\z}\frac{{\rm Im}\,G(\z)}{(\z-z)P_0(\z)}-\frac{1}{\pi}\int\dd{\z}\frac{R(\z,z){\rm Im}\,G(\z)}{P_0(\z)P_1(\z)}\,,
\label{ImGimprov}
}
where in the numerator, we have taken $P_1(z)$ inside the integration and replaced it with $P_1(\z-(\z-z))=P_1(\z)-(\z-z)R(\z,z)$, in which $R(\z,z)$ is a polynomial in both $\z$ and $z$. The integral in the second term is convergent since $R$ has degree one less than $P_1$. Notice that the last term is a polynomial in $z$, hence its effect in \cref{1dispsub} is simply to change $Q(z)$. 

\bigskip
Let us move on to multiple variables. In this case one has to distinguish between the asymptotic behavior of $z_1$ and $z_2$. As an example, assume $\tG(z_1,z_2)$ goes to zero for $z_1\to\infty$, while approaching a constant for $z_1\to\infty$ for fixed $z_2$. In that case, single subtraction is required for $z_1$, not for $z_2$. More generally, we consider $\tG(z_1,z_2)/(P_1(z_1)P_2(z_2))$ with polynomials $P_1(z_1)$ and $P_2(z_2)$ of appropriate degree, e.g. $n_1$ and $n_2$ respectively such that the ratio goes to zero at infinity.\footnote{Notice that the most general polynomial $P(z_1,z_2)$ is not necessarily separable in the variables. We simplify the discussion by taking it to be of the form $P_1(z_1)P_2(z_2)$.} The logic is the same as before, although the algebra is slightly involved. We begin by writing a dispersion relation for $z_1$ followed by $z_2$. The presence of the subtraction points necessarily implies terms like $\tG(z_{a_1},z_2)$ and $\tG(z_1,z_{a_2})$, where $z_{a_1}$ and $z_{a_2}$ are respectively zeros of $P_1$ and $P_2$ that lie in the UHP. This time we write a single dispersion relation for each term. We also perform the complex conjugation trick (as in \cref{2dcauchy2}) appropriately in every step to get the imaginary part in the integrand, which involves with zeros $z_{b_1}$ and $z_{b_2}$ in the LHP. Finally, we obtain 
\eq{
\spl{
\tG(z_1,z_2)=&Q_{12}(z_1,z_2)+Q_1(z_1,z_2)+Q_2(z_1,z_2)\\&-\frac{iP_1(z_1)P_2(z_2)}{2\pi^2}\int\dd{\z_1}\dd{\z_2}\frac{{\rm Im}\,\tG(\z_1,\z_2)}{(\z_1-z_1-i\epsilon)(\z_2-z_2-i\epsilon)P_1(\z_1)P_2(\z_2)}\,,
}\label{2dispsub}}
in which $Q_1(z_1,z_2)$ is a polynomial in $z_1$, of degree $n_1-1$, but not necessarily in $z_2$
\eq{
Q_1(z_1,z_2)=\frac{P_1(z_1)P_2(z_2)}{\pi}\sum_{a_1}\frac{1}{P_1'(z_{a_1})(z_1-z_{a1})}\int\dd{\z_2}\frac{{\rm Im}\,\tG(z_{a1},\z_2)}{(\z_2-z_2-i\epsilon)P_2(\z_2)}\,,
}
with $z_{a_1}$ being zeros of $P_1$ in the UHP. Similar expression holds for $Q_2(z_1,z_2)$, with the role of the two variables $z_1$ and $z_2$ exchanged
\eq{
Q_2(z_1,z_2)=\frac{P_1(z_1)P_2(z_2)}{\pi}\sum_{a_2}\frac{1}{P_2'(z_{a_2})(z_2-z_{a2})}\int\dd{\z_1}\frac{{\rm Im}\,\tG(\z_1,z_{a_2})}{(\z_1-z_1-i\epsilon)P_1(\z_1)}\,,
}
which is polynomial in $z_2$, of degree $n_2-1$, but not in $z_1$. Finally, $Q_{12}(z_1,z_2)$ is a polynomial in both $z_1$ and $z_2$, of degree $n_1-1$ and $n_2-1$ respectively,  
\eq{
\spl{
Q_{12}(z_1,z_2)=P_1(z_1)P_2(z_2)\Bigg[&\sum_{a_1,a_2}\frac{\tG(z_{a_1},z_{a_2})}{P_1'(z_{a_1})(z_1-z_{a_1})P_2'(z_{a_2})(z_2-z_{a_2})}\\&+\sum_{b_1,b_2}\frac{\tG(z_{b_1}^*,z_{b_2}^*)^*}{P_1'(z_{b_1})(z_1-z_{b_1})P_2'(z_{b_2})(z_2-z_{b_2})}\\
&+\sum_{a_1,b_2}\frac{\tG(z_{a_1},z_{b_2}^*)^*}{P_1'(z_{a_1})(z_1-z_{a_1})P_2'(z_{b_2})(z_2-z_{b_2})}\\
&+\sum_{b_1,a_2}\frac{\tG(z_{b_1}^*,z_{a_2})^*}{P_1'(z_{b_1})(z_1-z_{b_1})P_2'(z_{a_2})(z_2-z_{a_2})}\Bigg]\,,
}
}
in which $z_{a_1}$ and $z_{a_2}$ are respectively zeros of $P_1$ and $P_2$ in the UHP, and $z_{b_1}$ and $z_{b_2}$ are respectively zeros of $P_1$ and $P_2$ in the LHP. Notice that neither of $Q_{12}$, $Q_{1}$, and $Q_{2}$ are seperable in $z_1$ and $z_2$. The expression in \cref{2dispsub} suggests that for generic choice of subtractions points, we need $\tG\big(\{z_{a_1},z_{b_1}^*\},\{z_{a_2},z_{b_2}^*\}\big)$, the imaginary part on the real plane ${\rm Im}\,\tG(\z_1,\z_2)$, and on surfaces ${\rm Im}\,\tG(z_{a_1},\z_2)$ and ${\rm Im}\,\tG(\z_1,z_{a_2})$ in $\mathbb{C}^2$. If we take the polynomials $P_1$ and $P_2$ to have real coefficients, then the zeros appear in complex conjugate pairs. In contrast to the single-variable case, the polynomial $Q_{12}$ and the mixed terms $Q_1$ and $Q_2$ will in general have nonzero imaginary part.

We can simplify \cref{2dispsub} by choosing all the zeros in the LHP. Therefore, the mixed terms $Q_1$ and $Q_2$ will be absent and the polynomial term $Q_{12}$ will contain only the sum over zeros in the LHP, i.e. $z_{b_1}$ and $z_{b_2}$. In this case both $Q_{12}$ and the double-integral term in \cref{2dispsub} will be complex functions. This form is conceptually appealing since it shows that, similar to the single-variabel case, we can reconstruct the function in the PUHP knowing the imaginary part on the real plane in addition to the value at the $n_1\times n_2$ subtraction points. A special case, in which the polynomials are real is that $z_{b_1}=x_{b_1}-i\epsilon$ and $z_{b_2}=x_{b_2}-i\epsilon$ for real $x_{b_1}$ and $x_{b_2}$.

Similar to the un-subtracted case, the imaginary part must satisfy a set of integral constraints. One expects this from \cref{imcomut}, which holds regardless of the asymptotic behavior of the imaginary part. The easiest way to obtain the constraints is to take derivatives of $\tG(z_1,z_2)$ with respect to $z_1$ and $z_2$, for sufficient number of times to obtain the appropriate asymptotic behavior.\footnote{Equivalently, we choose polynomials with higher order zeros, i.e. $P(z)=(z-z_0)^n$.} The resulting imaginary part must then satisfy the integral constraints from an un-subtracted dispersion relation given before in \cref{imcons}. More specifically, for $n_1$ and $n_2$ subtractions, we must have
\eq{
\p_{z_1}^{n_1}\p_{z_2}^{n_2}{\rm Im}\,\tG(z_1,z_2)=-\frac{1}{\pi^2}{\rm PV}_{z_1,z_2}\int\dd{\z}_1\dd{\z}_2\frac{\p_{\z_1}^{n_1}\p_{\z_2}^{n_2}{\rm Im}\,\tG(\z_1,\z_2)}{(\z_1-z_1)(\z_2-z_2)}\,,
\label{imconssubz1z2}
}
or in terms of Hilbert transform, $\mathbb{H}_{z_1}{\rm Im}\,\tG^{(n_1,n_2)}(\z_1,z_2)=\mathbb{H}_{z_2}{\rm Im}\,\tG^{(n_1,n_2)}(z_1,\z_2)$.

\section{Linear maps without rotational symmetry}\label{AppMapsNonRot}

In this appendix we provide an explicit construction of the linear maps between the $\td$-dimensional ${\rm PUHP}$ and the set $\mathbb{R}^{\td-1,1}+i P$, $P\subset \mathring{{\rm FLC}}$. 

We begin by considering $S^{\td-2}$ in $\vec{k}_I$-space and choose $\td$ unit vectors $\{\vec{n}_i\}$ which identify the vertices of a regular $(\td-1)$-simplex inscribed in the unit sphere. A concrete way to proceed is to pick the first member of the family to be $\vec{n}_1=(1,0,...,0)$ and then construct the other vectors by adding progressively a new component in the remaining directions. Being more explicit, we choose
\begin{equation}
\begin{aligned}
   & \vec{n}_2=(a_{2,1},a_{2,2},0,...,0)\\
   &\vec{n}_3=(a_{3,1},a_{3,2},a_{3,3},0,...,0)\\
   &...\\
   &\vec{n}_{\td-1}=(a_{\td-1,1},...,a_{\td-1,\td-1})
    \end{aligned}
\end{equation}
and determine the components $a_{j,l}$ enforcing the conditions
\begin{equation}
    \vec{n}_i \cdot \vec{n}_j =
    \begin{cases}
        1 & \text{if } i = j \\
        -1/(\td-1) & \text{if } i \neq j
    \end{cases}\>,
\end{equation}
which guarantee maximal symmetry and uniform angular separation between all vectors. The last member of the family will be automatically given by $\vec{n}_{\td}=-\sum_{l=1}^{\td-1}\vec{n}_l$, as the simplex is centered at the origin of $\mathbb{R}^{\td-1}$. This choice ensures that all edges of the simplex have equal length, and its vertices partition the sphere into $\td$ congruent spherical regions. 

We then extend the family $\{\vec{n}_i\}$ to a set of $\td$-vectors in Minkowski space
\begin{equation}
    v_i=\frac{1}{\td}\begin{bmatrix} 1 \\ \vec{n}_i/\alpha\end{bmatrix}\>,\>\alpha>1.
\end{equation}
with $\alpha>1$. Notice that 
\begin{equation}
    \eta_{\mu \nu}v^{\mu}_i  v^{\nu}_j=\frac{1}{\td^2}\begin{cases}
        -\frac{\alpha^2-1}{\alpha^2}& \text{if } i = j \\
        -1+\frac{1}{\alpha^2(\td-1)}& \text{if } i \neq j \\
    \end{cases}\>,
\end{equation}
so every $v_i$ is future-directed and timelike. As mentioned in the main text, the directions $v_i$ identify the edges of a hyperpyramid whose constant-$\omega_I$ sections are given by regular $(\td-1)$-simplexes inscribed in the sphere $S^{\td-2}_{\omega_I/\alpha}$ centered at $(\omega_I,\vec{0})$. The parameter $\alpha$ controls the spread of the hyperpyramid inside $\mathring{\rm{FLC}}$. Notice that there is nothing special about the specific construction of the set $\{\vec{n}_i\}$: acting with an $SO(\td-1)$ transformation on each vector $\vec{n}_i$ yields an equally good family of candidates. Thus, we can finally write the family of maps from the tube $\mathbb{R}^{\td-1,1}+iP$ to the $\td$-dimensional ${\rm PUHP}$ as
\begin{equation}
    \omega = \frac{1}{\td} \sum_{i=1}^{\td} z_i, \quad
    \vec{k} = \frac{1}{\td\alpha}\sum_{i=1}^{\td} R^a_{\>b}\,n^b_i \, z_i, \quad R \in SO(\td-1).
\end{equation}
Here $R^a_{\>b}$ explicitly parametrizes the freedom to rotate the hyperpyramid along the $\omega_I$-axis (or equivalently, to perform an $SO(\td-1)$ rotation of each constant-$\omega_I$ section of the hyperpyramid). 
The total parameter space of the family thus consists of:
\begin{itemize}
    \item One real parameter $\alpha>1$ controlling the opening angle of the hyperpyramid.
    \item $\frac{(\td-1)(\td-2)}{2}$ parameters describing the spatial rotation $R \in SO(\td-1)$.
\end{itemize}
By varying these $ (4+\td^2-3\td)/2$ parameters, one can map the $d$-dimensional ${\rm PUHP}$ to different hyperpyramids contained within $\mathring{\rm FLC}$, ultimately covering the entire interior of the analyticity region. 
\fg{
\centering
\adjustbox{valign=c}{\begin{tikzpicture}[scale=1.75, >=latex, line join=round, line cap=round]

\coordinate (O)  at (0,0,0);
\coordinate (k1) at (2,0,0);
\coordinate (k2) at (0,2,0);
\coordinate (k3) at (0,0,3);

\draw[dashed] (O) circle [x radius=1, y radius=1];

\coordinate (A) at (1,0,0);
\coordinate (B) at (-0.33,0.94,0);
\coordinate (C) at (-0.33,-0.47,0.82);
\coordinate (D) at (-0.33,-0.47,-0.82);

\draw[thick] (A) -- (B) -- (C) -- cycle;
\draw[] (A) -- (D) -- (C) -- cycle;
\draw[] (D) -- (B) -- (A) -- cycle;
\draw[] (B) -- (C) -- (D) -- cycle;

\fill[green!20, opacity=0.25] (A) -- (B) -- (C) -- cycle;
\fill[green!20, opacity=0.25] (A) -- (D) -- (C) -- cycle;
\fill[green!20, opacity=0.25] (D) -- (B) -- (A) -- cycle;
\fill[green!20, opacity=0.25] (B) -- (C) -- (D) -- cycle;

\draw[dashed] (O) -- (1,0,0);
\draw[dashed] (1,0,0) -- (1.5,0,0);
\draw[dashed] (O) -- (0,0.725,0);
\draw[dashed] (0,0.725,0) -- (0,1.05,0);
\draw[dashed] (O) -- (0,0,1);
\draw[dashed] (0,0,1) -- (0,0,1.9);

\draw[dashed] (O) circle [x radius=1.5, y radius=1.5];

\coordinate (E) at (1.5,0,0);
\coordinate (F) at (-0.5,1.41,0);
\coordinate (G) at (-0.5,-0.71,1.22);
\coordinate (H) at (-0.5,-0.71,-1.22);

\draw[thick] (E) -- (F) -- (G) -- cycle;
\draw[] (E) -- (H) -- (G) -- cycle;
\draw[] (H) -- (F) -- (E) -- cycle;
\draw[] (F) -- (G) -- (H) -- cycle;

\fill[yellow!20, opacity=0.25] (E) -- (F) -- (G) -- cycle;
\fill[yellow!20, opacity=0.25] (E) -- (H) -- (G) -- cycle;
\fill[yellow!20, opacity=0.25] (H) -- (F) -- (E) -- cycle;
\fill[yellow!20, opacity=0.25] (F) -- (G) -- (H) -- cycle;

\draw[->, thick] (0,0,1.9) -- (k3) node[left] {$k^3_{I}$};
\draw[->, thick] (0,1.05,0) -- (k2) node[left] {$k^2_{I}$};
\draw[->, thick] (1.5,0,0) -- (k1) node[right] {$k^1_{I}$};

\node[gray!70!black] at (0.5,-0.95,0.2) {$S^{2}$};
\node[gray!70!black] at (1.5,1.2,0.2) {$S^2_{3/2}$};

\end{tikzpicture}}
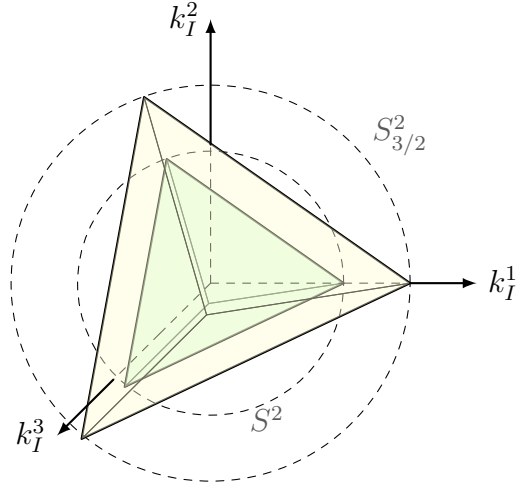
\caption{Comparison between constant-$\omega_I$ sections in $\td=4$. The constant-$\omega_I$ sections of the hyperpyramids are regular tetrahedrons inscribed in $S^2_{\omega_I/\alpha}$. Setting $\omega_I=2$, the green and yellow tetrahedrons are obtained by the choices $\alpha=2$ and $\alpha=4/3$ respectively.}
\label{PyramidsApp}
}

\vfill\footnotesize
\bibliographystyle{klebphys2}
\bibliography{refs}

\end{document}